\documentclass[a4,12pt]{article}
\pdfoutput=1

\usepackage{jheppub}
\usepackage{caption}
\usepackage{subcaption}
\usepackage{lscape}
\usepackage{makecell}
\usepackage{amsmath,amssymb,amsfonts, graphics,graphicx,amsfonts,amscd,epsf,epsfig,dsfont,color,braket}

\newcommand{\be}{\begin{equation}}
\newcommand{\ee}{\end{equation}}
\newcommand{\nn}{\nonumber}
\newcommand{\p}{\partial}
\newcommand{\Tr}[1]{\:{\rm Tr}\,#1}

\allowdisplaybreaks[4]

\title{Nonperturbative test of the Maldacena-Milekhin conjecture for the BMN matrix model}

\author[a]{Stratos Pateloudis,}
\author[b]{Georg Bergner,}
\author[a]{Norbert Bodendorfer,}
\author[c]{Masanori Hanada,}
\author[d,e,f,g]{Enrico Rinaldi,}
\author[a]{Andreas Sch\"{a}fer}

\affiliation[a]{University of Regensburg, Institute of Theoretical Physics,Universit\"{a}tsstrasse 31, D-93053 Regensburg, Germany}
\affiliation[b]{University of Jena, Institute for Theoretical Physics,Max-Wien-Platz 1, D-07743 Jena, Germany}
\affiliation[c]{Department of Mathematics, University of Surrey, Guildford, Surrey, GU2 7XH, United Kingdom}
\affiliation[d]{Physics Department, University of Michigan, Ann Arbor, MI 48109, United States}
\affiliation[e]{Theoretical Quantum Physics Laboratory, Cluster of Pioneering Research, RIKEN, Wako, Saitama 351-0198, Japan}
\affiliation[f]{Interdisciplinary Theoretical \& Mathematical Science Program (iTHEMS), RIKEN, Wako, Saitama 351-0198, Japan}
\affiliation[g]{Center for Quantum Computing (RQC), RIKEN, Wako, Saitama 351-0198, Japan}

\abstract{
	We test a conjecture by Maldacena and Milekhin for the ungauged version of the Berenstein-Maldacena-Nastase (BMN) matrix model
	by lattice Monte Carlo simulation. The numerical results reproduce the perturbative and gravity results in the limit of large and small flux parameter, respectively, and are consistent with the conjecture.}

\begin{document}

\maketitle

\section{Introduction}\label{sec:intro}
The holographic principle claims that quantum gravity can be described by a dual non-gravitational theory. 
AdS/CFT duality~\cite{Maldacena:1997re} provides us with concrete realizations of the holographic principle. 
But in fact, neither AdS nor CFT is crucial. AdS/CFT duality is a special case of gauge/gravity duality that admits non-AdS/non-CFT duality.
As yet other examples of the duality emerge, it became natural to ask if the gauge-singlet constraint is crucial on the QFT side. 
Indeed, this is something that gained more popularity since the appearance of the Sachdev-Ye-Kitaev (SYK) model~\cite{SYpaper,Kitaev:2015}. 
The SYK model is described by quantum mechanics of $N$ fermions with random coupling between them and it does not have a gauge symmetry. 
Despite the lack of gauge symmetry, fermion bilinears that look like ``singlets" appear to have natural dual descriptions in an emergent two-dimensional spacetime~ \cite{Witten:2016iux}.
The same story holds for the Gurau-Witten tensor models~\cite{Gurau:2011xp,Witten:2016iux} in which the gauge-singlet constraint can be either imposed or not. 

Maldacena and Milekhin~\cite{Maldacena:2018vsr} discussed whether the gauge-singlet constraint is necessary for more traditional types of gauge/gravity duality. 
They considered the D0-matrix model~\cite{Banks:1996vh,deWit:1988wri} and its ungauged version, and conjectured that the difference between the gauged model (with singlet constraint) and the ungauged model (without singlet constraint) is exponentially small at low temperature. In particular, they conjectured that the same gravity dual describes the low-energy dynamics of the gauged and ungauged models. Numerical simulation of the D0-matrix model~\cite{Berkowitz:2018qhn} provided results consistent with the conjecture.

There is a natural generalization of the D0-brane matrix model keeping maximal\\ supersymmetry, called the BMN matrix model~\cite{Berenstein:2002jq}. Maldacena and Milekhin considered the ungauged version of the BMN matrix model as well. In this paper, we test their conjecture for the BMN matrix model by employing numerical methods.
In addition to the original motivation coming from the Gurau-Witten tensor model, yet another motivation, in this case, comes from quantum simulations~\cite{Gharibyan:2020bab}. The gauge-invariant Hilbert space consisting only of singlet states is rather complicated, and merely writing down the orthonormal basis is already a difficult task. This problem can be avoided by introducing the extended Hilbert space that contains non-singlet states. A potential worry arises if the additional non-singlet degrees of freedom introduce other unexpected technical issues. Specifically, if many light modes emerge, they can easily be excited and lead to a large error. If the conjecture by Maldacena and Milekhin is correct, such light modes do not exist and hence the use of the extended Hilbert space can be rather straightforward.

The BMN matrix model contains the 't Hooft coupling constant $\lambda$, the deformation parameter $\mu$, and the effective dimensionless coupling constant is $g_{\rm eff}=\lambda/\mu^3$. This, together with temperature, spans the phase diagram of the model, describing different regimes. For works regarding the phase diagram see e.g \cite{Bergner:2021goh, Dhindsa:2022vch, Schaich:2022duk}.  In particular, we will be interested in the strong coupling regime where a dual description by weakly-coupled gravity exists.

The paper is organized as follows: in Sec.~\ref{sec:action} we are defining the models, while in Sec.~\ref{sec:gravityduals} we describe the gravity duals. In Sec.~\ref{MMconj} we give more details for the gauged and ungauged versions of the models and we discuss the conjecture. The numerical analysis and the main results for the BMN model~\cite{Berenstein:2002jq} follow in Sec.~\ref{sec:numeric_analysis}. We discuss observables obtained from the partition function and their difference between the gauged and ungauged versions.  In Sec.~\ref{sec:conclusions} we conclude, while supplementary results are contained in the appendices.

\section{Maldacena-Milekhin conjecture}
In this section, we review the conjecture made by Maldacena and Milekhin~\cite{Maldacena:2018vsr}. Firstly, we specify the theories under consideration in Sec.~\ref{sec:action}. 
Then, in Sec.~\ref{sec:gravityduals}, we introduce the dual gravity description. 
The details of the conjecture are described in Sec.~\ref{MMconj}. 
\subsection{Matrix models under consideration}\label{sec:action}
\subsubsection{BFSS Matrix Model}
\hspace{0.51cm}
We start by discussing the BFSS matrix model on an Euclidean circle with the \\circumference $\beta$.  
For bosonic fields, we always impose the periodic boundary condition. 
When the boundary condition for the fermion is anti-periodic, $\beta$ is the inverse of the temperature, $\beta=1/T$. 

This model consists of nine $N\times N$ bosonic hermitian matrices $X_M$ ($M=1,\cdots,9$), sixteen 
fermionic matrices $\psi_\alpha$ ($\alpha=1,\cdots,16$) and the gauge field $A_t$. Moreover $\gamma_{\alpha\beta}^M (M=1,\cdots, 9)$ are the $16\times 16$ left-handed parts of the gamma matrices in $(9+1)$-dimensions. This theory arises as a dimensional reduction of $(9+1)$-dimensional super Yang-Mills theory or $(3+1)$-dimensional maximal super Yang-Mills theory to $(0+1)$ dimension.

Both $X_M$ and $\psi_\alpha$ are in the adjoint representation of the $U(N)$ gauge group, and the covariant derivative $D_t$ acts on them as 
$D_tX_M = \partial_t X_M -i[A_t,X_M]$ and $D_t\psi_\alpha = \partial_t\psi_\alpha -i[A_t,\psi_\alpha]$. 
The action is given by
\be
S_{BFSS}=
\frac{N}{\lambda}\int_0^\beta dt\ {\rm Tr}\left\{ \frac{1}{2}(D_t X_M)^2 -\frac{1}{4}[X_M,X_N]^2 +\frac{1}{2}\bar{\psi}^\alpha\gamma^{10} D_t\psi_\alpha -\bar{\psi}^\alpha\gamma^M_{\alpha\beta}[X_M,\psi^\beta]
\right\}. 
\ee
The equations of motion for the gauge field  $A_t$ give rise to the Gauss constraint
\be \label{singletconstraint}
G=\frac{i N}{2\lambda} (2[D_tX_M, X_M]+[\psi_\alpha,\psi_\alpha])=0.
\ee 
In the operator formalism, the gauge-singlet constraint on physical states emerges due to the integration over $A_t$. See Sec.~\ref{sec:singlet-constraint}.
\subsubsection{BMN Matrix Model}
\hspace{0.51cm}
The plane-wave deformed theory, which is called the BMN matrix model \cite{Berenstein:2002jq}, is given by \footnote{Our normalization for mass is different from Refs.~\cite{Berenstein:2002jq,Costa:2014wya} by a factor 3.}
\begin{eqnarray}
S_{BMN}=S_{BFSS}+\Delta S, 
\end{eqnarray}
where 
\begin{align}\nn
\Delta S\label{BMNaction}
&=
\frac{N}{\lambda}\int_0^\beta dt\ {\rm Tr}\Big\{
\frac{\mu^2}{2}\sum_{i=1}^3X_i^2
+\frac{\mu^2}{8}\sum_{a=4}^9X_a^2 
+
i\sum_{i,j,k=1}^3\mu\epsilon^{ijk}X_iX_jX_k+\frac{3i\mu}{4}\bar{\psi}^\alpha\gamma_{123\alpha}{}^\beta\psi_\beta
\Big\} .
\nonumber\\
\end{align}
The extra terms appearing in the action are mass terms for bosons, fermions and interaction terms. We can take $\gamma_{123}$ as\footnote{This is the $16\times 16$ representation in ten dimensions. In general, the $\gamma^I, I=1,\cdots,10$ matrices are $16\times16$ sub-matrices of the $32\times 32$ ten-dimensional Gamma matrices $\Gamma^I$.} 
\be 
\gamma_{123}=\begin{pmatrix}
	-i\mathbf{1}_2\otimes\mathbf{1}_4&0\\
	0&i\mathbf{1}_2\otimes\mathbf{1}_4
\end{pmatrix},
\ee 
which simplifies the fermionic mass term to (see also Appendix \ref{sec:Hamiltonian_splitting})
\be 
\frac{3i\mu}{4}\bar{\psi}^\alpha\gamma_{123}\psi_\alpha=\frac{3\mu}{2}\bar{\psi}^\alpha\psi_\alpha.
\ee 
In addition, these new terms result in a new class of vacua labelled by representations of the $SU(2)$ group. In other words, matrices that minimize the potential in addition to the trivial ones (i.e, $X_i=0=X_a=\psi_\alpha$) can be written in the form 
\be \label{newvacua}
\psi_\alpha=0,\qquad X_a=0\ \ {\rm for}\ \ a = 4,\cdots, 9, \qquad X_i=\mu J_i\ \ {\rm for}\ \  i=1,2,3,
\ee   
where $J_i$ are the generators of ${\rm SU}(2)$.
In the limit $\mu\to 0$, the deformation terms vanish and one expects the above model to converge to the BFSS model. This however assumes that there is no phase transition between the models and indeed evidence until now supports this idea. 
Note also that the singlet constraint \eqref{singletconstraint} is not affected by the deformation. 
One can construct an effective, dimensionless coupling constant via
\be \label{eq:dimensionless_coupling_mu}
g_{\rm eff}^{(\mu)}:=\frac{\lambda}{\mu^3},~~[\lambda]=({\rm energy})^3,~~[\mu]=({\rm energy})^1. 
\ee 
we can also define another dimensionless quantity
\be \label{eq:dimensionless_coupling_r}
g_{\rm eff}^{(r)}:=\frac{\lambda}{r^3},~~[\lambda]=({\rm energy})^3,~~[r]=({\rm energy})^1. 
\ee 
These $g_{\rm eff}^{(\mu)}$ and $g_{\rm eff}^{(r)}$ control the regimes of validity of the two descriptions which can be  put into the following language
\begin{itemize}
	\item Supergravity region: $g_{\rm eff}^{(\mu)},g_{\rm eff}^{(r)}\gg 1$\
	\item Matrix model region: $g_{\rm eff}^{(\mu)},g_{\rm eff}^{(r)}\ll1$
\end{itemize} 

\begin{figure}[h!]
		\centering
		\includegraphics[scale=0.4]{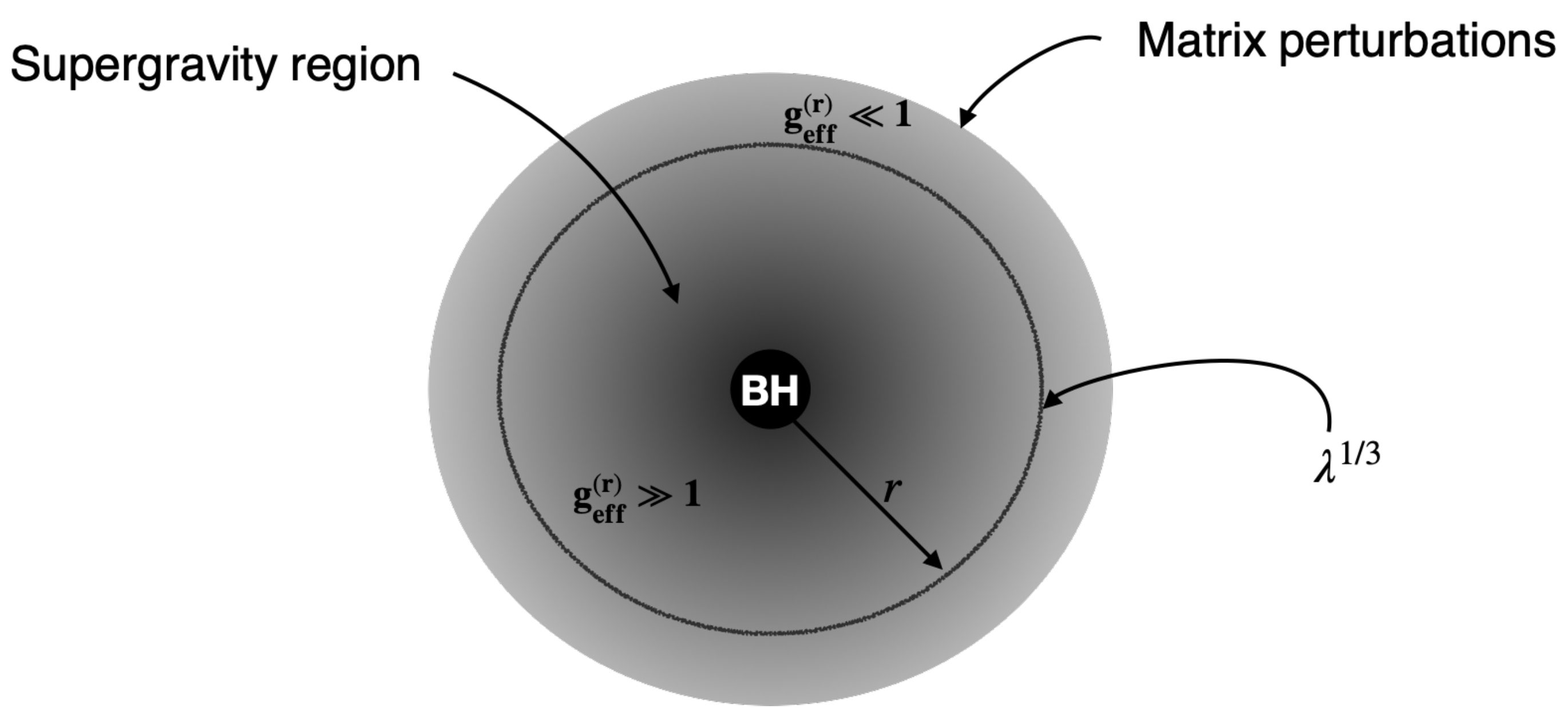}
		\caption{A pictorial representation indicating the gravity and matrix model side. The parameter that controls the different regimes is the effective, dimensionless couplings $g_{\text{eff}}^{(\mu)}=\frac{\lambda}{\mu^3}$ and $g_{\text{eff}}^{(r)}=\frac{\lambda}{r^3}$. Near the boundary the supergravity description is not valid while one can apply matrix perturbation theory with expansion parameters $g_{\text{eff}}^{(\mu)}, g_{\text{eff}}^{(r)},$. The supergravity solution can be trusted up to $r\lesssim\lambda^{1/3}$. This figure is based on \cite{Maldacena:2018vsr}.}
		\label{validfig}
	\end{figure}

Note that this $r$ appearing in \eqref{eq:dimensionless_coupling_r} controls the regimes of the duality in a similar spirit with the holographic spatial dimension appearing in the usual $AdS/CFT$ (see Fig.~\ref{validfig}). In particular, as we will see in the gravity analysis this $r$ corresponds to energy scales in supergravity but let us briefly clarify this point. It will turn out (see  e.g Eq.~\eqref{BFSSdual}) that $r$ controls the size of the eight-sphere $\mathbf{S^8}$ in the 't Hooft large $N$ limit. Small $r$ means small energies deep in the bulk (even inside the black hole) and very large $r$ means near the "boundary" where the matrix model perturbation is performed (see again Fig.~\ref{validfig}). The trustable region for supergravity is discussed below Eq.~\eqref{eq:S8_curvature}. 

Being in the canonical ensemble, we study (thermodynamic) features of the models at a specific temperature, which results in specific energy of the system via $E=E(T)$. 
In this paper, we are mainly interested in the strong coupling limit of the model, which according to \eqref{eq:dimensionless_coupling_mu}, \eqref{eq:dimensionless_coupling_r}, and \eqref{eq:dimensionless_coupling_E} results in low energies and via the canonical ensemble at low temperatures. The temperature in the matrix models sets the circumference ($\beta=1/T$) of the Euclidean circle on which we put our models and is also connected with the Hawking temperature on the gravity side. We will comment more on this when we discuss the gravity dual picture in \eqref{eq:BFSS_energy}.  In both theories, since conformal symmetry is absent, the 't Hooft coupling $\lambda=g_{YM}^2N$  can be set to 1 by proper rescaling of time $t$ and matrices.
In other words, all dimensionful quantities can be made dimensionless by multiplying with appropriate powers of $\lambda$ (e.g, $\tau=T/\lambda^{1/3}$).  
In simulations we set $\lambda=1$. 
	
\subsubsection{Ungauged matrix model}
\hspace{0.51cm}
If we turn-off the gauge field $A_t$ in the BFSS or BMN matrix model, we obtain the \textit{ungauged} matrix model. Specifically, we define the ungauged action $S_{\rm ungauged}$ as 
\begin{align}
S_{\rm ungauged}[X,\psi]=S_{\rm gauged}[X,A_t=0,\psi]. 
\end{align} 
The thermal  partition functions of gauged and ungauged models are defined by  
\begin{align}
Z_{\rm gauged}&=\int\mathcal{D}[A_t]\mathcal{D}[X_M]\mathcal{D}[\psi_\alpha]e^{-S_{\rm gauged}[X,A_t,\psi]},
\label{eq:Z_gauged_path_integral}
\\
Z_{\rm ungauged}&=\int\mathcal{D}[X_M]\mathcal{D}[\psi_\alpha]e^{-S_{\rm ungauged}[X,\psi]}. 
\label{eq:Z_ungauged_path_integral}
\end{align}
Here, the time direction is Wick-rotated to the Euclidean signature and compactified to circumference $\beta=T^{-1}$. 
We impose the periodic boundary condition for bosonic fields $X$ and $A_t$, and antiperiodic boundary condition for fermionic fields $\psi$. 	
\subsubsection{Gauge-singlet constraint}\label{sec:singlet-constraint}
\hspace{0.51cm}
Let us discuss the meaning of the ungauging in terms of quantum states in the Hilbert space. 
By using the gauge-invariant Hilbert space $\mathcal{H}_{\rm inv}$ and the extended Hilbert space $\mathcal{H}_{\rm ext}$, thermal partition functions are written as (see e.g., Appendix~A.2 in Ref.~\cite{Rinaldi:2021jbg}) 
\begin{align}\label{eq:Z_gauged_H_ext}
Z_{\rm gauged}(\beta)
&=
\frac{1}{V_{{\rm SU}(N)}}\int_{{\rm SU}(N)} dg
{\rm Tr}_{\mathcal{H}_{\rm ext}}\left(\hat{g}e^{-\beta\hat{H}}\right) \\
\label{eq:Z_gauged_H_inv}&=
{\rm Tr}_{\mathcal{H}_{\rm inv}}\left(e^{-\beta\hat{H}}\right)
\end{align} 
and
\begin{align}\label{eq:Z_ungauged_H}
Z_{\rm ungauged}(\beta)
=
{\rm Tr}_{\mathcal{H}_{\rm ext}}\left(e^{-\beta\hat{H}}\right). 
\end{align} 
In \eqref{eq:Z_gauged_H_ext}, $V_{{\rm SU}(N)}$ is the volume of the SU($N$) gauge group, $\int_{{\rm SU}(N)} dg$ is the integral with the Haar measure, and $\hat{g}$ is the operator acting on the Hilbert space as the SU($N$) transformation corresponding to the group element $g\in{\rm SU}(N)$. By construction, $\frac{1}{V_{{\rm SU}(N)}}\int_{{\rm SU}(N)} dg\hat{g}$ acts as the projector from $\mathcal{H}_{\rm ext}$ to $\mathcal{H}_{\rm inv}$. 
In fact, \eqref{eq:Z_gauged_H_ext} is directly related to the path-integral formulation with gauge field $A_t$ described by the canonical partition function \eqref{eq:Z_gauged_path_integral}. The projection operator tells us that we should count gauge-equivalent states only once, and hence, we obtain \eqref{eq:Z_gauged_H_inv} from \eqref{eq:Z_gauged_H_ext}. 
The operator $\hat{g}$ is the counterpart of the Polyakov loop in the path-integral formalism. Integration over SU($N$) is the remnant of the path integral with respect to $A_t$.

\subsection{The gravity duals}\label{sec:gravityduals}

\subsubsection*{Gravity dual of gauged matrix models}
Let us start with the 't Hooft large-$N$ limit of the gauged D0-brane matrix model without deformation. We take the effective dimensionless coupling constant $\lambda T^{-3}$ to be large (which is equivalent to low temperature). 

The BFSS gravity dual at strong coupling is believed \cite{Itzhaki:1998dd} to be the charged black zero brane in type IIA supergravity formed by $N$ coincident D0-branes. The metric is given as 
\begin{align}
\label{BFSSdual}
\frac{ds^2}{\alpha'}&=-H(r)^{-1/2}f(r)dt^2+H(r)^{1/2}\left(\frac{dr^2}{f(r)}+r^2d\Omega_8^2\right),\\
\nn H(r)&= \frac{240\pi^5 \lambda}{r^7},~~\lambda=g_{YM}^2N,\\
\nn e^\phi&=\frac{(2\pi)^2}{240\pi^5}\frac{1}{N}\left(\frac{240\pi^5 \lambda}{r^3}\right)^{\frac{7}{4}},\\
\nn f(r)&=1-\left(\frac{r_0}{r}\right)^7.
\end{align}
There is a killing horizon at $g_{tt}=0$ that sets the temperature of the black IIA brane to 
\be \label{eq:temperature}
T=\frac{7}{4\pi\sqrt{240\pi^5 \lambda}}r_0^{\frac{5}{2}}.
\ee 
Knowing the temperature one can pursue a thermodynamic analysis and compare with the relevant quantities of the matrix model \cite{Itzhaki:1998dd, KlebanovEntropyOfNear, HyakutakeQuantumNear, HyakutakeQuantumMwave}. In particular the energy of the system is 
\be \label{eq:BFSS_energy}
E\simeq 7.41 N^2\lambda^{-3/5}T^{\frac{14}{5}},
\ee 
and the entropy is given by the first law as
\be 
S\simeq 11.53N^2\lambda^{-3/5}T^{\frac{9}{5}}.
\ee 
For the BFSS model the dimensionless effective coupling is constructed as 
\be \label{eq:dimensionless_coupling_E}
g_{\rm eff}^{(E)}=\frac{\lambda}{(E/N^2)^3},
\ee 
in addition to $g_{\rm eff}^{(r)}$, such that the supergravity region is given at $g_{\rm eff}^{(E)}\gg1$, e.g at low energies and via \eqref{eq:BFSS_energy} at low temperatures.
One could use $g_{\rm eff}^{(T)}=\frac{\lambda}{T^3}$ as well.

 Comparison between the BFSS matrix model and the black zero-brane was explored numerically using Monte-Carlo simulations for the internal energy of the theory accessing in this way in a non-perturbative fashion gauge/gravity duality and introducing $\alpha'$ corrections to eq.~ \eqref{eq:BFSS_energy} (see Refs.~\cite{Anagnostopoulos:2007fw,Catterall:2008yz} for the first simulations and Ref.~\cite{Berkowitz:2016jlq} for large-$N$ and continuum limit).  The trustable supergravity region demands the effective curvature of the eight-sphere to be small, which is given by the inverse sphere radius ($\mathcal{R}$) as
\be \label{eq:S8_curvature}
\kappa=\frac{1}{\alpha '\mathcal{R}}=H(r)^{-1/2}r^{-2}.
\ee  
As long as the effective curvature $\kappa$ is small, we can trust the supergravity description \eqref{BFSSdual} which is the case for $r\lesssim \lambda^{1/3}$ (see Fig.~\ref{validfig})).  In addition, we have to ensure that the dilaton on the horizon is small
\be 
e^\phi\Big|_{r=r_0}\ll 1,
\ee 
 which via \eqref{BFSSdual} and \eqref{eq:temperature} results in $\frac{\lambda}{T^3}\lesssim N^{10/7}$. Considering always the large $N$ limit the latter condition is always satisfied and combining all the conditions we have 
 \be 
 1\ll\frac{\lambda}{T^3}\ll N^{\frac{10}{7}}.
 \ee 

\subsubsection*{Gravity dual of ungauged matrix models}
The idea behind gauging a symmetry or not in this particular example might become more intuitive if we present it in the language of string theory. The gauge invariant (physical) states are singlets. These states correspond to closed strings and are constructed by acting with combinations of matrix operators on the vacuum state upon taking the trace over the gauge group indices\footnote{Here the gauge group is $SU(N)$ and by taking the trace we mean summing over the $SU(N)$ indices $\Tr{\hat{X}_I}=\sum_{i=1}^N\left(\hat{X}_I\right)^i_{}{}^{}_i$, 
	$\Tr{\hat{X}_I\hat{X}_J}=\sum_{i=1}^N\sum_{j=1}^N\left(\hat{X}_I\right)^i_{}{}^{}_j\left(\hat{X}_I\right)^j_{}{}^{}_i$, etc. 
	This is different from the trace over the Hilbert space, $\Tr_{\mathcal{H}_{\rm inv}}$ or $\Tr_{\mathcal{H}_{\rm ext}}$. 
}, i.e 
\be 
\ket{{\rm physical}}=\Tr\left({\hat{X}_{I_1}\hat{X}_{I_2}\cdots \hat{X}_{I_L}}\right)\ket{\text{\rm vacuum}}.
\ee  
On the other hand, the ungauged model allows some room for non-singlets without a trace 
\be 
\ket{\rm{physical}}={\hat{X}_{I_1}\hat{X}_{I_2}\cdots \hat{X}_{I_L}}\ket{{\rm vacuum}}.
\ee  
Even though the singlet sectors of both theories are the same, in the ungauged model there is a new sector that hosts non-singlets. The latter can intuitively be realized as an arbitrary long open string made out of $L$ bits whose endpoints can reach the boundary. Let us now discuss the continuum picture, e.g $L\to\infty$. In the bulk, this configuration of long strings is described by the gravity dual of non-supersymmetric Wilson loops and the difference with the usual supersymmetric Wilson loop is the fact that in the latter case the string obeys Neumann boundary conditions while in the former Dirichlet boundary conditions \cite{Maldacena:2018vsr}. This means that the tip of the non-singlet string can freely move in the bulk and reach even to the boundary. To be able to compare with gravity, we should ask what would be a natural cut-off such that we could approximate the energy of this massive string with supergravity. This is dictated by the validity of \eqref{BFSSdual}. Recalling that $[r]=\rm (energy)^1$ we may demand that $E_{\rm min}\sim\lambda^{1/3}$. This indeed would be a natural cut-off because when we have a pair of a string and an anti-string, we can arbitrarily lower their energy (length) by placing both its endpoints on the boundary. On the other hand, we cannot approximate its energy from the supergravity side in this case because the latter is not valid at the boundary. Hence, a natural non-zero value of minimal energy calculable on the gravity side is
\be 
E_{\rm min}\propto C_{\rm adj} \lambda^{1/3},
\ee   
with $C_{\rm adj}$ being a number of order one. The idea then is that the fate of non-singlet adjoint strings is to end on the boundary minimizing in this way its energy. Therefore, contributions from the energies of non-singlet states should be negligible in the large $N$ and low-temperature limit and the gravitational dual of the non-singlet strings in this regime is the same as that of the singlets in \eqref{BFSSdual}. This is the conjecture made in \cite{Maldacena:2018vsr} which we focus on in what follows. 

\subsection{Gauged vs ungauged: the conjecture}\label{MMconj}
The temporal component of gauge field $A_t$ is not dynamical. Its role is to impose the gauge-singlet constraint. 
Matrix models do not have spatial gauge field components $A_x,A_y,\cdots$, and hence, the only effect of a gauge symmetry is to impose the gauge-singlet constraint. The ungauged version of the model does not obey the singlet constraint. The $SU(N)$ symmetry is treated as a global symmetry of the system. 


In \cite{Maldacena:2018vsr}, Maldacena and Milekhin considered the BFSS matrix model ($\mu=0$) and BMN matrix model ($\mu>0$), and claimed that 
gauged and ungauged versions are essentially the same at large $N$ and strong coupling, 
in the sense that the contribution of the non-singlet sector in the partition function is negligible in the large $N$ limit.  The partition functions of the two models are given as 
\eqref{eq:Z_gauged_path_integral}, \eqref{eq:Z_gauged_H_ext}, 
\eqref{eq:Z_gauged_H_inv} for the gauged model, and  \eqref{eq:Z_ungauged_path_integral}, \eqref{eq:Z_ungauged_H} for the ungauged model.

Let us denote the difference of the gauged and ungauged free energies in the large $N$ limit as
\begin{align}
\Delta \mathcal{F}
&=-\log Z_{\text{gauged}}(\beta)+\log Z_{\text{ungauged}}(\beta)
\nonumber\\
&=\beta \mathcal{F}_{\rm ungauged}-\beta \mathcal{F}_{\rm gauged}
\nonumber\\
&=
N^2g(\beta)
\end{align}
up to $\frac{1}{N}$-corrections, 
where $g(\beta)$ is a function that depends on the parameter regimes of the model\footnote{We remind that we are measuring everything in terms of the t'Hooft coupling $\lambda=g_{YM}^2N$  which we set to one from now onwards.  In a general case it would also have a $\lambda$ dependence as $g(\lambda^{1/3}\beta)$.}. The factor $N^2$ comes from the 't Hooft counting.  

\noindent
Ref.~\cite{Maldacena:2018vsr} discussed the free energy difference in two different regimes and focused on the strong coupling regime.  The BMN model has two parameters that control the regimes of the system, that is $\mu$ and $T$.

\subsubsection*{The high temperature and weak coupling regime}
In this limit, the difference of the free energies is  
\be 
\beta\Delta \mathcal{F}
\simeq N^2\log(\mu\beta),~~~\frac{\lambda}{T^3}\ll 1,\quad\frac{\mu^3}{\lambda}\gg 1.
\ee 
This is the weak coupling regime at high temperatures and large $\mu$. At these temperatures, we do not have a bulk dual so we will not be interested in this regime. 

\subsubsection*{The low-temperature and strong coupling regime}
Under the presence of a gravitational dual the free energy difference is
\be \label{DeltafreeE}
\beta\Delta \mathcal{F}
\simeq N^2n_{\rm adj}e^{-C_{\rm adj}/T}, \quad\frac{\lambda}{T^3}\gg 1,\quad\frac{\mu}{T}\ll 1.
\ee 
This limit is a bit subtle. Note that for the BMN model it is not enough to consider just small temperatures, but one also has to consider the $\mu\to0$ limit to compare with gravity. The reason is that even though for finite $\mu$ there is a gravitational dual description it is given by a deformed geometry~\cite{Costa:2014wya}.

\noindent
From \eqref{DeltafreeE} we get the difference of energies $E=\frac{\partial(\beta \mathcal{F})}{\partial\beta}$ as
\be\label{DeltaE}
\Delta E=E_{\rm ungauged}-E_{\rm gauged}=N^2n_{\rm adj}C_{\rm adj}e^{-C_{\rm adj}/T}+\cdots, 
\ee
to lowest order in temperature and $\mu=0$. The consensus built in \cite{Maldacena:2018vsr} and \cite{Berkowitz:2018qhn} is that $n_{\rm adj}$ is the degeneracy of the lightest mode and $C_{\rm adj}$ its energy. This result seems now to be understood for the $\mu=0$ case.  Indeed at $\mu=0$, the factor $n_{\rm adj}$ is an $\mathcal{O}(1)$ integer, and $C_{\rm adj}$ is an $\mathcal{O}(1)$ positive number. Results of numerical simulation at $\mu=0$~\cite{Berkowitz:2018qhn} are consistent with $n_{\rm adj}=2$ and $C_{\rm adj}\simeq 1$.

\subsubsection*{The low-temperature and weak coupling regime}
The BMN model can be also weakly coupled at low temperatures, in contrast to the BFSS model. This is due to the effective coupling \eqref{eq:dimensionless_coupling_mu}. In the large $\mu$ region the BMN model admits a perturbative analysis \cite{Kim:2002if,   Dasgupta2002, Dasgupta:2002ru} with perturbation parameter \eqref{eq:dimensionless_coupling_mu}. The spectrum is discussed in Appendix \ref{sec:Hamiltonian_splitting} and the lightest mode is the one created by $B^\dagger_a$ acting on $\ket{0}$. Whether we take the trace or not corresponds to the decision between the gauged or ungauged theory.  In the latter case, there is no trace and the lightest mode is created by $B^\dagger_a\ket{0}$ with energy \eqref{eq:lowest_adjoint}. This is what we called $C_{\rm adj}$ above and we have six of them because we have six harmonic oscillators (one for each direction in the $SO(6)$ part) which gives the degeneracy. Also, in the low-temperature region, we can still use an exponential ansatz and we expect the perturbative result of the lightest mode to be given by 
\be \label{eq:lightest_mode}
E_{\rm adj}=N^2\cdot 6\cdot \frac{\mu}{2}e^{-\frac{\mu}{2}\frac{1}{T}}, \quad\frac{\lambda}{T^3}\gg1,\quad \frac{\mu^3}{\lambda}\gg 1.
\ee 
This is what we expect in the perturbative, low-temperature regime for the lightest mode of the theory. In fact, if we wish to study heavier modes or if we wish to be more precise we should consider the full $U(1)$ sector
given by 
\be \label{eq:full_U(1)}
\frac{E_{U(1)}}{N^2}=6\cdot\frac{\mu}{2}\cdot e^{-\frac{\mu}{2}\frac{1}{T}}+8\cdot\frac{3\mu}{4}\cdot e^{-\frac{3\mu}{4}\frac{1}{T}}+3\mu e^{-\frac{\mu}{T}}, \quad\frac{\lambda}{T^3}\gg1,\quad \frac{\mu^3}{\lambda}\gg 1,
\ee 
(see Table~\ref{vacuadegentable}) but as we discuss later on and we show in Fig.~\ref{fig:pert_C-D} at very low temperatures, the difference between the contribution of the lightest mode and that of the full $U(1)$ sector is negligible (at finite $\mu$).

\section{Numerical analysis}\label{sec:numeric_analysis}
In this section, we summarize the numerical analysis. We extrapolate the results to large $N$ and continuum limit to eliminate lattice artifacts and finite $N$ corrections. 
\subsection{Lattice regularization}\label{subsec:lattice-regularization}

Below, we explain the details of the lattice regularization. The action is the same as the one used in Ref.~\cite{Berkowitz:2016jlq}, except that also the deformation terms are added (see also \cite{Bergner:2021goh}).
\subsubsection{Gauge fixing}
\hspace{0.51cm}
The action of the gauged BMN matrix model given in \eqref{BMNaction} is invariant under the SU($N$) gauge transformation.
For numerical efficiency, we took the static diagonal gauge,
\begin{eqnarray}
A_t=\frac{1}{\beta}\cdot{\rm diag}(\alpha_1,\cdots,\alpha_N),
\qquad
-\pi<\alpha_i\le\pi.
\end{eqnarray}
Associated with this gauge fixing, we added the Faddeev-Popov term
\begin{eqnarray}
S_{\rm F.P.}
&= &
-
\sum_{i<j}2\log\left|\sin\left(\frac{\alpha_i-\alpha_j}{2}\right)\right|
\label{eq:Faddeev-Popov}
\end{eqnarray}
to the action.
\subsubsection{Lattice action}

We regularized the gauge-fixed continuum theory by introducing a lattice with $L$ sites and spacing $a$. The time parameter takes the values $t=a,2a,\cdots,La$.  Breaking the action into the bosonic part ($S_b$), the fermionic part ($S_f$), 
the Faddeev-Popov term $S_{\rm F.P.}$
and the mass deformation parts ($\Delta S_{b}$ and $\Delta S_{f}$), the respective lattice action is
\begin{eqnarray}
S_{b}
&= &
\frac{N}{2a}\sum_{t}\sum_{I=1}^9{\rm Tr}\left(D_+X_I(t)\right)^2
-
\frac{Na}{4}\sum_t\sum_{I,J=1}^9{\rm Tr}[X_I(t),X_J(t)]^2.
\end{eqnarray}
\begin{eqnarray}
S_{f}
=
iN\sum_{t}\Tr\bar{\psi}(t)
\left(
\begin{array}{cc}
0 & D_+\\
D_- & 0
\end{array}
\right)
\psi(t)
-
aN\sum_{t}\sum_{I=1}^9\bar{\psi}(t)\Gamma^I[X_I(t),\psi(t)],
\end{eqnarray}
\begin{eqnarray}
\Delta S_b
=
aN\sum_{t}\Tr\left\{
\frac{\mu^2}{2}\sum_{i=1}^3X_i(t)^2
+
\frac{\mu^2}{8}\sum_{a=4}^9X_a(t)^2
+
i\sum_{i,j,k=1}^3\mu\epsilon^{ijk}X_i(t)X_j(t)X_k(t)
\right\}
\nonumber\\
\end{eqnarray}
and
\begin{eqnarray}
\Delta S_f
=
\frac{3i\mu}{4}\cdot aN\sum_{t}\Tr\left(
\bar{\psi}(t)\gamma^{123}\psi(t)
\right),
\end{eqnarray}
where
\begin{eqnarray}
D_\pm\psi(t)
\equiv \mp\frac{1}{2}U^2\psi(t\pm 2a)\left(U^\dagger\right)^2
\pm 2U\psi(t\pm a)U^\dagger
\mp\frac{3}{2}\psi(t)
=
aD_t\psi(t) + O(a^3).
\nonumber\\
\end{eqnarray}
Here, $U={\rm diag}(e^{i\alpha_1/L},e^{i\alpha_2/L}\cdots,e^{i\alpha_N/L})$,
$-\pi\le \alpha_i<\pi$.
The Faddeev-Popov term $S_{\rm F.P.}$ is given in \eqref{eq:Faddeev-Popov}.

For the ungauged theory, we set $U$ to identity and remove the Faddeev-Popov term. 

\subsection{Simulation strategy}\label{subsec:strategy}
The BMN matrix model has a few nice features that make the numerical simulation easier than the BFSS matrix model. 
A technical challenge for the latter is the existence of flat directions. To tame them, we have to take $N$ very large. 
In the BMN matrix model, the flat directions are lifted due to the mass term in the flux deformation. 
Therefore, we can study the BMN matrix model at relatively small values of $N$. 
Furthermore, the condition number of the Dirac operator decreases as $\mu$ becomes large. This makes simulations more tractable in this regime. 

Although the flat directions are absent in the BMN matrix model, there is a somewhat related issue associated with the existence of multiple vacua.  
In this paper, we study physics around the trivial background ($X_i=0=X_a,\psi_\alpha=0$). 
For small $\mu$ and $T$, tunnelings between the trivial background and fuzzy spheres ($X_a=0=\psi_\alpha, X_i=\mu J_i$) can take place frequently. 
The potential barrier between them depends on  $\mu$ and  $N$. More precisely, by investigating the stability of a minimum of the potential $\frac{\p V}{\p X}=0$ and relating $X\sim r$ one can find that the barrier between the two backgrounds scales as $\sim\mu^4N^4$ (see Appendix \ref{app:SO3_potential}). Therefore, for very small $\mu$ and fixed $N$, the trivial background configuration can tunnel to a fuzzy sphere background and a distinction between the two is not possible.
On the other hand, at relatively large values of $N$, the simulation remains in the trivial background.

The fluctuation of each matrix entry is roughly given by\footnote{The value $0.6N^{-1/2}$ is the standard deviation. The order-one factor can be determined numerically, from the expectation value of $\sum_{I=1}^9{\rm Tr}X_I^2$, which is approximately $3.5N$~\cite{Berkowitz:2016jlq}. There are $9N^2$ matrix entries and  hence $\sqrt{\frac{3.5N}{9N^2}}\approx 0.6 N^{-\frac{1}{2}}$.}  $0.6N^{-1/2}$. Demanding that the radius of the maximum fuzzy sphere given by $r_{\rm max}=\mu\sqrt{s(s+1)}$ with spin $s=(N-1)/2$ is less than the matrix fluctuations  gives a constraint on values of $\mu$ 
\be 
\mu\leq\frac{1.2}{\sqrt{N(N^2-1)}},
\ee 
for which we cannot distinguish a fuzzy sphere configuration and matrix entry fluctuations at finite $N$.  For the values of $\mu$ we have used this constraint is always satisfied so we do not have to worry about it.  
It is furthermore essential in our simulations to consider a large enough number of lattice points $L$ to avoid lattice artefacts.

Hence, the strategy we used is the following:
\begin{itemize}
	\item For fixed $\mu$ we perform a series of simulations at different temperatures as well as varying $N$ and $L$ checking that we always stay in the trivial background of the confined phase.
	
	\item We extrapolate to the large $N$ and continuum limit of the theory for our observables.	
	
	\item
	We calculate the difference between gauged and ungauged observables such as the energy and those defined in \eqref{eq:R2} and \eqref{eq:F2}, for different temperatures at fixed $\mu$ and then fit exponentials of the form \eqref{DeltafreeE}, \eqref{R2exponential}, \eqref{F2exponential} respectively.
	
	\item  Finally, we take the limit $\mu\to 0$ to cross-check the results with Ref.~\cite{Berkowitz:2018qhn}. 
\end{itemize}

\begin{figure}[htbp]
	\begin{center}
		\rotatebox{0}{
			\scalebox{0.25}{
				\includegraphics{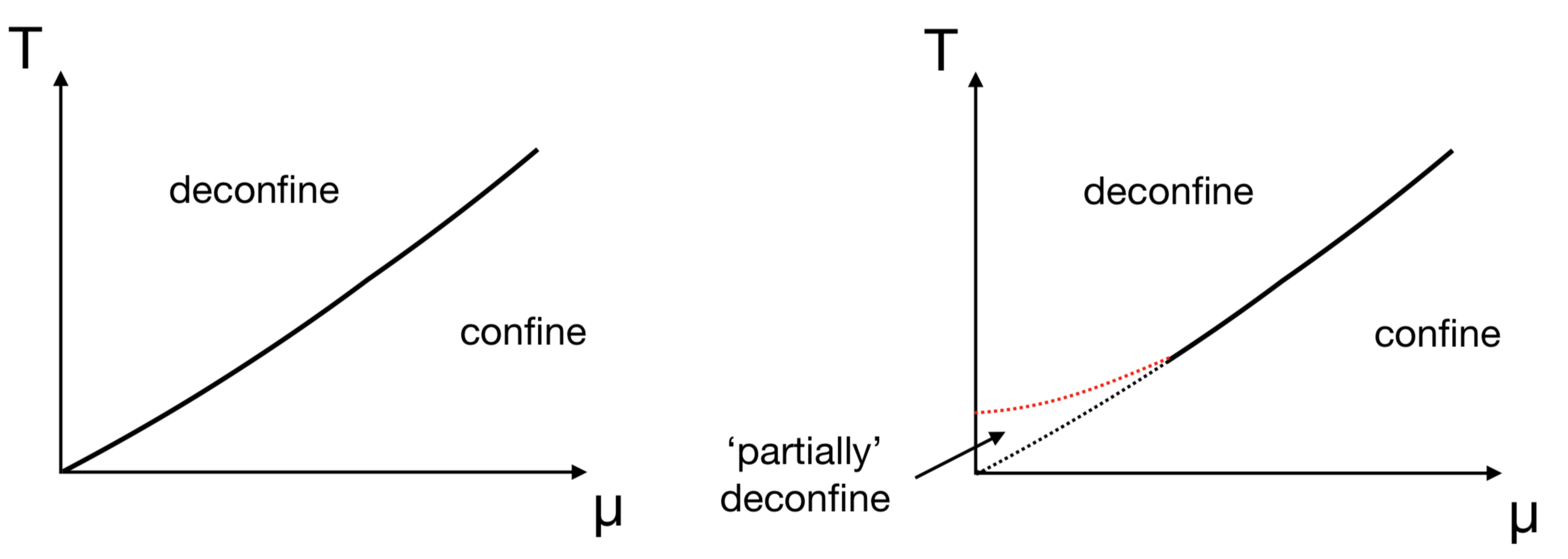}}}
	\end{center}
	\caption{
		Two kinds of conjectured phase diagrams of the gauged BMN matrix model. Most likely, the left one is correct. This figure is taken from \cite{Bergner:2021goh}.
	}
	\label{fig:conjectured-phase-diagram-BMN}
\end{figure}

In the past, there were two possible kinds of conjectured phase diagrams of the gauged BMN matrix model at $N=\infty$ (Fig.~\ref{fig:conjectured-phase-diagram-BMN}). 
The large-$\mu$ region admits perturbative calculations\cite{Furuuchi:2003sy,Spradlin:2004sx}
and the transition is found to be of first order. 
The small-$\mu$ region has been studied by using the dual gravity description~\cite{Costa:2014wya}
and recently via Monte-Carlo simulations~\cite{Bergner:2021goh}. A first-order transition has been established both analytically and numerically, while in addition a surprising possibility to study aspects of M-theory, like the Schwarzschild black hole has been established. 
Numerical simulation applies to the intermediate-$\mu$ region as well, and the results are consistent with a first-order phase transition~\cite{Bergner:2021goh}. 
Therefore, the left diagram of Fig.~\ref{fig:conjectured-phase-diagram-BMN} is most likely the correct one.
In this paper, we do not need to know the order of the phase transition as it does not affect our argument. 

According to the Maldacena-Milekhin conjecture, the gauged and ungauged theories are exponentially close at low temperatures. 
Therefore, we will focus on the confined phase. 
In it, observables like e.g the energy, are independent of temperature up to $1/N$ corrections. 
We can determine these temperature-independent values by taking the large-$N$ and continuum limit at some fixed value of $T$. 
To determine these values reliably, we will study different values of $T$ at each $\mu$.

To extrapolate to large $N$ and continuum regime, we use the fit ansatz
\begin{eqnarray}\label{energylargeNcontfit}
\frac{E(L,N)}{N^2}
=
\varepsilon_{0,0}
+
\frac{\varepsilon_{1,0}}{N^2}
+
\frac{\varepsilon_{0,1}}{L}
+
\frac{\varepsilon_{1,1}}{N^2L}
+
\cdots.  
\end{eqnarray}
This ansatz contains $1/L$ corrections to the continuum limit and $1/N^2$ corrections to the large $N$ limit. In practice we have considered only terms up to $\varepsilon_{0,1}$ to avoid over-fitting and considering the large $L$ values of our simulations. The extrapolations of other observables are of the same form.

The values of the flux that have been used are in the range $\mu=2,3,4,5$ while the range of temperatures is adjusted for each $\mu$ with  $T\in[0.2,1.1]$. In addition, the size of the matrices and the lattice spacing runs over $N=8, 12, 16$ and $L=12, 24, 48, 96$, respectively.  We extrapolate to the large $N$ continuum limit of the theory as explained above. This procedure is repeated both for the gauged and ungauged data.

\subsection{Energy of the system}\label{subsec:energy}
\begin{figure}[htbp]
	\centering
	\includegraphics[scale=0.4]{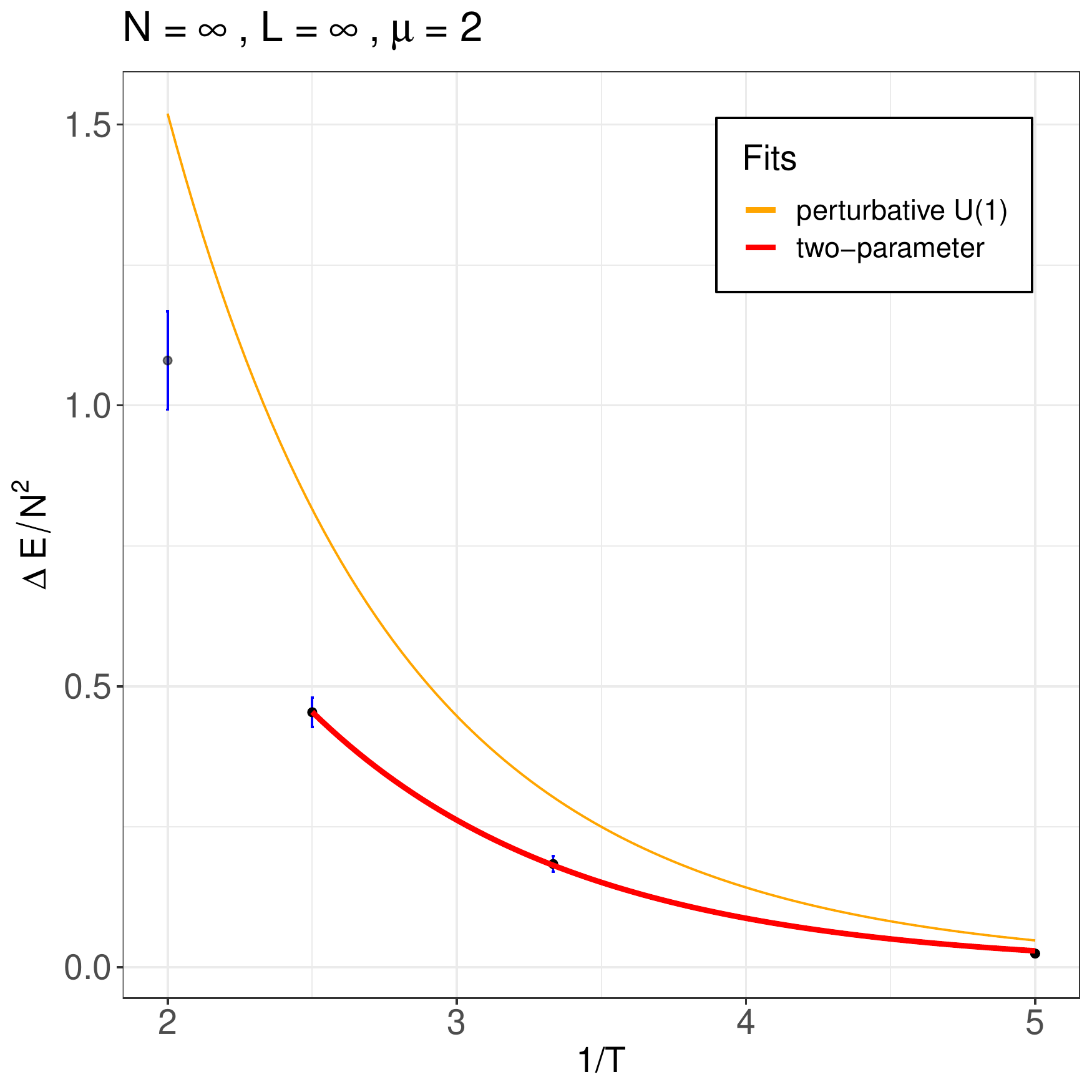}
	\includegraphics[scale=0.4]{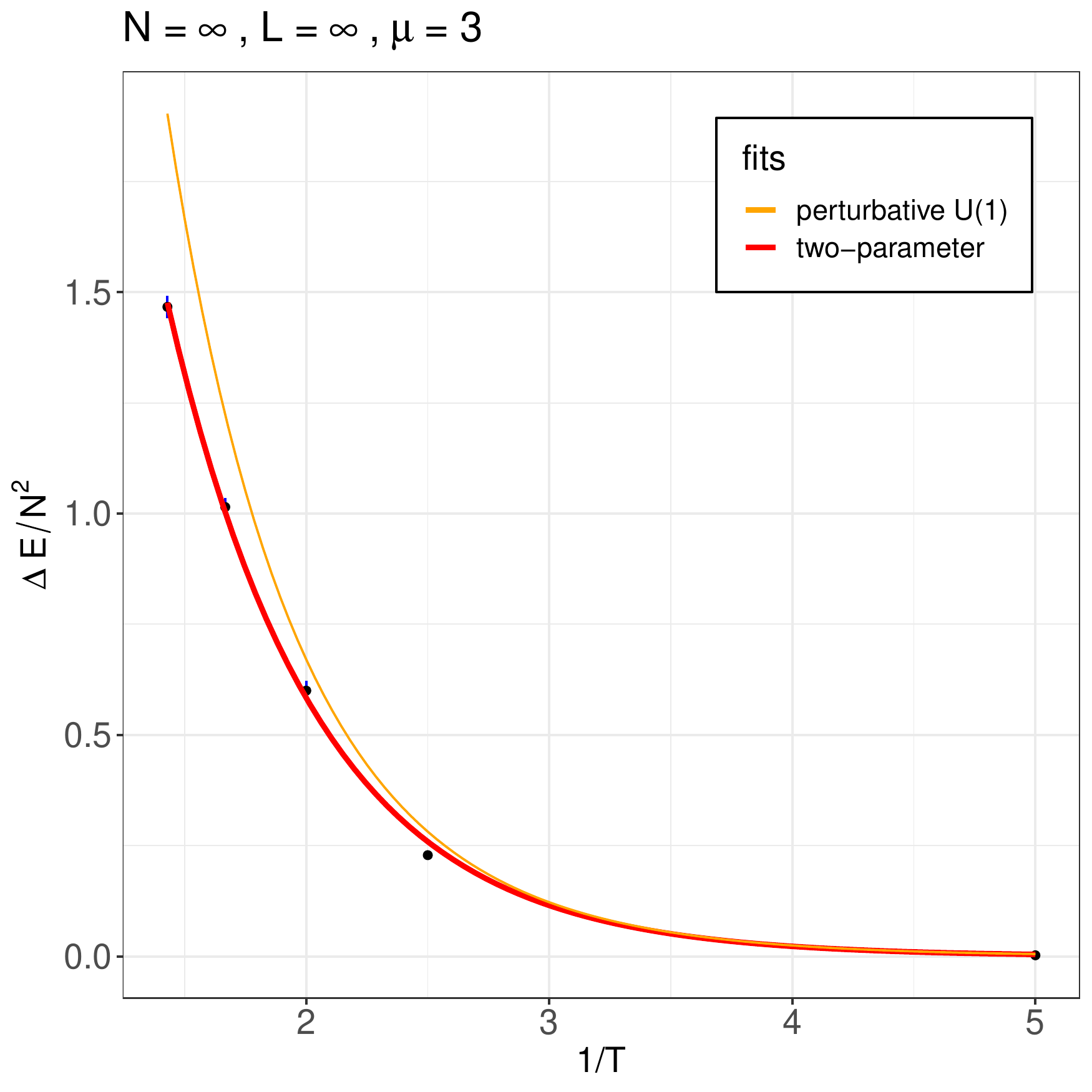}\\
	\includegraphics[scale=0.4]{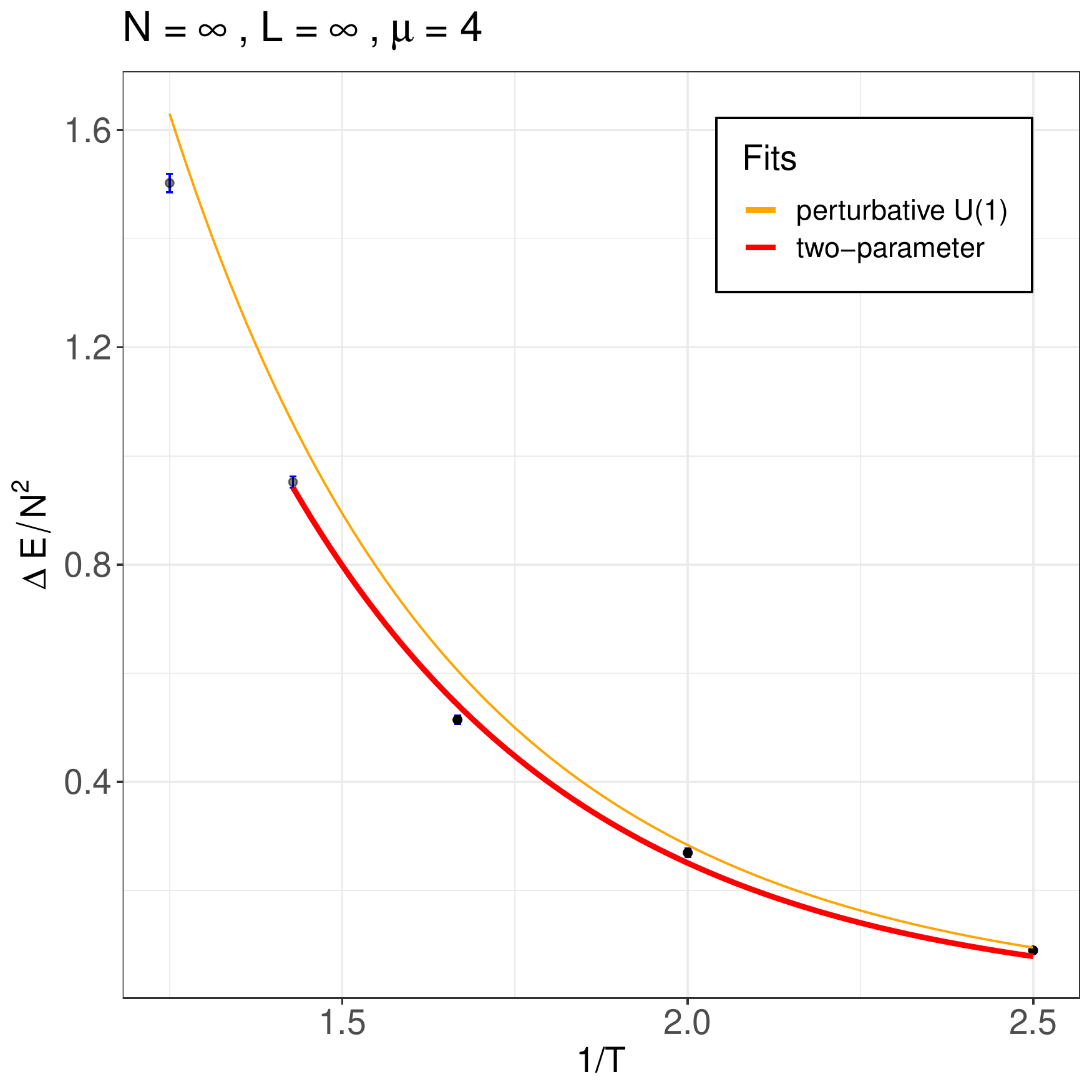}
	\includegraphics[scale=0.4]{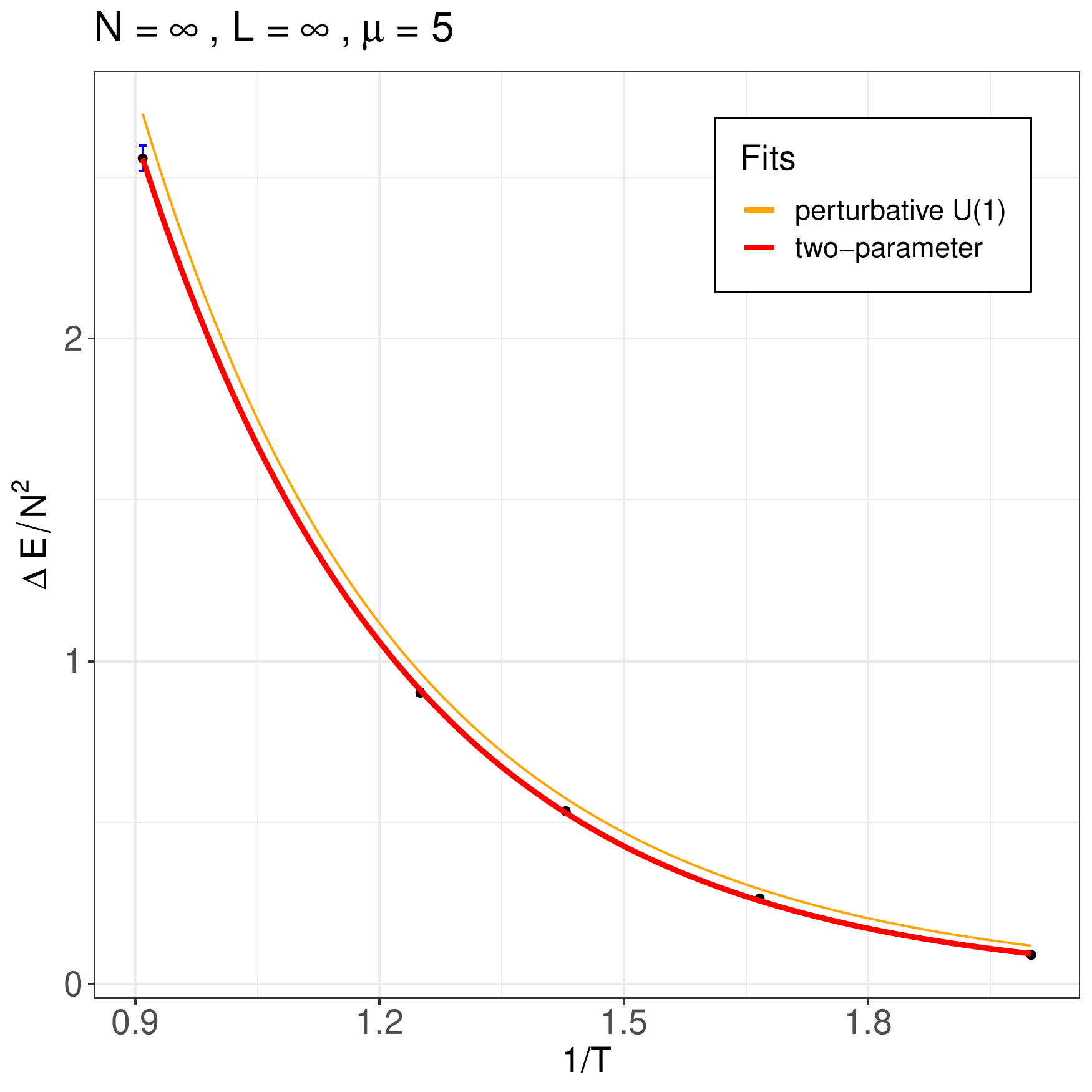}
	\caption{The large $N$ and continuum fits of the difference of energies \eqref{DeltaE} for different $\mu$ and the relevant fits. The two-parameter fit is given in \eqref{fit-E-2-parameter}. The perturbative curve is given by the contribution of the full $U(1)$ sector \eqref{eq:full_U(1)}. Errors are contained for all data points but are small.}
	\label{energy_g_vs_u}
\end{figure}

\begin{figure}[htbp]
	\centering
	\includegraphics[scale=0.4]{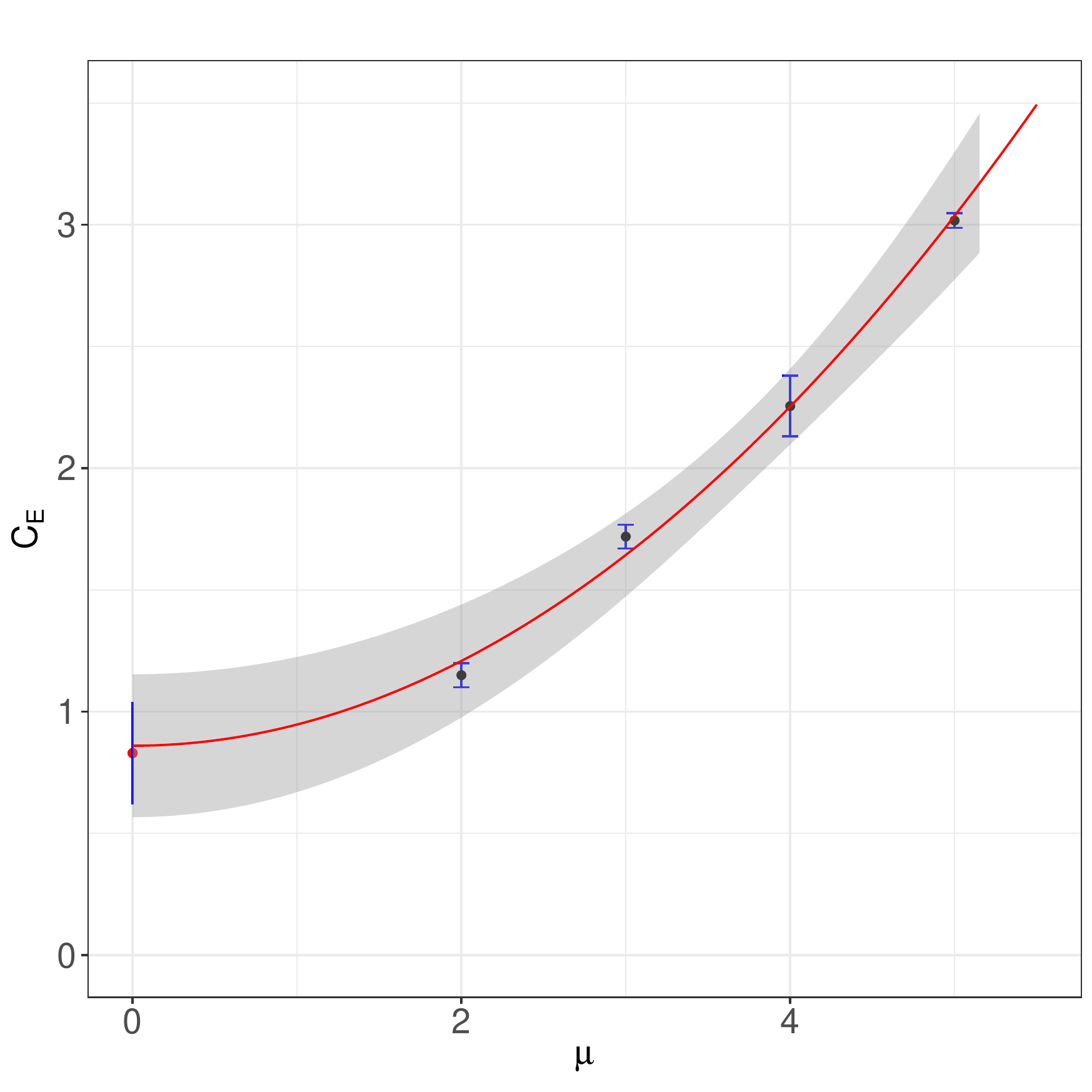}
	\includegraphics[scale=0.4]{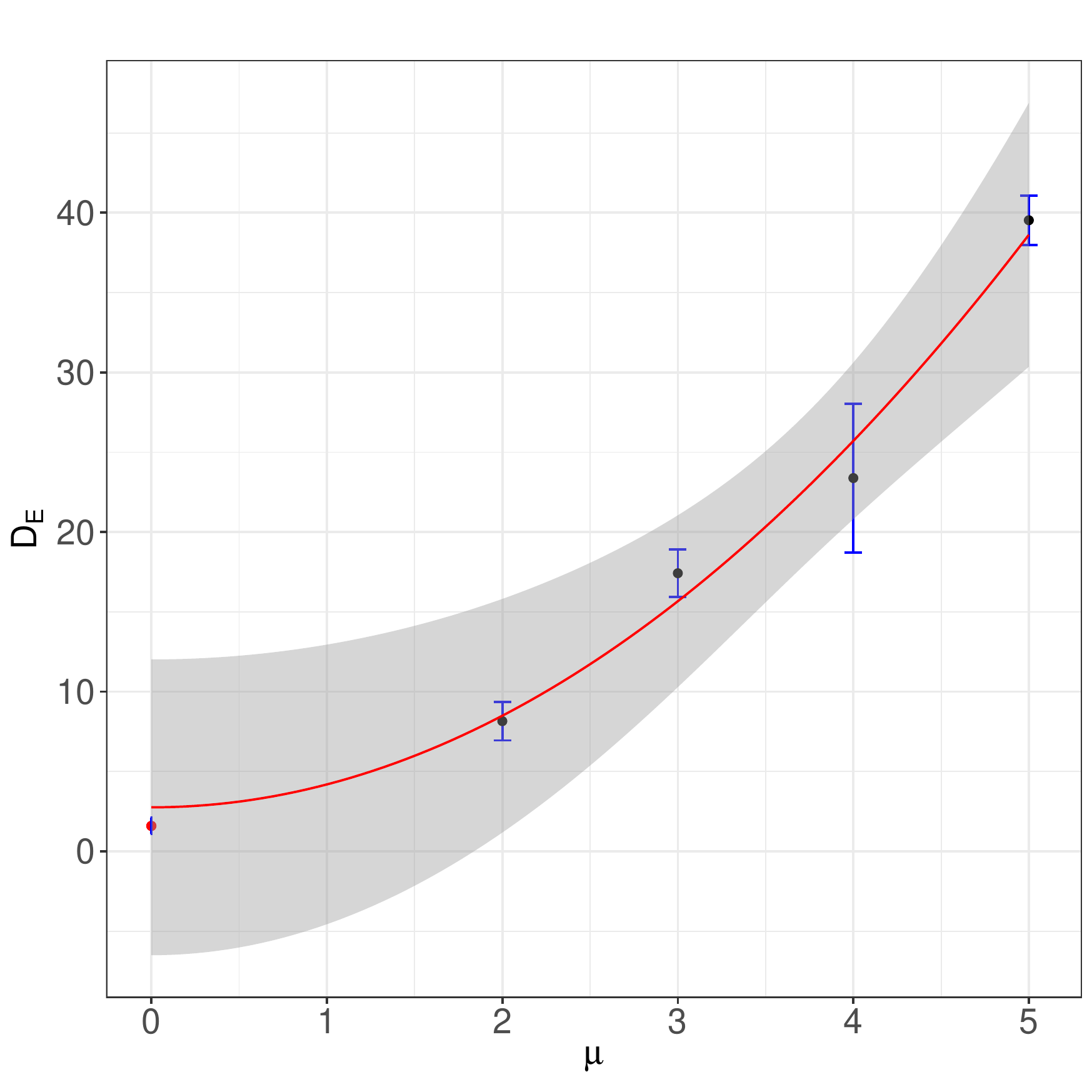}
	\caption{The dependence of the energy parameters on $\mu$ . [Left]: The fit is done by $C_E=0.860(68)+0.087(4)\mu^2$. [Right]: The fit curve is again given by $D_E=2.8(2.2)+1.44(14)\mu^2$ .Points at $\mu=0$ are from \cite{Berkowitz:2018qhn}. The extrapolated values agree within error bars with the points at $\mu=0$. }
	\label{fig:C_E_D_E_extrapolations}
\end{figure}

\begin{figure}[ht!]
	\centering
	\includegraphics[scale=0.5]{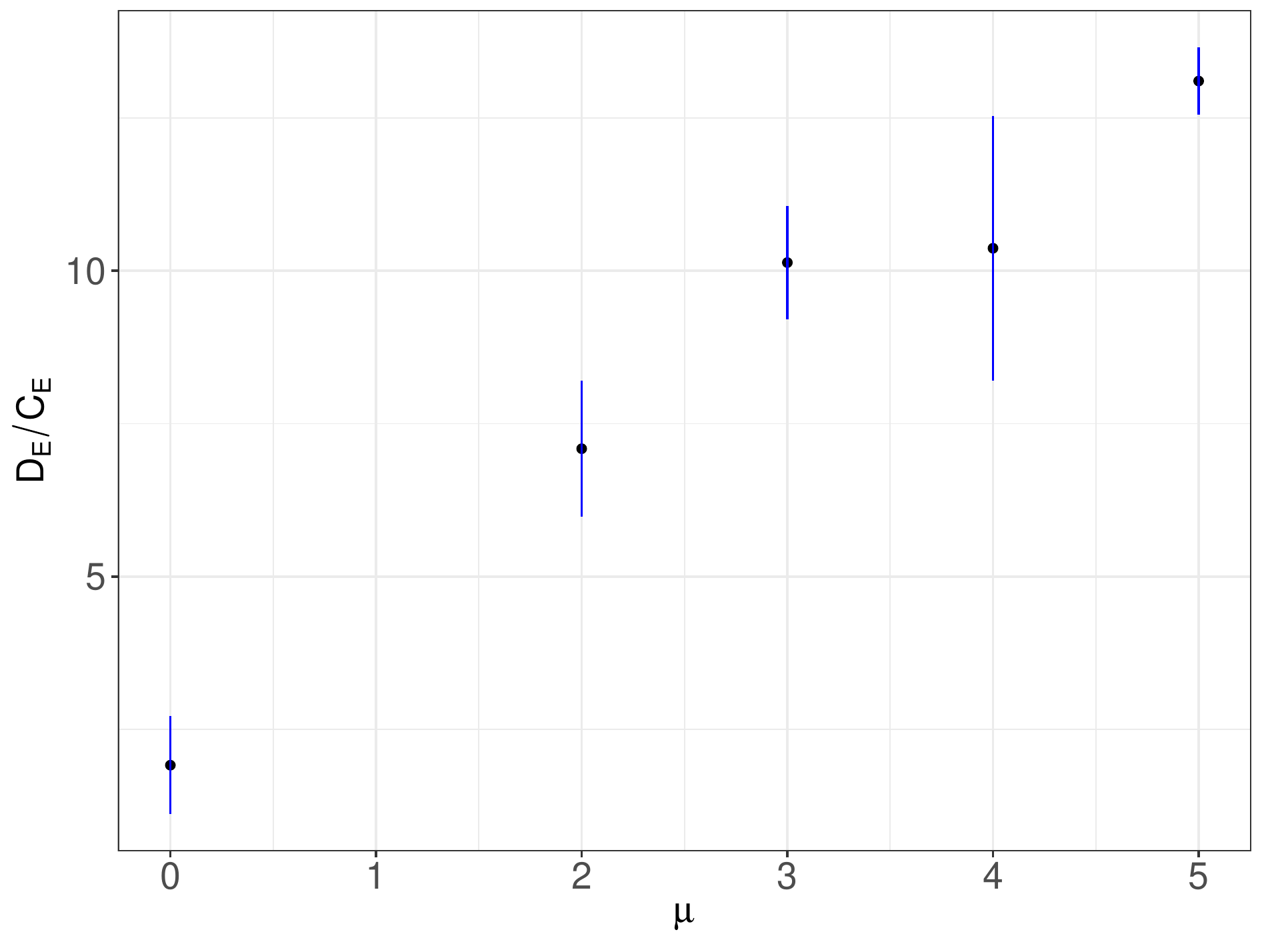}
	\caption{The change of ratio $D_E/C_E$ with respect to $\mu$. At $\mu=0$ we show the point measured with a two parameter fit in \cite{Berkowitz:2018qhn}, i.e $\frac{D_E}{C_E}\big|_{\mu=0}=1.91(78)$. }
	\label{fig:nadj}
\end{figure}

 In the BFSS limit $\mu=0$, Maldacena and Milekhin conjectured \eqref{DeltaE}, which we repeat here:
\be
\Delta E=E_{\rm ungauged}-E_{\rm gauged}= N^2n_{\rm adj}C_{\rm adj}e^{-C_{\rm adj}/T}+\cdots.  
\nonumber
\ee
Numerical simulation in Ref.~\cite{Berkowitz:2018qhn} suggests $n_{\rm adj}=2$ and $C_{\rm adj}\simeq 1$. 
We studied the behavior of $\Delta E$ at finite $\mu$ and low temperature. 
The results are given in Fig.~\ref{energy_g_vs_u} with perturbative results valid at large $\mu$ and two kinds of numerical fits. The punchline at this stage is that there is an exponential decay with $1/T$, which verifies the predicted functional form \eqref{DeltafreeE}. 
We have done fits with a different number of free parameters to check the reliability of our fit ansatz.
The red curves in Fig.~\ref{energy_g_vs_u} are fits of the two free parameters $D_{E,2}(\mu)$ and $C_{E,2}(\mu)$ according to
\be \label{fit-E-2-parameter}
\Delta E (\mu, T)=N^2D_{E,2}(\mu) e^{-C_{E,2}(\mu)/T}. 
\ee 
The fit results are shown in Table~\ref{C-D-table-E}. We have extrapolated $C_{E,2}(\mu)$ to $\mu=0$ assuming the quadratic dependence  $C_{E}(\mu)=C_{E}(0)+A\mu^2$.  As long as $\mu$ is small enough we can assume that an expansion of this form is valid.\footnote{Note that the linear term is neglected since the sign of $\mu$ does not play an important role. Similar to the sign of a mass, changing the sign of $\mu$ leads to a physically equivalent BMN model. This symmetry is broken by lattice artifacts but gets restored when we are considering the continuum limit.} The result we get is $C_{E}(0)=0.860(68)$,  which is consistent with the value $0.83(21)$ obtained in Ref.~\cite{Berkowitz:2018qhn} and the other parameter is estimated as $A=0.087(04)$.
On the other hand, it is difficult to extrapolate $D_E(\mu)$ to $\mu=0$; the quadratic ansatz gives 2.755 $\pm$ 2.153, which has a very large error within numerical accuracy (see Fig.~\ref{fig:C_E_D_E_extrapolations}).  In addition, we can consider the ratio between $C_{E}$ and $D_{E}$, and as we can see in Fig.~\ref{fig:nadj} it shows an increasing trend with $\mu$. 


The ratio $n_E\equiv D_{E}/C_{E}$ changes as we vary $\mu$. At $\mu=0$, simulations in Ref.~\cite{Berkowitz:2018qhn} estimated that the numerical value is $\left.n_E\right|_{\mu=0}=1.91(78)$ and our findings for the BMN matrix model is consistent with this.

\begin{center}
	\begin{tabular}{|c|cc|}
		\hline
		{}& \multicolumn{2}{c|}{\text{two-parameter fit}}  \\ 
		\hline
		$\mu$ &$C_{E}$& $D_{E}$\\
		\hline 
		0  & 0.860 $\pm$ 0.068 & 2.755 $\pm$ 2.153\\ 
		2   & 1.150 $\pm$ 0.049 & 8.154 $\pm$ 1.200\\
		3   & 1.719 $\pm$ 0.049 & 17.42 $\pm$ 1.48  \\
		4  & 2.255 $\pm$ 0.124 & 23.38 $\pm$ 4.65 \\
		5   & 3.017 $\pm$ 0.030 & 39.53 $\pm$ 1.55\\
		\hline
	\end{tabular}
	\captionof{table}{We fit $\Delta E$ by using the two-parameter ansatz \eqref{fit-E-2-parameter}. 
		The $\mu=0$ value is the extrapolation of the fits to the $\mu\to 0$ limit.} 
	\label{C-D-table-E}
\end{center}

\subsubsection*{Validity of perturbative results}
The perturbative prediction is expected to hold in the limit $\mu\to\infty$. Our non-perturbative methods allow us to check how much their range of validity extends towards lower $\mu$. The perturbative estimates are shown in Fig.~\ref{energy_g_vs_u} in comparison to the numerical data. As expected, one observes a convergence of perturbative and non-perturbative data in the low temperature limit. For large $\mu$ this range of convergence should extend up to higher temperatures.

It is instructive to discuss in more detail the perturbative predictions. We will focus in this discussion on the largest $\mu$ ($\mu=5$), which should provide the best convergence to the numerical data. In the large $\mu$ limit, the Hamiltonian of BMN decouples in two parts (see Appendix \ref{sec:Hamiltonian_splitting}). In the energy plots in Fig.~\ref{energy_g_vs_u} the full $U(1)$ sector \eqref{eq:full_U(1)} is taken into account, but we can also investigate the relevance of the different contributions and compare the full $U(1)$ sector with the lightest mode \eqref{eq:lightest_mode}.

At very low temperatures and finite $\mu$ the contribution of the full $U(1)$ sector almost coincides with the lightest mode \eqref{eq:lightest_mode} and we cannot distinguish them practically.  

\begin{figure}[ht!]
	\centering
	\includegraphics[scale=0.48]{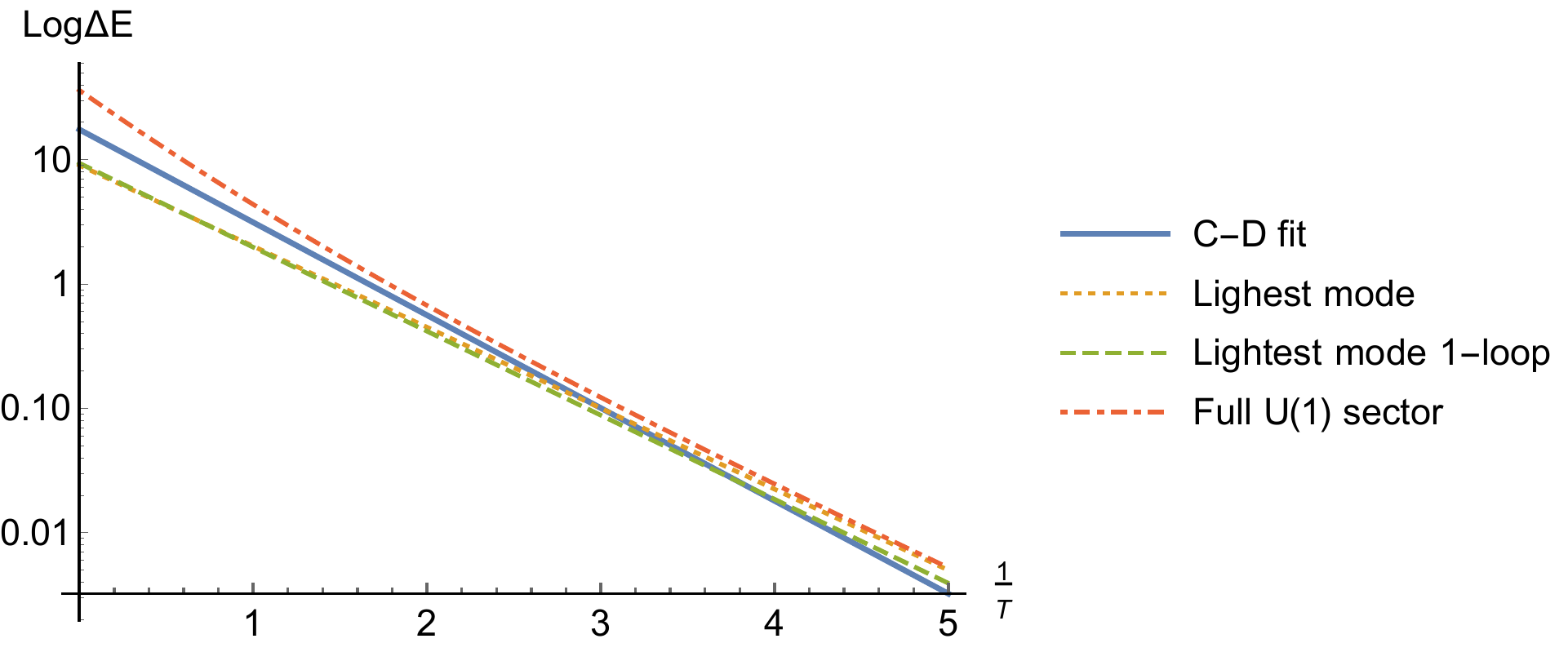}
	\includegraphics[scale=0.3]{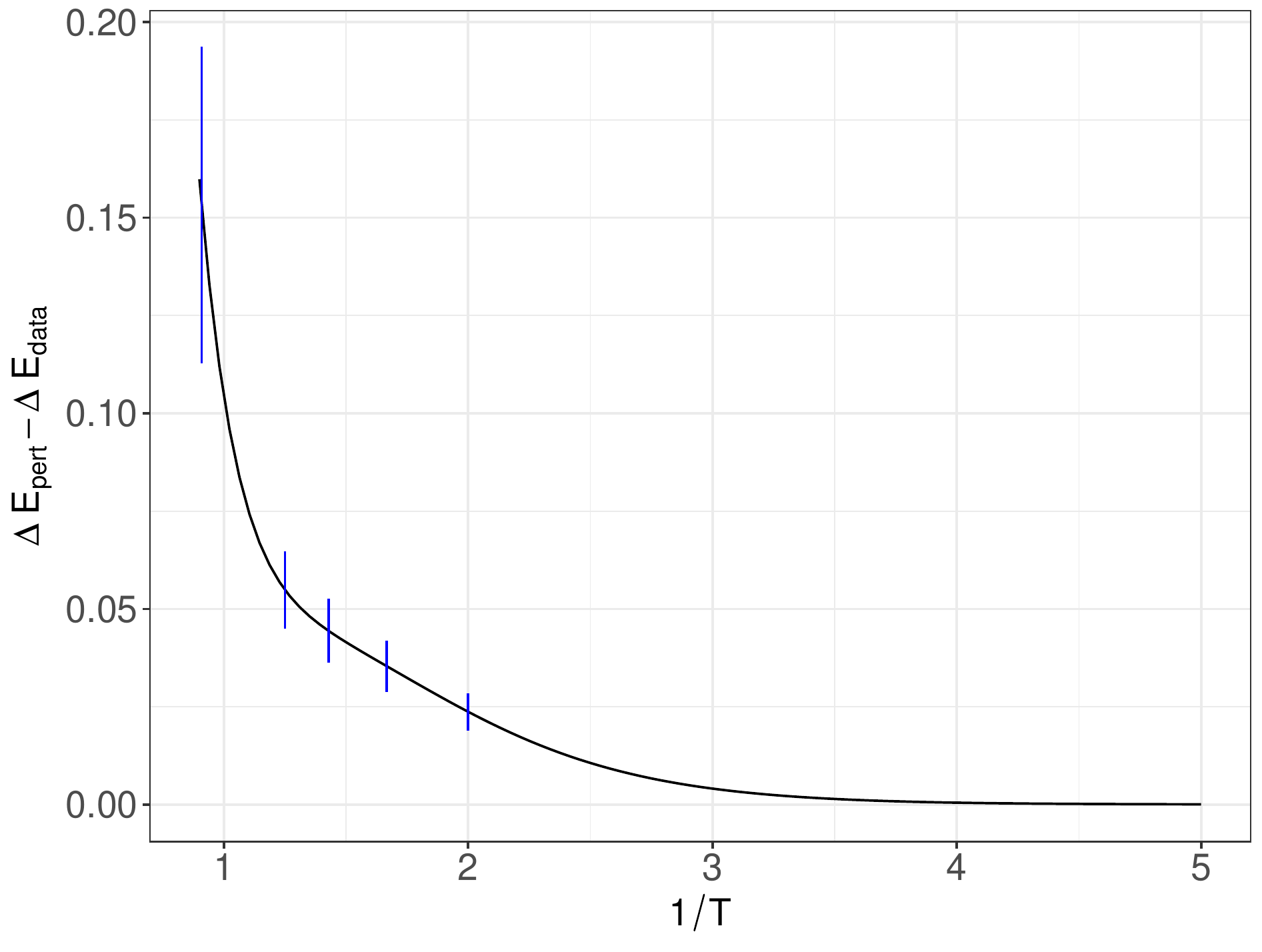}
	\caption{[Left]: Various perturbative results for the $U(1)$ sector of the model compared with the two-parameter fit data for $\mu=5$ with logarithmic scaling on the $y$-axis. There is a crossing of the two-parameter data with the lightest mode \eqref{eq:lightest_mode} and its one-loop correction \eqref{eq:lightest_mode_one-loop} but no crossing for the full $U(1)$ perturbative sector \eqref{eq:full_U(1)}. [Right]: The difference of the perturbative result \eqref{eq:full_U(1)} and the two-parameter fit with respect to temperature (as the black curve) including the error bars of the data points for $\mu=5$. We observe an indistinguishability between the actual data and the perturbative result using the full $U(1)$ sector within the error bars of the data. At smaller temperatures the difference approaches zero.}
	\label{fig:pert_C-D}
\end{figure}

We observe that at finite $\mu$ the perturbative result of the lightest mode and the exponential fits from the numerical data cross at finite values of temperature as shown in Fig.~\ref{energy_g_vs_u}. There is, consequently, no asymptotic convergence in the small temperature limit. When we include the one-loop correction given by \cite{Kim:2003rza, Maldacena:2018vsr}
\be \label{eq:lightest_mode_one-loop}
E_{\rm SO(6)}=\frac{\mu}{2}+\frac{\lambda}{2\mu^2}+\cdots,
\ee 
we do not expect things to change by much because the one-loop corrections in the gauge and ungauged models cancel each other \cite{Maldacena:2018vsr}. Indeed, we checked that this is the case and at small temperatures, the difference between the two is suppressed. Fig.~\ref{fig:pert_C-D} is an additional illustration of these findings.

However, in view of the overall behavior we still conclude that the full $U(1)$ sector \eqref{eq:full_U(1)} provides a cleaner description of the data since there is an asymptotic convergence instead of a crossing with the fitted non-perturbative result. Only a small shift smaller than our statistical uncertainty remains in this case. In the parameter region we studied, considering only the lightest mode shows a large deviation from the numerical data at lager temperatures. Except for the region where the fit and the lightest mode contribution cross, the full $U(1)$ contribution is closer to the data and it seems to capture the asymptotic behaviour better. Therefore we decided to show the contribution of the full $U(1)$ sector in Fig.~\ref{energy_g_vs_u}. If we insist that the $U(1)$ sector is protected from finite $\mu$ corrections \cite{Dasgupta2002, Dasgupta:2002ru, Kim:2002if, Kim:2002zg} we could carry these results to finite $\mu$ and the slight shift from the actual data could be related with the $SU(N)$ sector where we do not precisely know what to expect at finite $\mu$. This would require, however, a further detailed analysis. Such investigation can reveal more about the fine print of the perturbative calculations. Overall we can, however, conclude that our results reproduce correctly the perturbative predictions.

\subsection{Other observables}
\label{sec:otherobservables}
\begin{figure}[htbp]
	\centering
	\includegraphics[scale=0.4]{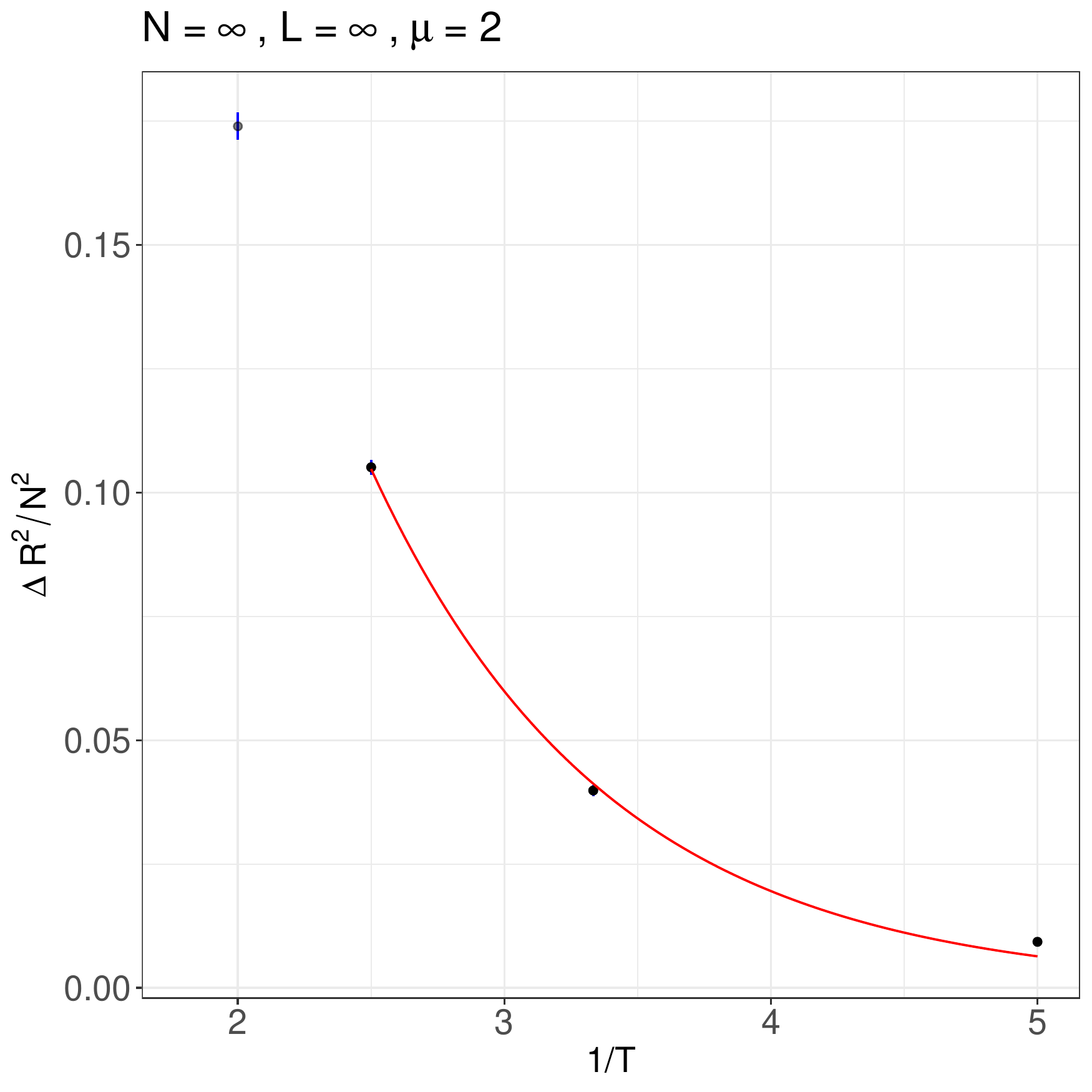}
	\includegraphics[scale=0.4]{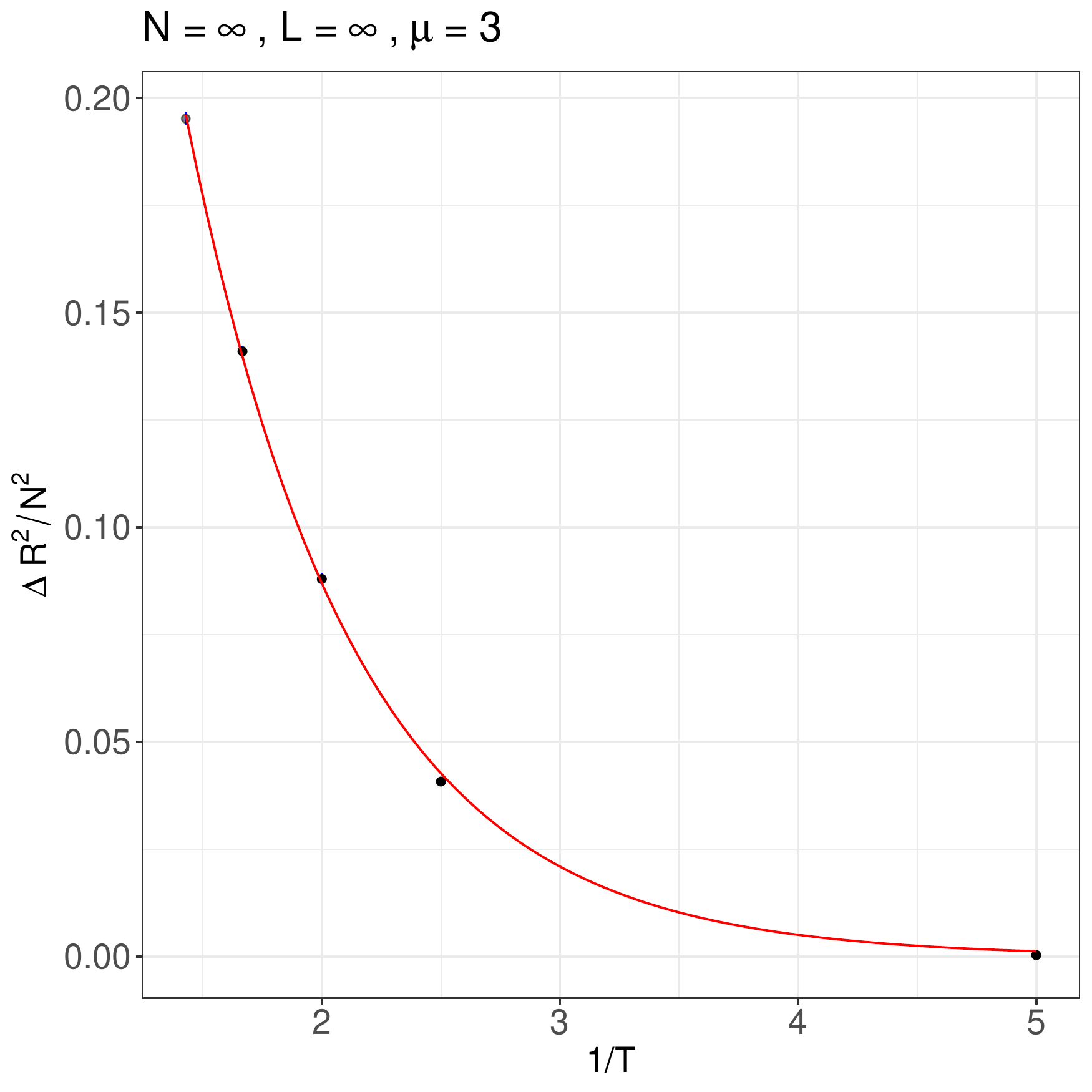}\\
	\includegraphics[scale=0.4]{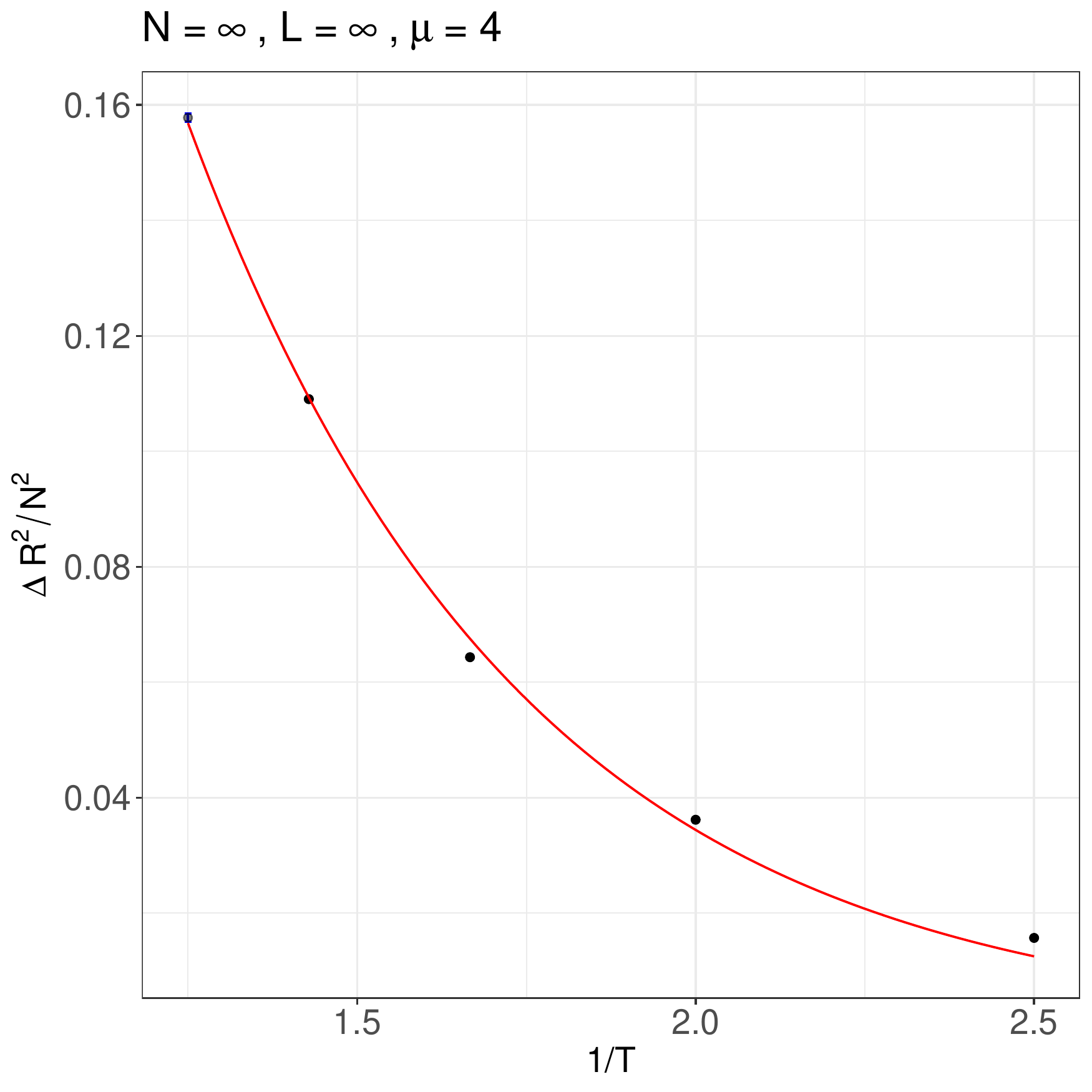}
	\includegraphics[scale=0.4]{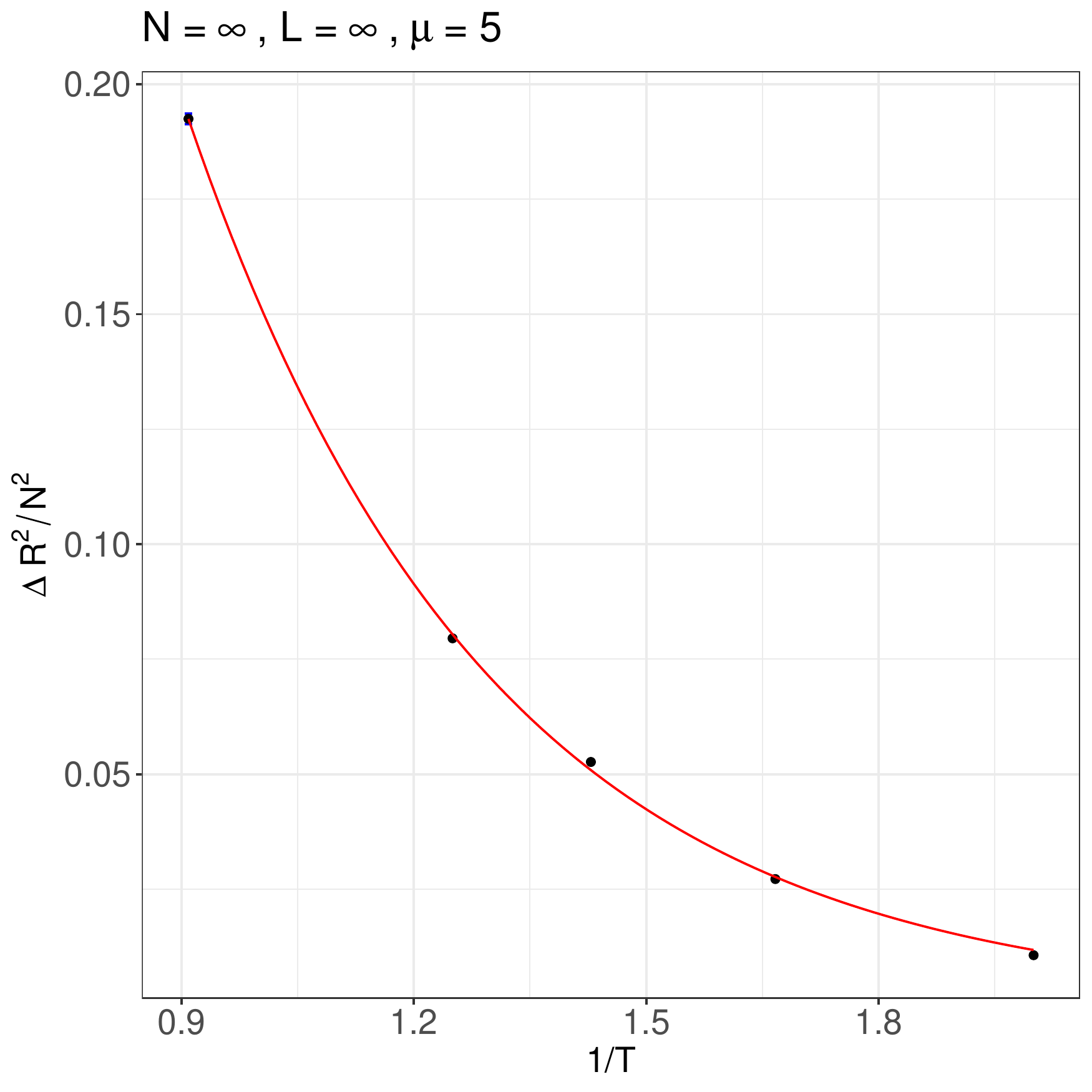}
	\caption{The large $N$ and continuum exponential  fits for $\Delta R^2$ and different $\mu$. The fitting parameters are shown in Table~\ref{C-D-table-RF}. The errors are also included in the plot but are very small.}
	\label{Strx2}
\end{figure}

The same analysis can be done for other  gauge-invariant quantities such as the sum of traces of the squared matrices defined via 
\be \label{eq:R2}
R^2\equiv\frac{1}{N\beta}\int dt\left(\sum_{I=1}^9 \Tr{(X_I)^2}\right),
\ee 
and the commutator  
\be \label{eq:F2} 
F^2\equiv -\frac{1}{N\beta}\int dt\left(\sum_{I,J=1}^9\Tr{[X_I,X_J]^2}\right).
\ee 
Similar to the energy, we may take the difference between the gauged and ungauged results which we expect to be exponentially close to each other in the limit of \eqref{DeltafreeE}. The justification for this comes from the partition functions, which are close to each other. Therefore we expect the relevant observables to scale as 
\begin{align}
\label{R2exponential}\Delta R^2&=R^2_{\rm{ungauged}}-R^2_{\rm{gauged}}=N^2 D_R e^{-C_R/T},\\
\label{F2exponential}\Delta F^2&=F^2_{\rm{ungauged}}-F^2_{\rm{gauged}}=N^2 D_F e^{-C_F/T}
\end{align} 
with a priori unknown parameters $D_R$, $D_F$, $C_R$ and $C_F$. We report the results in Table~\ref{C-D-table-RF} while the respective diagrammatic fits are shown in Figs.~\ref{log_g_vs_u_F2} and \ref{Strx2}. We have not shown the exponential fit for the observable $F^2$ because it is similar as can be seen from the logarithmic plots in appendix \ref{app:logplots}. It is important to note that there is no prediction whatsoever for the parameters $C_{R,F}$ and $D_{R,F}$. In addition, there are no constraints between  $C_R, D_R$ and $C_F, D_F$ so we always use a two-parameter fit.

\begin{figure}[h!]
	\centering
	\includegraphics[scale=0.4]{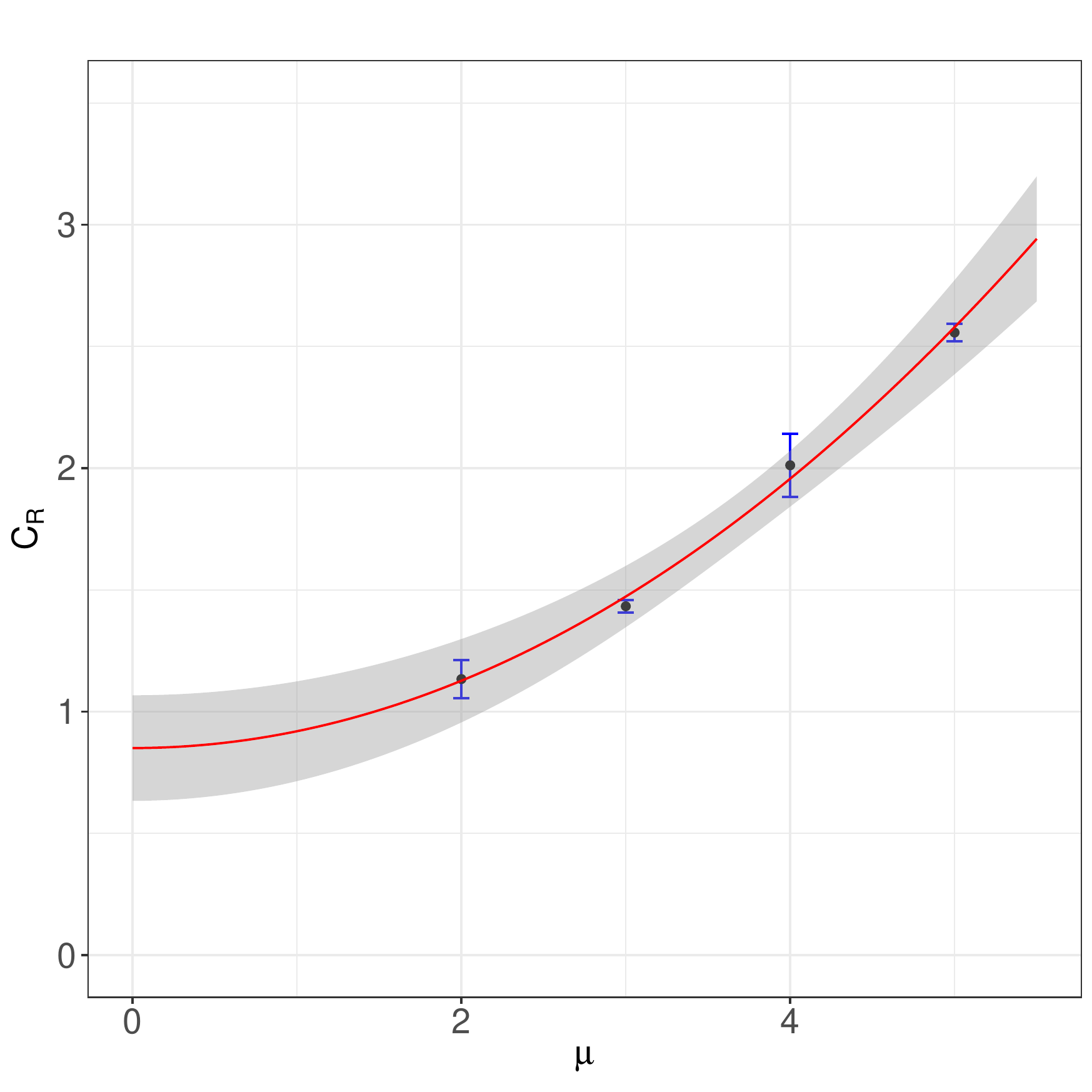}\quad\quad
	\includegraphics[scale=0.4]{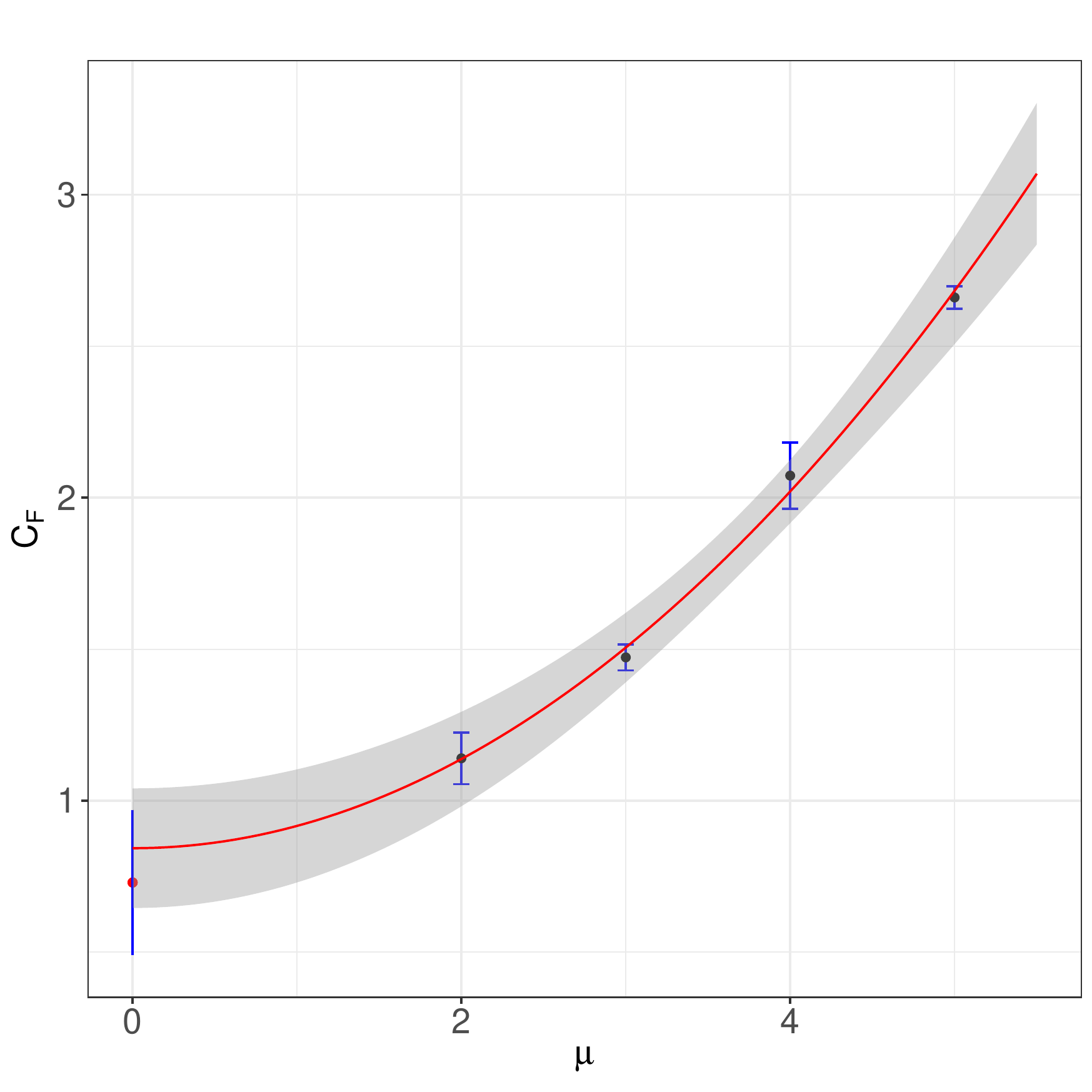}
	\caption{The behaviour of the exponential parameter ($C_{R,F}$)  from \eqref{R2exponential}, \eqref{F2exponential} with respect to $\mu$. [Left]: the fit for $C_R$ is given by the equation $C_R=0.834(58)+0.070(4)\mu^2$. [Right]: the fit for $C_F$ is given by the equation $C_F=0.825(59)+0.074(4)\mu^2$. The left most point is the BFSS point taken from \cite{Berkowitz:2018qhn} (see also Table:~\ref{table:BFSS_data}).}
	\label{CtraceX2}
\end{figure}

\begin{figure}[h!]
	\centering
	\includegraphics[scale=0.38]{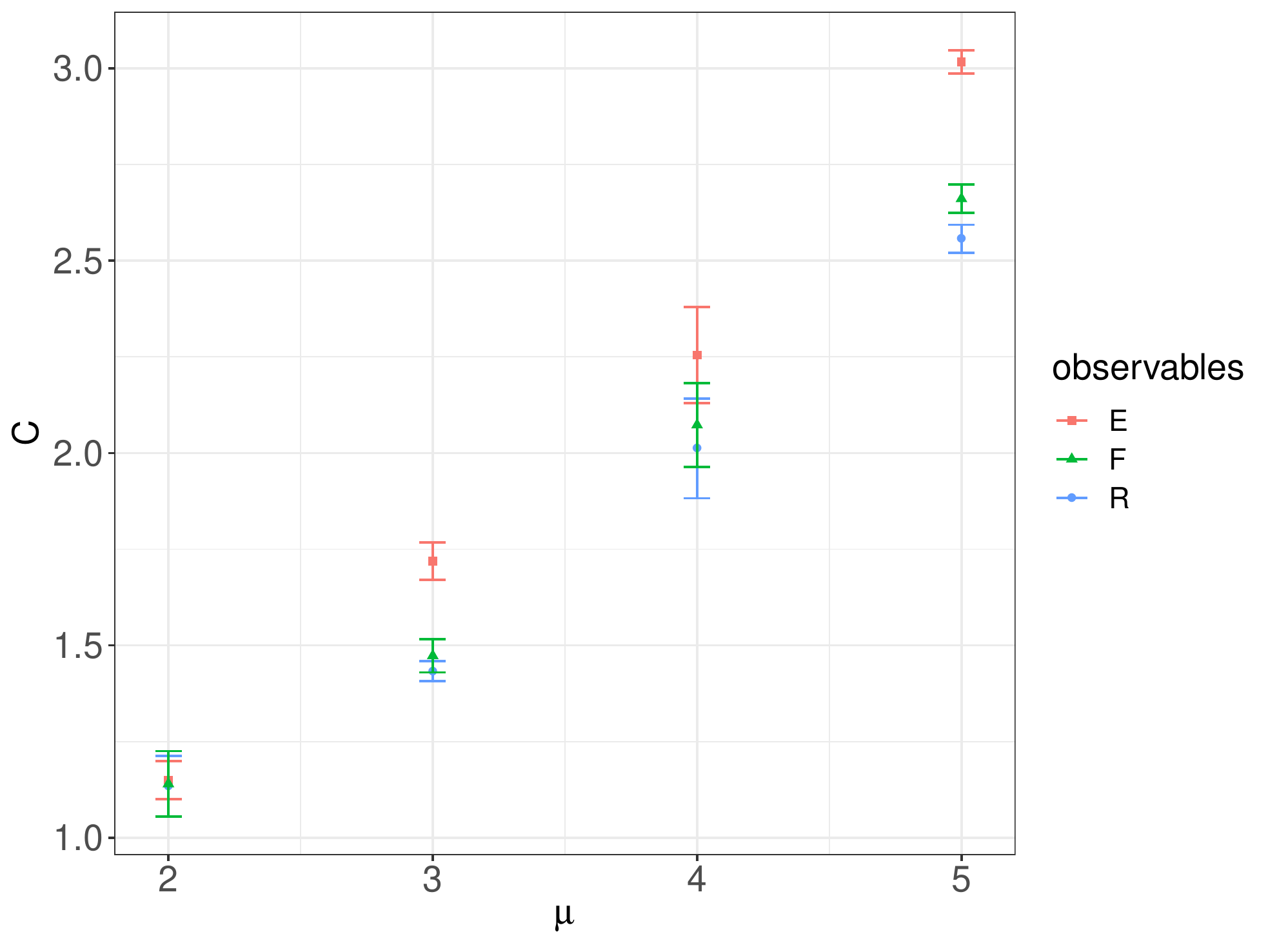}
	\includegraphics[scale=0.38]{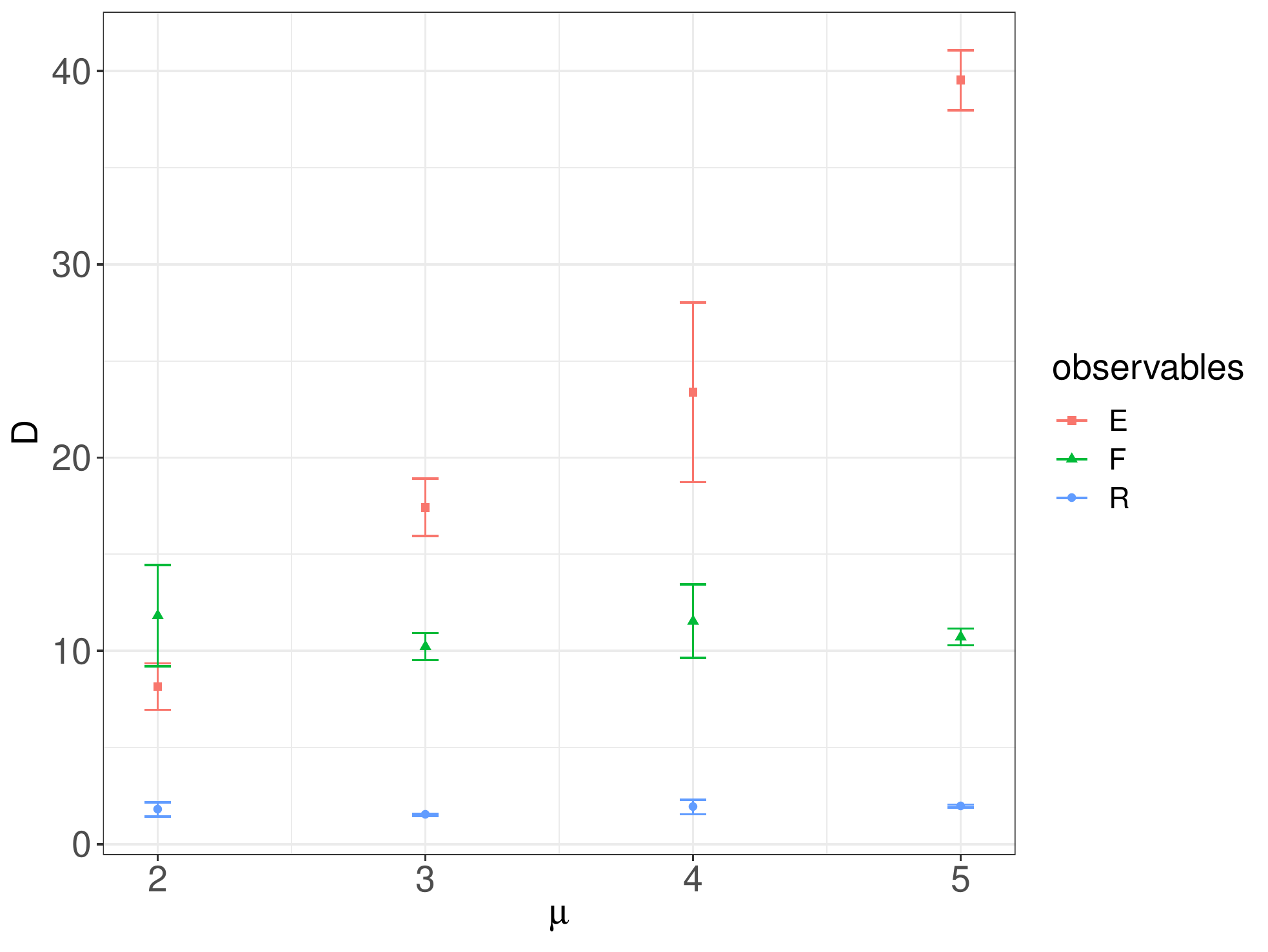}
	\caption{The change of parameters with respect to $\mu$.  [Left]: Values of $C_E, C_F, C_R$ labelled as $E,F,R$ respectively. We expect these values to be approximately the same when the partition functions of both models are exponentially close to each other. We see that this happens as $\mu\to 0$ and indeed we expect this to be the case in this limit as we can recall from  eq. \eqref{DeltafreeE}. [Right]: Values of $D_E, D_F, D_R$. We observe that $D$ for  $\Tr{X_I^2}$ and $\Tr[X_I,X_J]^2$ observables do not change with respect to $\mu$ but for energy does. This is probably due to different degeneracies of the energy eigenstate of the system.}
	\label{CD_observables}
\end{figure}

Due to finite $\mu$ deformations, we expect the situation to be different than for the BFSS model. However, in the $\mu\to 0$ limit one should recover the BFSS values. Since there are only data available for the $F^2$ term from \cite{Berkowitz:2018qhn},  we can compare the commutator, and indeed for $C_F$ it converges to its BFSS value as $\mu\to0$ (right panel of Fig.~\ref{CtraceX2}). On the other hand, for $D_F$ we observe a discrepancy with the BFSS value. It is likely that we need to study smaller values of $\mu$ to observe convergence with the BFSS results regarding $D_F$. In addition, the small discrepancy at intermediate $\mu$ for $C_{E,R,F}$ and $D_{E,R,F}$ could probably be an effect caused by finite $\mu$ contributions. The important result is that $C_{E,R,F}$ agree nicely when $\mu=0$ as we can see from the left panel of Fig.~\ref{CD_observables}. Since, no correlation is assumed between $D_{R,F}$ and $C_{R,F}$ we can not claim the same for $D_{R,F}$. We can expect that this should be the case when we are studying particularly small $\mu$ values, but on the other hand, it is quite challenging to do a large $N$, continuum, gauged and ungauged study and remain in the confined phase, given the current resources.   

\begin{figure}[ht!]
	\centering
	\includegraphics[scale=0.33]{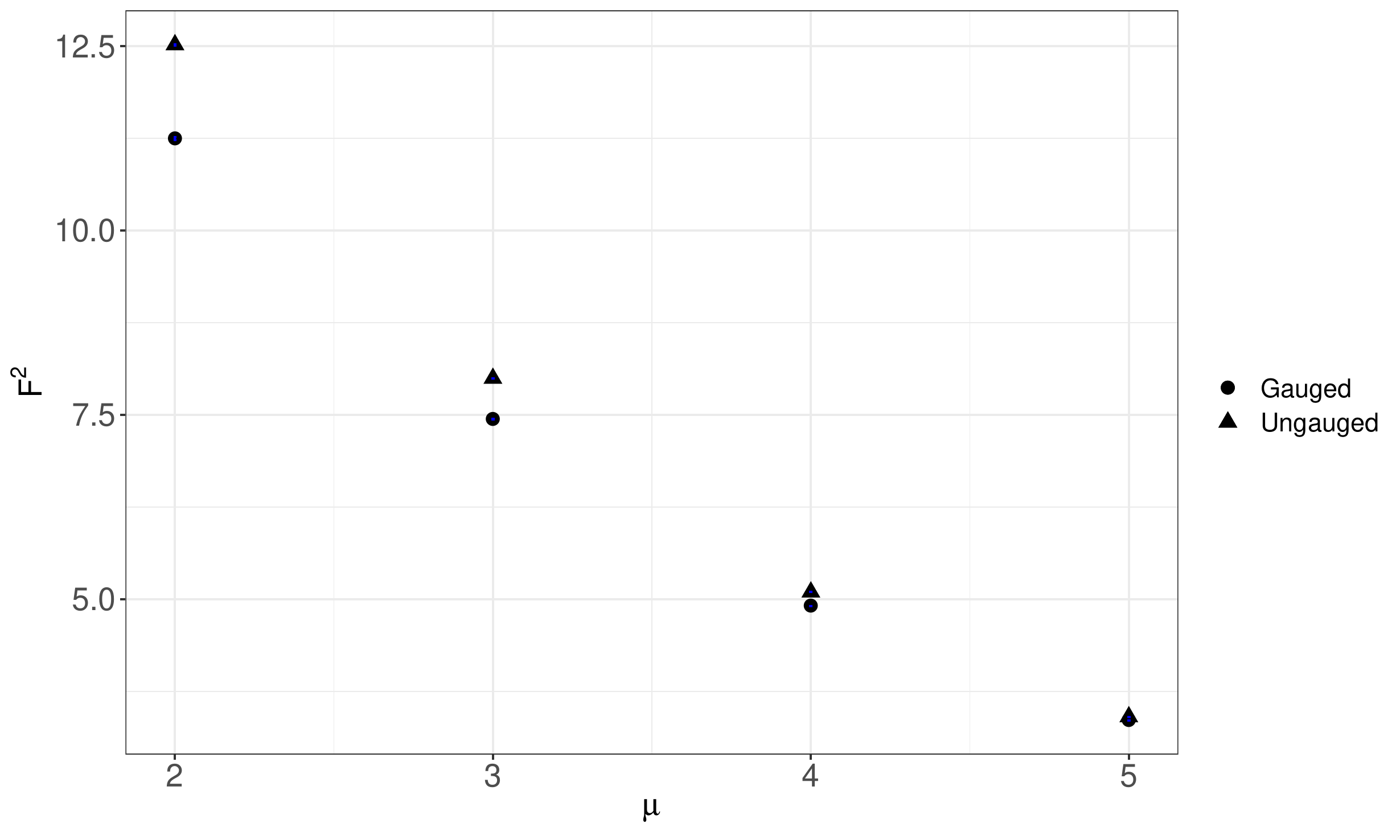}
	\includegraphics[scale=0.47]{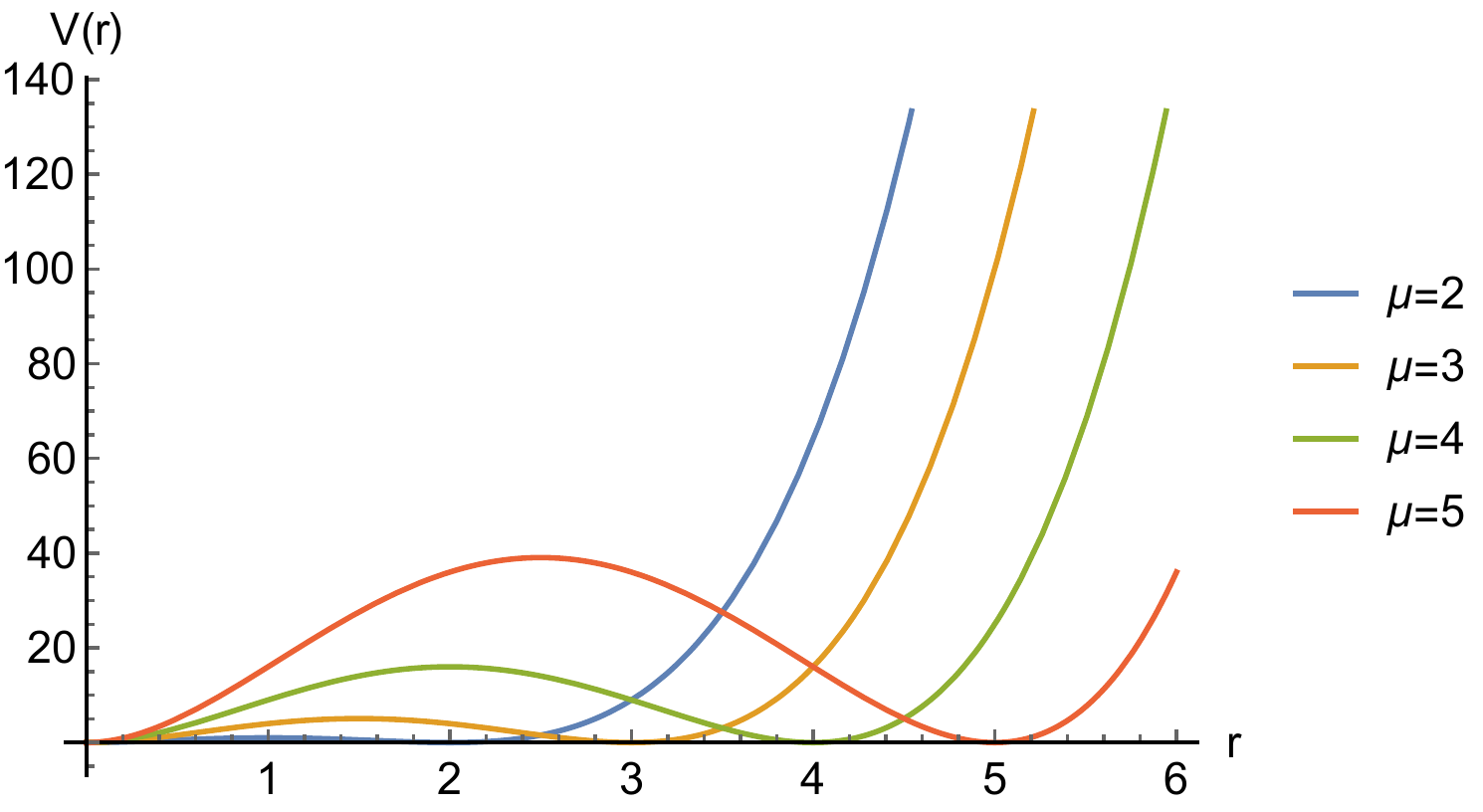}
	\caption{[Left]: The observable $F^2$ (errors included but are very small) with respect to $\mu$ for the gauged and ungauged data at temperature $T=0.5$ in the large $N$ and continuum limit.  [Right]: The $SO(3)$ potential corresponding to extrema of the $SO(3)$ action with respect to $\mu$ in the positive-$r$ axis. This potential is symmetric with respect to the $y$-axis. As we increase $\mu$ we see that when we are in the trivial background $r=0$ the simulation becomes more and more stable, such that we observe a decrease of $F^2$ as we increase $\mu$. The same scaling behaviour holds also for $R^2$.}
	\label{fig:SO(3)_comparison}
\end{figure}

In Fig.~\ref{fig:SO(3)_comparison} we show that our results are consistent with theoretical expectations. Finite $\mu$ effects could be observed even though we are putting the system in the trivial background $X^I=0, I=1,\cdots, 9$. This can be already observed in Monte Carlo histories (see for example Fig.~\ref{fig:MC_histories}) and by studying the $SO(3)$ potential of the action. We remind that on top of the classical trivial background there are quantum fluctuations that give actually some non-zero expectation values for the matrices. The classical potential of the bosonic $SO(3)$ part of the action is found to be (see Appendix \ref{app:SO3_potential} and specifically eq.~ \eqref{eq:potential_profile})
\be \label{eq:SO3_BMN_pot}
V(r)\sim r^2\left(r-\mu\right)^2.
\ee 
 On top of that, we have quantum fluctuations on the $r$ value resulting in fluctuations on $V(r)$. For the case of $\mu=0$, we immediately get the BFSS result $V(r)\sim r^4$ which, in the matrix model language is given by the observable $F^2$ since considering quantum fluctuations the matrices do not strictly commute. We may now ask how this observable behaves for $\mu\neq0$ and the answer is clear from \eqref{eq:SO3_BMN_pot}. Bigger values of $\mu$ result in smaller values of $V(r)$ because the effect of finite $\mu$ is to confine the simulation more into the classical vacuum $r=0$. For finite $\mu$, the matrices (to be more precise their eigenvalues) are grouped around $r=0$ while for $\mu=0$ they can spread in the whole range of the potential barrier and this makes (some of) the eigenvalues escape to infinity resulting in the flat direction problem. Indeed, this is the behaviour we observe in the left panel of Fig.~\ref{fig:SO(3)_comparison} both for gauged and ungauged models, and the theoretical explanation is because for bigger values of $\mu$ the fluctuations of the potential become smaller due to a bigger potential wall created by finite $\mu$. This is shown in the right panel of Fig.~\ref{fig:SO(3)_comparison} and it affects both the observables $\{R^2,F^2\}$ such that they decrease as we increase $\mu$.  At the same time, the observable $F^2$ for the gauged and ungauged models studied at $\mu=0$ in \cite{Berkowitz:2018qhn} is of order $\mathcal{O}(15)$ always while here we see clearly a decreasing trend as we increase $\mu$. 

Another point of view that arrives at the same conclusion is to consider the large $\mu$ limit. In this case, all the bosonic matrices can be written in terms of harmonic oscillators, and they all scale as $1/\mu$ (for the $SO(3)$ part) and $2/\mu$ (for the $SO(6)$ part) as it can be seen from eq.'s~ \eqref{eq:oscillators_1} and \eqref{eq:oscillators_2}.

An interesting puzzle to understand is the role of the unstable solution of the $SO(3)$ potential given by $r=\mu/2$ in our conventions (see again Appendix~ \ref{app:SO3_potential}). It is not known in the literature how this term appears in the simulations and how it affects them. Thus, we do not exclude the possibility that the simulation reaches frequently this solution altering the results non-trivially at finite $\mu$. 

Nevertheless, finite $\mu$ effects change the observables non-trivially as we saw, and indeed a better understanding of the smaller $\mu$ region will be important in the future since currently, it is out of reach for such a precise analysis.

\begin{center}
	\begin{tabular}{|c|c|c|}
		\hline
		$\mu$ & $C_R$ & $D_R$ \\
		\hline 
		0 &0.834 $\pm$ 0.058  & 1.542 $\pm$ 0.187 \\
		2 & 1.119 $\pm$ 0.091 & 1.718 $\pm$ 0.416  \\
		3 & 1.423 $\pm$ 0.026  & 1.499 $\pm$ 0.063 \\
		4 & 2.022 $\pm$ 0.075 & 1.965 $\pm$ 0.202  \\
		5 & 2.561 $\pm$ 0.033 & 1.975 $\pm$ 0.066  \\
		\hline
	\end{tabular}
	\begin{tabular}{|c|c|c|}
		\hline
		$\mu$ & $C_F$ & $D_F$ \\
		\hline 
		0 &0.825 $\pm$ 0.059  & 11 $\pm$ 0.7 \\
		2 &1.125 $\pm$ 0.098  & 11.370 $\pm$ 2.939 \\
		3 &1.457 $\pm$ 0.042 & 9.983 $\pm$ 0.677 \\
		4 &2.088 $\pm$ 0.065  & 11.900 $\pm$ 1.061 \\
		5 &2.663 $\pm$ 0.032  & 10.750 $\pm$ 0.344 \\
		\hline
	\end{tabular}
	\captionof{table}{ We fit $\Delta R^2$ and $\Delta F^2$ by using the two-parameter ansatz \eqref{R2exponential} and \eqref{F2exponential}, respectively. The $\mu=0$ value is the extrapolation of the fit to the $\mu\to 0$ limit.} 
	\label{C-D-table-RF}
\end{center}

\begin{center}
	\begin{tabular}{|c|c|c|c|}
		\hline
		\multicolumn{4}{|c|}{BFSS values for two-parameter fits} \\
		\hline
		$C_E$ & $D_E$  & $C_F$ & $D_F$ \\
		\hline 
		0.83 $\pm 0.21$ & 1.59 $\pm$ 0.51 & 0.73 $\pm$ 0.24 & 1.93 $\pm$ 0.65 \\
		\hline
	\end{tabular}
	\captionof{table}{The available large $N$, continuum and two-parameter fit data for the BFSS model taken from \cite{Berkowitz:2018qhn}. } 
	\label{table:BFSS_data}
\end{center}

The dependence of the exponential parameters ($C_R,C_F$) on $\mu$ is given in Fig.~\ref{CtraceX2} and for $C_F$ it converges to the BFSS result  \cite{Berkowitz:2018qhn}. This provides additional evidence that the limit $\mu\to0$ is smooth and consistent but nonetheless considering smaller values of $\mu$ will be of much importance such that also $D_F$ will approach its known BFSS value. 

\section{Conclusions and Discussions}\label{sec:conclusions}
We have studied the gauged and ungauged BMN model at finite flux $\mu$ values and relatively small temperatures with high statistics.
We confirmed that the difference between the gauged and ungauged partition function is exponentially small, being proportional to $e^{-C_{adj}/T}$ also for finite $\mu$.
This is our main result.
This happens also in the case that the limit $\mu/T\ll1$ is not satisfied, namely at intermediate temperatures and finite $\mu$.
Of particular importance is the exponential decay of energies in Fig.~\ref{energy_g_vs_u} supporting the two different limits, namely the gravity limit and the perturbative regime as we vary $\mu$.
In particular, at higher $\mu$ values and relatively small temperatures, the system behaves in such a way that it converges to the perturbative result \eqref{eq:full_U(1)}.
On the other hand, when we gradually decrease $\mu$ towards $\mu\to0$ the system seems to approach the gravitational results $C_{\rm adj}=1$ and $n_E\to n_{\rm adj}=2$ obtained in \cite{Berkowitz:2018qhn}.
To verify this numerically for $n_E$ turned out to be difficult at this level due to fitting problems, namely if we insist on an even power expansion for $C_E$ and $D_E$ then the ratio is a rational function that does not have a rapidly converging expansion in (even powers of) $\mu$ for a large range of $\mu$.

On the other hand, we observed that the degeneracy of the energy states can change with $\mu$ as it is shown in Fig.~\ref{fig:nadj}.
It seems that by turning on $\mu$ there could be some massive modes that affect the simulation in a non-trivial way.
In particular, in Fig.~\ref{fig:nadj} we observe a consistent decrease of the ratio $n_E=D_E/C_E$ as we decrease $\mu$.
This ratio appears to be the degeneracy of the eigenstate of the system \cite{Maldacena:2018vsr} and it seems that it changes with $\mu$.  

This is an interesting and puzzling issue since the respective change is not known theoretically.
One possibility is that, at finite $\mu$, the contribution from second-lightest or even higher modes is not yet negligible in this parameter region.
We tried to fit $\Delta E$ by using two different excitations as $\Delta E =2N^2C_E(\mu)e^{-C_E(\mu)/T}+n'N^2C'_E(\mu)e^{-C'_E(\mu)/T}$ with a few different values of $n'$ (including $n'=6$, which is the number of lightest adjoint modes in the limit of $\mu\to\infty$), but we were not able to obtain reasonable fits. 

Indeed, it seems that we have a factor behaving like an effective degeneracy.
The cause of this behaviour is most likely finite $\mu$ effects whose form in the interacting gauged sector has not been studied extensively. 
It may be the case that several low-energy excited modes with slightly different energies are contributing. 
The precise details of this remain unknown to us but we hope that this will initialise a more systematic study of the precise finite $\mu$ contributions to the partition function of the model, perturbatively and (if possible) also non-perturbatively. 

That non-singlet modes are exponentially suppressed can make quantum simulation based on the extended Hilbert space simpler.
Suppose that Hamiltonian time evolution is performed on a quantum computer, as $|\Phi\rangle\to e^{-i\hat{H}t}|\Phi\rangle$.
If $|\Phi\rangle$ is gauge-invariant, $e^{-i\hat{H}t}|\Phi\rangle$ is also gauge-invariant, as long as the time evolution is exactly realized.
If $|\Phi\rangle$ is a specific gauge fixed state, e.g., a fuzzy sphere with a certain representation, then the gauge fixing will not be spoiled via the Hamiltonian time evolution.
However, if there were light non-singlet modes, small simulation errors could easily excite non-singlet modes and lead to a large deviation from the exact result.
Our findings in this paper suggests that we do not have to worry about such a possibility.  

\acknowledgments

The authors thank the ECT* for its hospitality during the workshop Quantum Gravity meets Lattice QFT where this work was initiated.
G. B. acknowledges support from the Deutsche Forschungsgemeinschaft (DFG) Grant No. BE 5942/3-1.
N. B. and S. P. were supported by an International Junior Research Group grant of the
Elite Network of Bavaria. E. R. is supported by Nippon Telegraph and Telephone Corporation (NTT) Research.
M. H. was supported by the STFC Ernest Rutherford Grant ST/R003599/1.
The numerical simulations were performed on ATHENE, the HPC cluster of the\\ Regensburg University Compute Centre. A.S. thanks the University of the Basque Country, Bilbao, for hospitality.

\vspace{5mm}
\noindent
{\large \textbf{Data management}}\\
No additional research data beyond the data presented and cited in this work are needed to validate the research findings in this work.
Simulation data will be publicly available after publication.

\appendix
\section{SO(3) potential profile}\label{app:SO3_potential}

We are interested in finding the profile of the $SO(3)$ classical potential in the BMN model. This feature also allows finding the probability for the trivial vacuum configuration to transition to a fuzzy sphere configuration. 

Let us then concentrate on the bosonic action of the BMN model which has the following form
\begin{align} \label{eq:BMN_bosonic}
S=&\frac{N}{\lambda}\int dt\Tr\Bigg[ \frac{1}{2}(D_tX^i)^2+\frac{1}{4}[X^i,X^j]^2-\mu^2 (X^i)^2-i\mu X^iX^jX^k\epsilon_{ijk}\Bigg].
\end{align}  
To find extrema of this potential let us consider an ansatz of the form $X^i(t)=\rho(t)J^i$ and substitute it into the Lagrangian. Using identities of $\epsilon_{ijk}$ and that for irreducible (considering actually the maximal) $SU(2)$ representations we have $\Tr{(J^i)^2}=\frac{N}{3}(N^2-1)$, we find 
\be 
\mathcal{L}=\frac{N^2(N^2-1)}{6\lambda}\left(\dot{\rho}^2-\rho^2\left(\rho-\mu\right)^2\right),
\ee 
with potential 
\be \label{eq:potential_profile}
V(\rho)=\frac{N^2(N^2-1)}{6\lambda}\rho^2\left(\rho-\mu\right)^2.
\ee 
We are considering the case where we have just one big fuzzy sphere, which is a configuration that can be distinguished by the simulation and specifically by the Myers observable (the cubic term in \eqref{BMNaction}).

Asking stability of this potential, we differentiate with respect to $\rho$ and we obtain three solutions 
\begin{itemize}
	\item $\rho=0$ which is stable,\
	\item $\rho=\mu$ which is stable,\
	\item $\rho=\frac{\mu}{2}$ which is unstable.
\end{itemize} 
For $\rho=0$ we have the trivial background solution, while for $\rho=\mu$ we have a fuzzy sphere of radius $\rho$. This solution is known as the fuzzy sphere background and in fact, $\rho$ gives us the size of the $SO(3)$ subpart of the spacetime. 
On the other hand, for $\rho=\frac{\mu}{2}$ we have an unstable solution whose interpretation we are not aware of. The maximum of the potential is given by this solution
\be 
V\left(\rho=\frac{\mu}{2}\right)=\frac{N^2(N^2-1)}{6\cdot 4\cdot 4\lambda}\mu^4.
\ee 
In the large $N$ limit, this potential barrier behaves like $\sim\mu^4N^4$, which means that the combination $\mu, N$ is what matters such that we keep under control the system. Specifically, in simulations, it is of much importance to keep this combination large to avoid undesirable tunnellings from the trivial backgrounds to fuzzy sphere backgrounds.

\section{Details for log plots}\label{app:logplots}
\begin{figure}[ht!]
	\centering
	\includegraphics[scale=0.4]{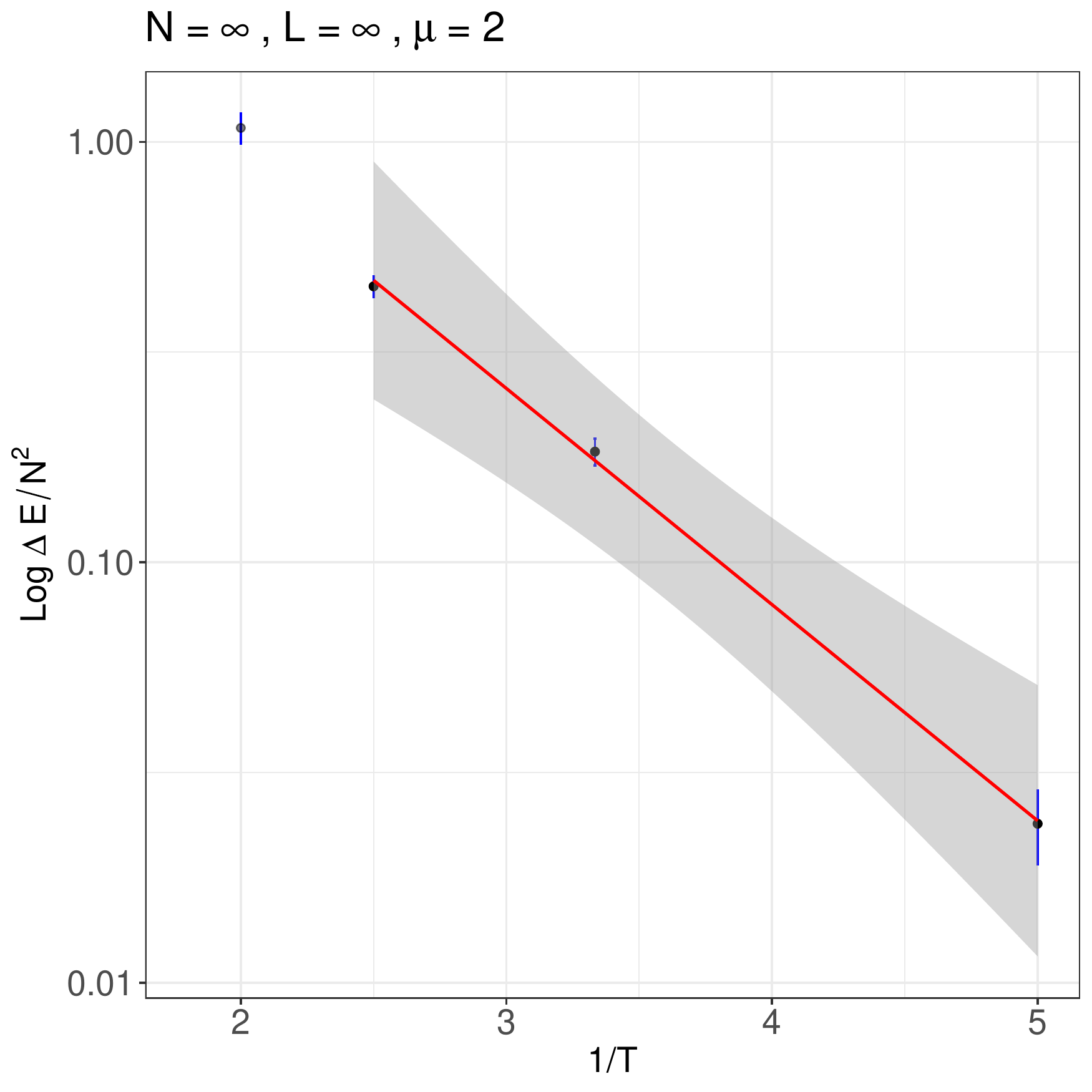}
	\includegraphics[scale=0.4]{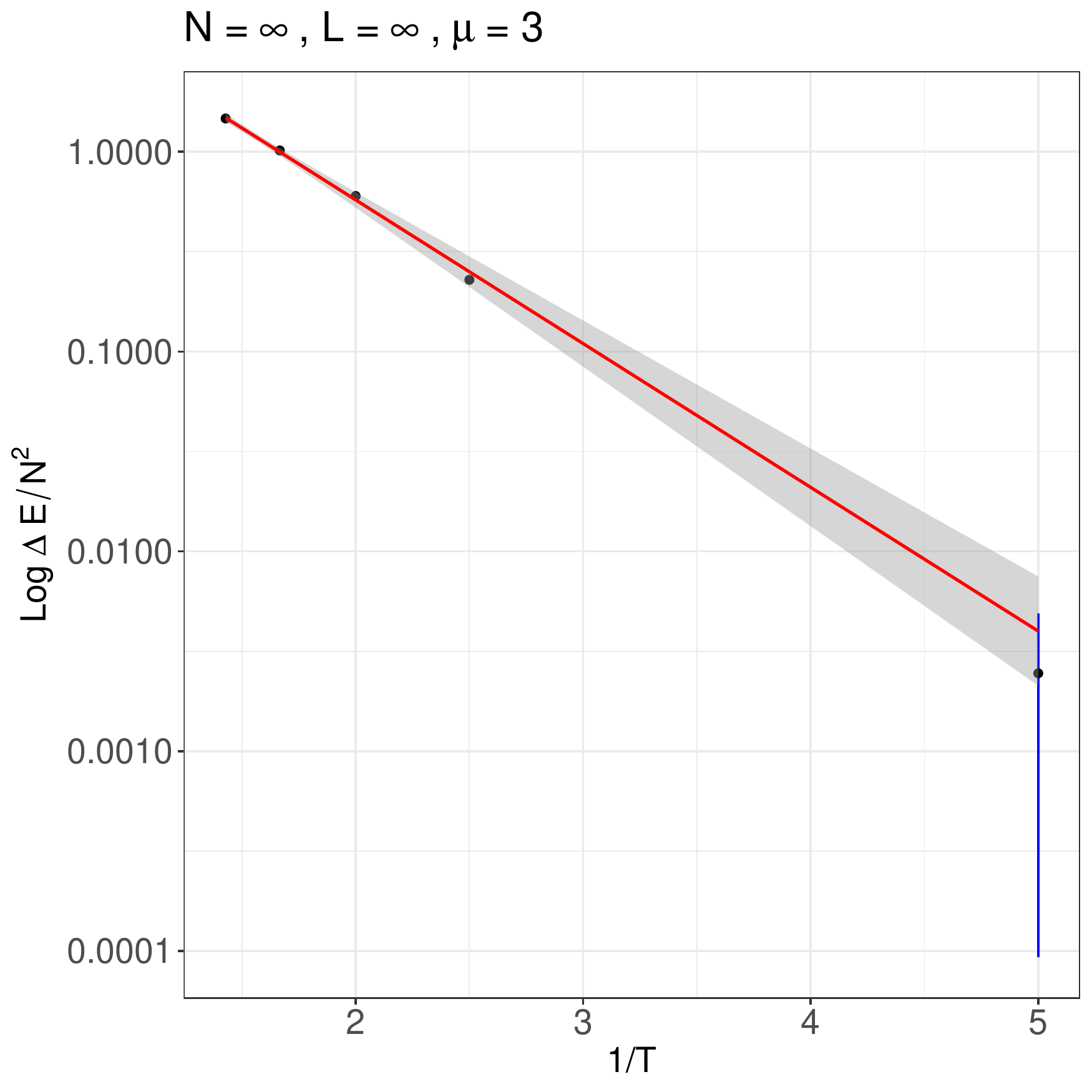}
	\includegraphics[scale=0.4]{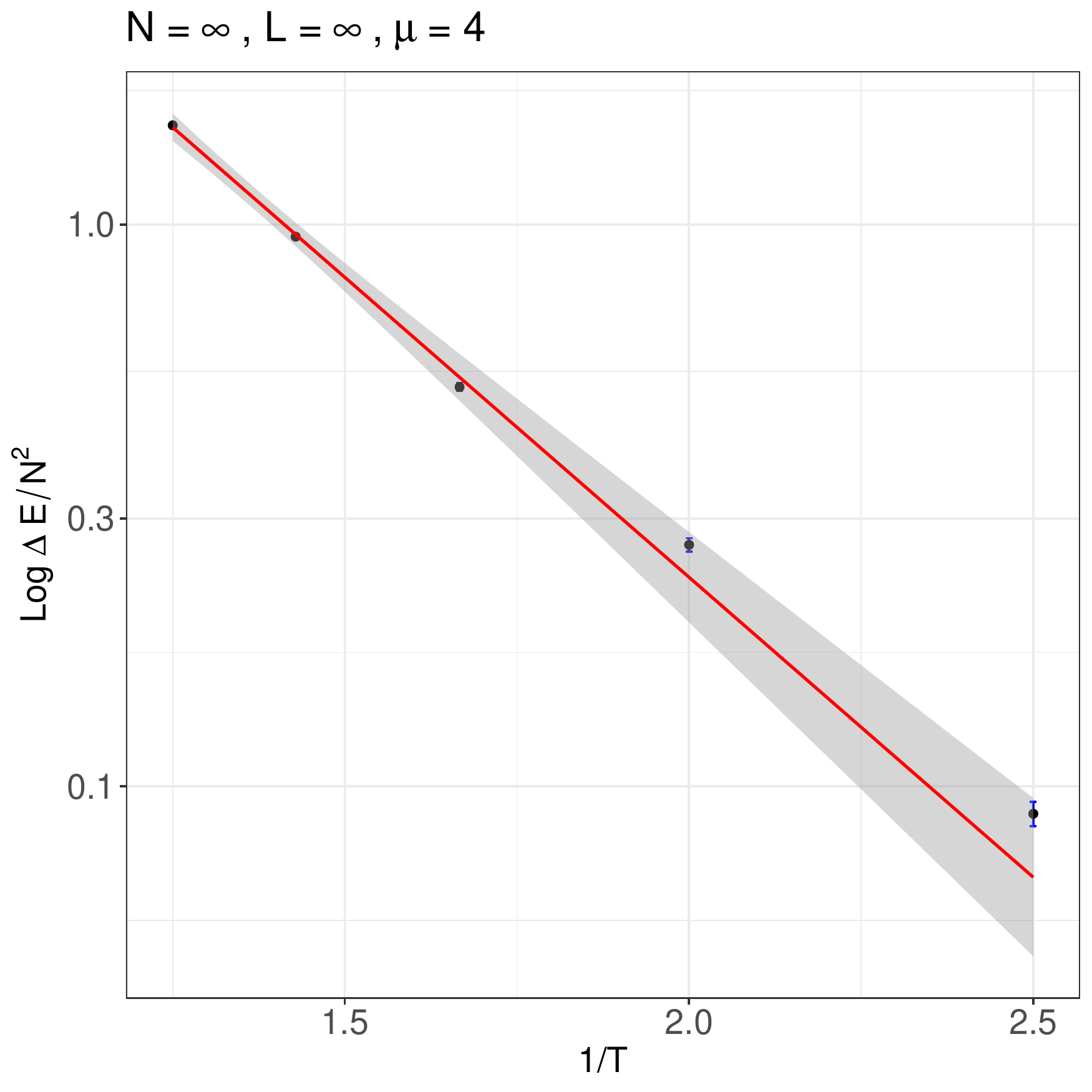}
	\includegraphics[scale=0.4]{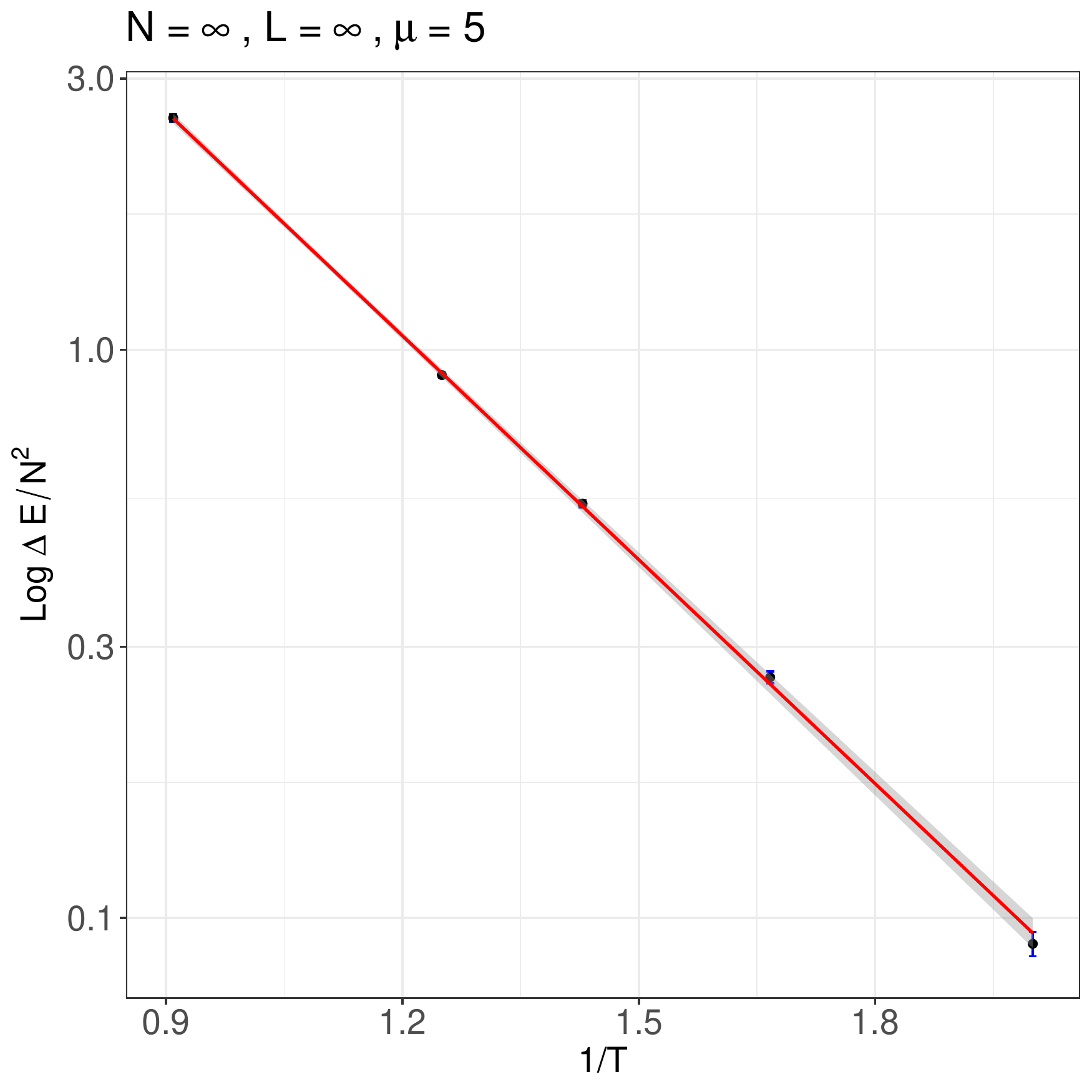}
	\caption{Large $N$ and continuum extrapolations of the logarithm of the energy difference and, fits for different $\mu$. }
	\label{log_g_vs_u}
\end{figure}

\begin{figure}[ht!]
	\centering
	\includegraphics[scale=0.4]{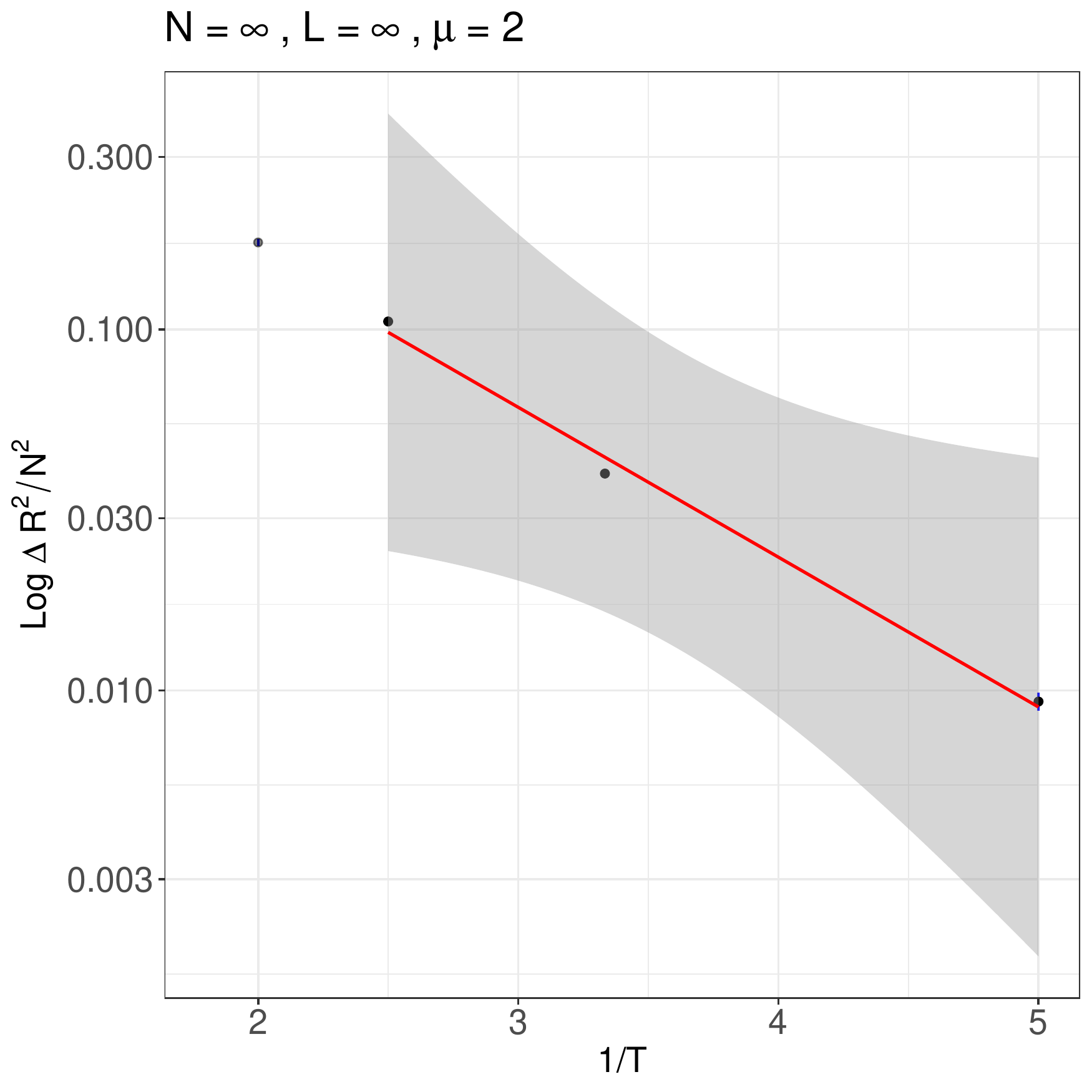}
	\includegraphics[scale=0.4]{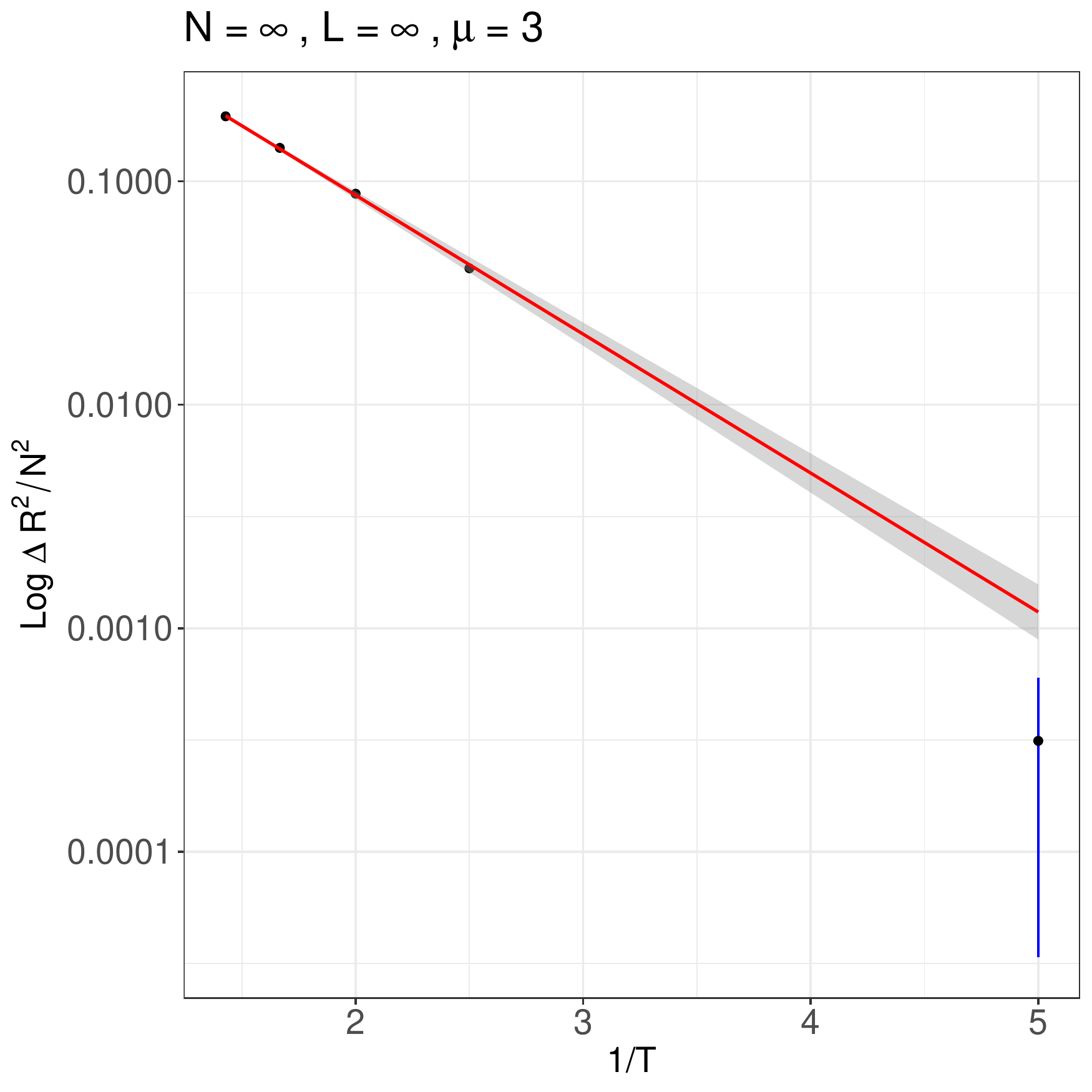}\\
	\includegraphics[scale=0.4]{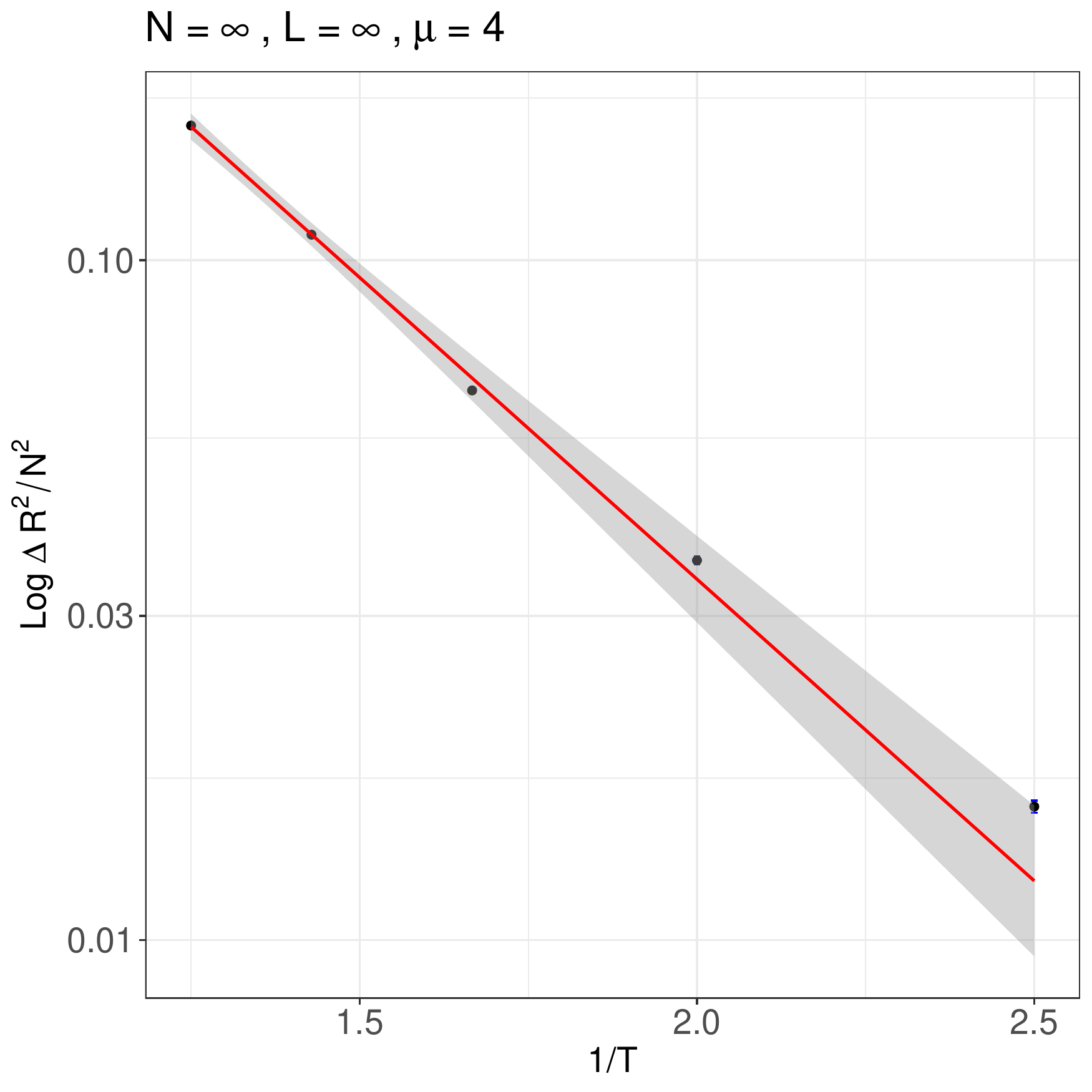}
	\includegraphics[scale=0.4]{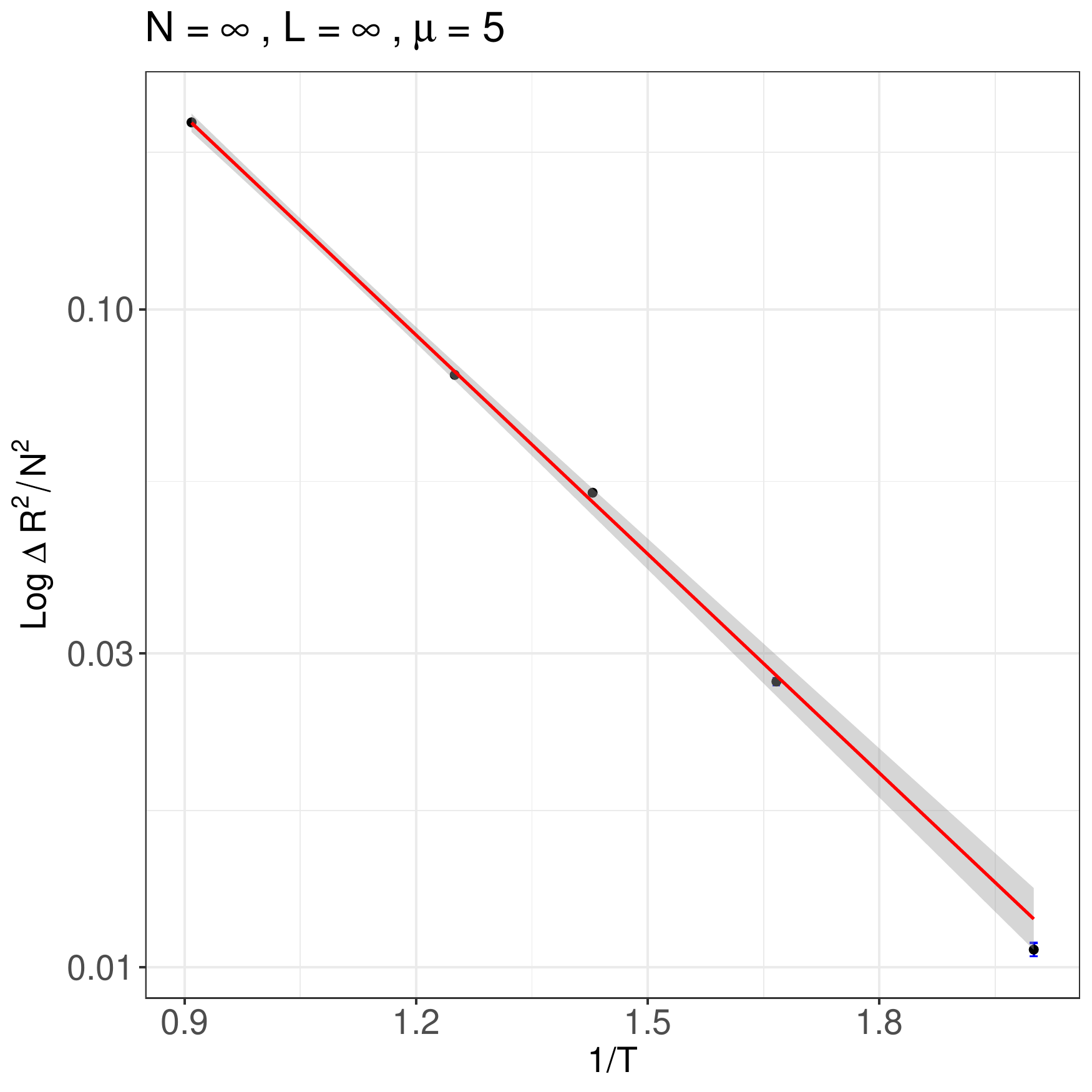}
	\caption{Large $N$ and continuum logarithmic plots for $\Delta R^2$ and different $\mu$. }
	\label{log_g_vs_uStrx2}
\end{figure}

\begin{figure}[ht!]
	\centering
	\includegraphics[scale=0.4]{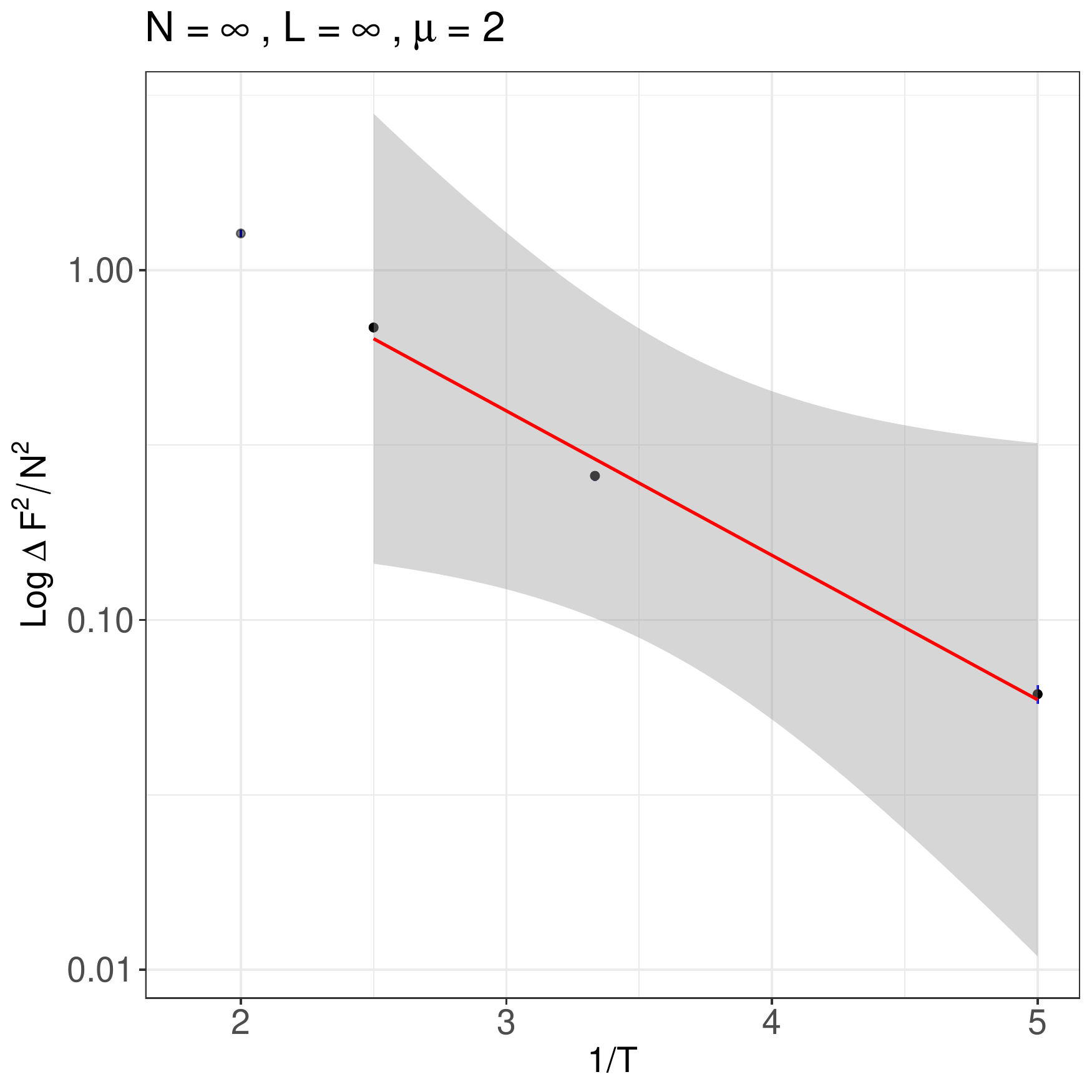}
	\includegraphics[scale=0.4]{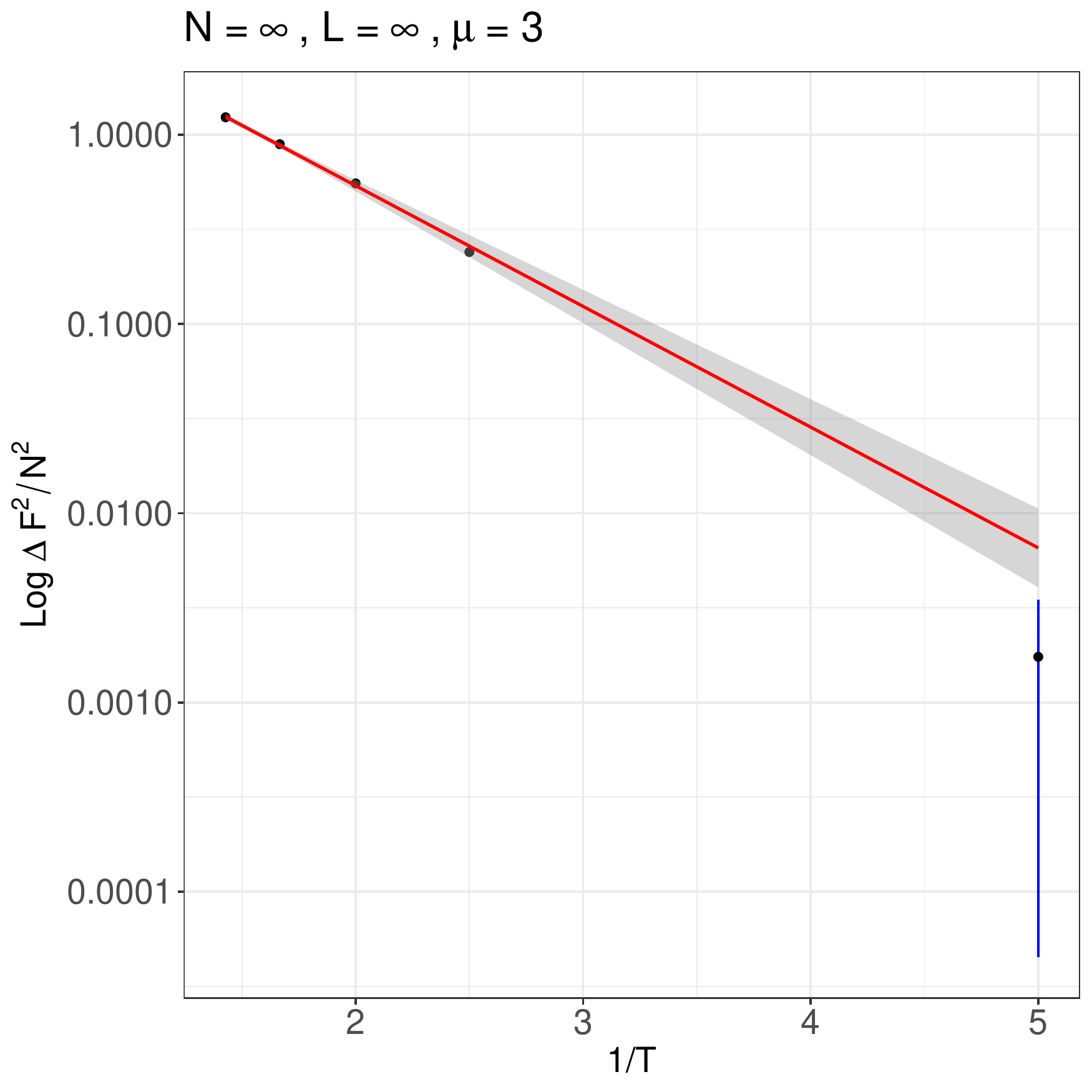}\\
	\includegraphics[scale=0.4]{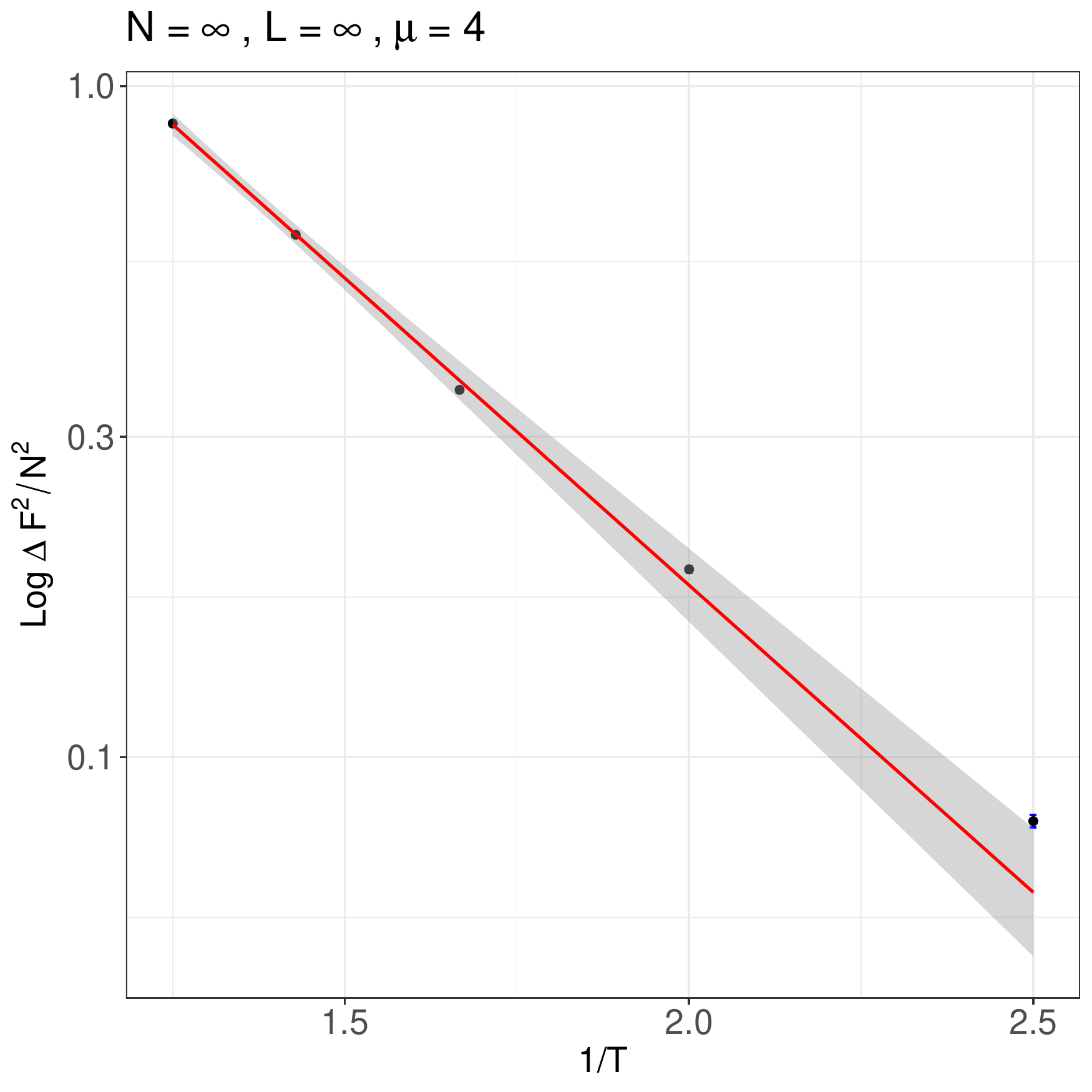}
	\includegraphics[scale=0.4]{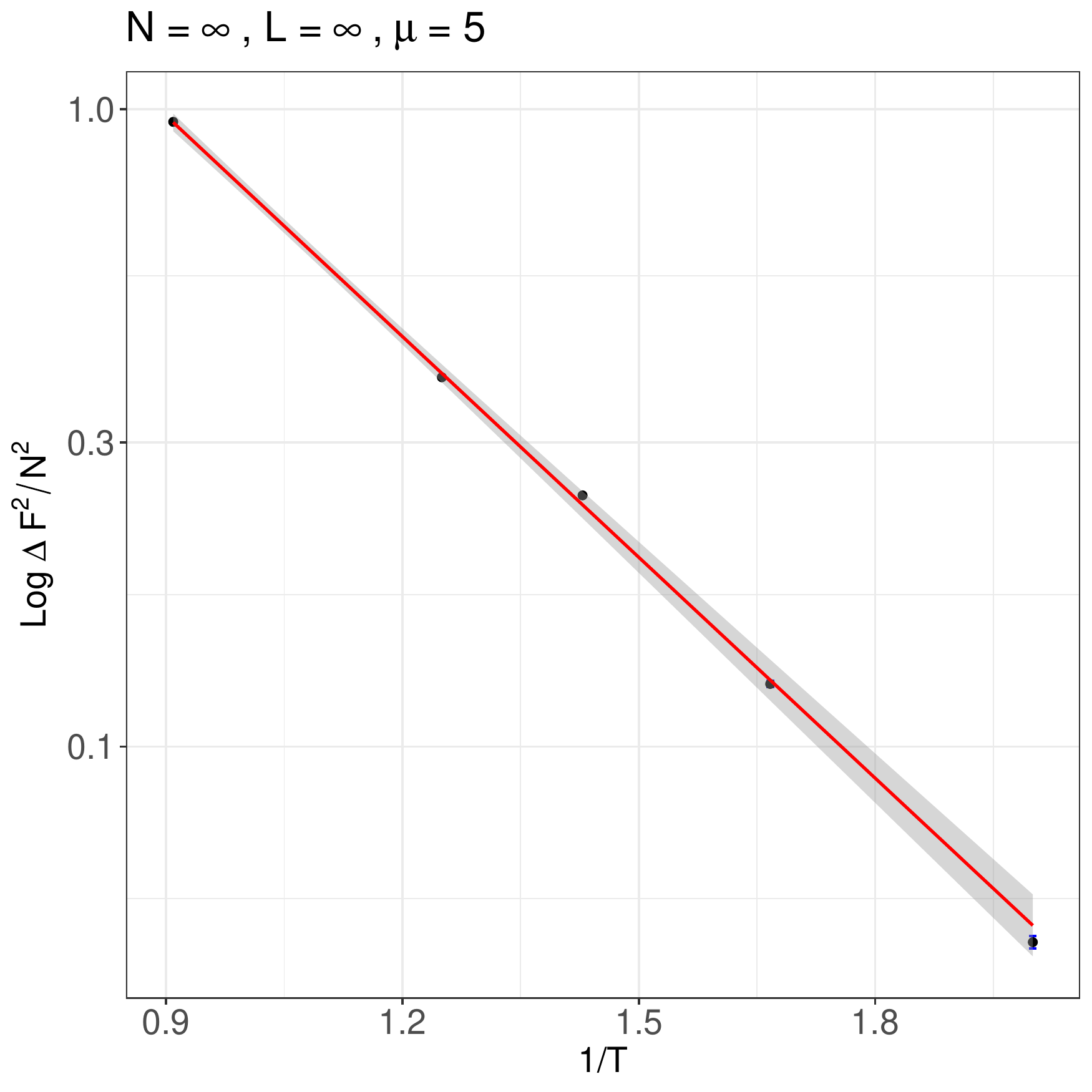}
	\caption{Large $N$ and continuum logarithmic plots for $\Delta F^2$ and different $\mu$. }
	\label{log_g_vs_u_F2}
\end{figure}

In this Appendix we also report the logarithmic plots for all observables $E, R^2, F^2$ in Figs. \ref{log_g_vs_u}, \ref{log_g_vs_uStrx2}, \ref{log_g_vs_u_F2} respectively.
\section{Monte Carlo histories}
\begin{figure}[h!]
	\centering
	\includegraphics[scale=0.21]{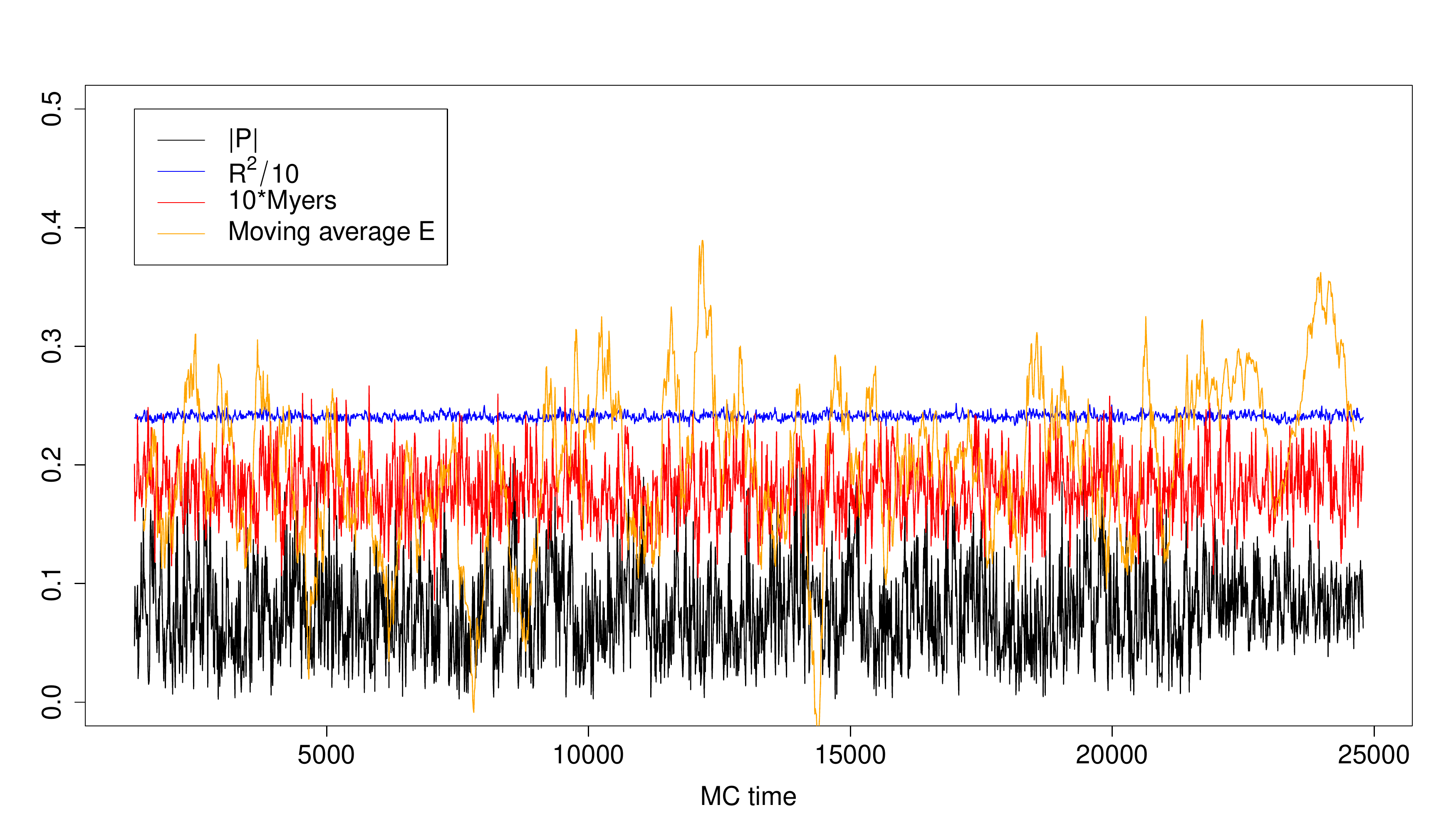}
	\includegraphics[scale=0.21]{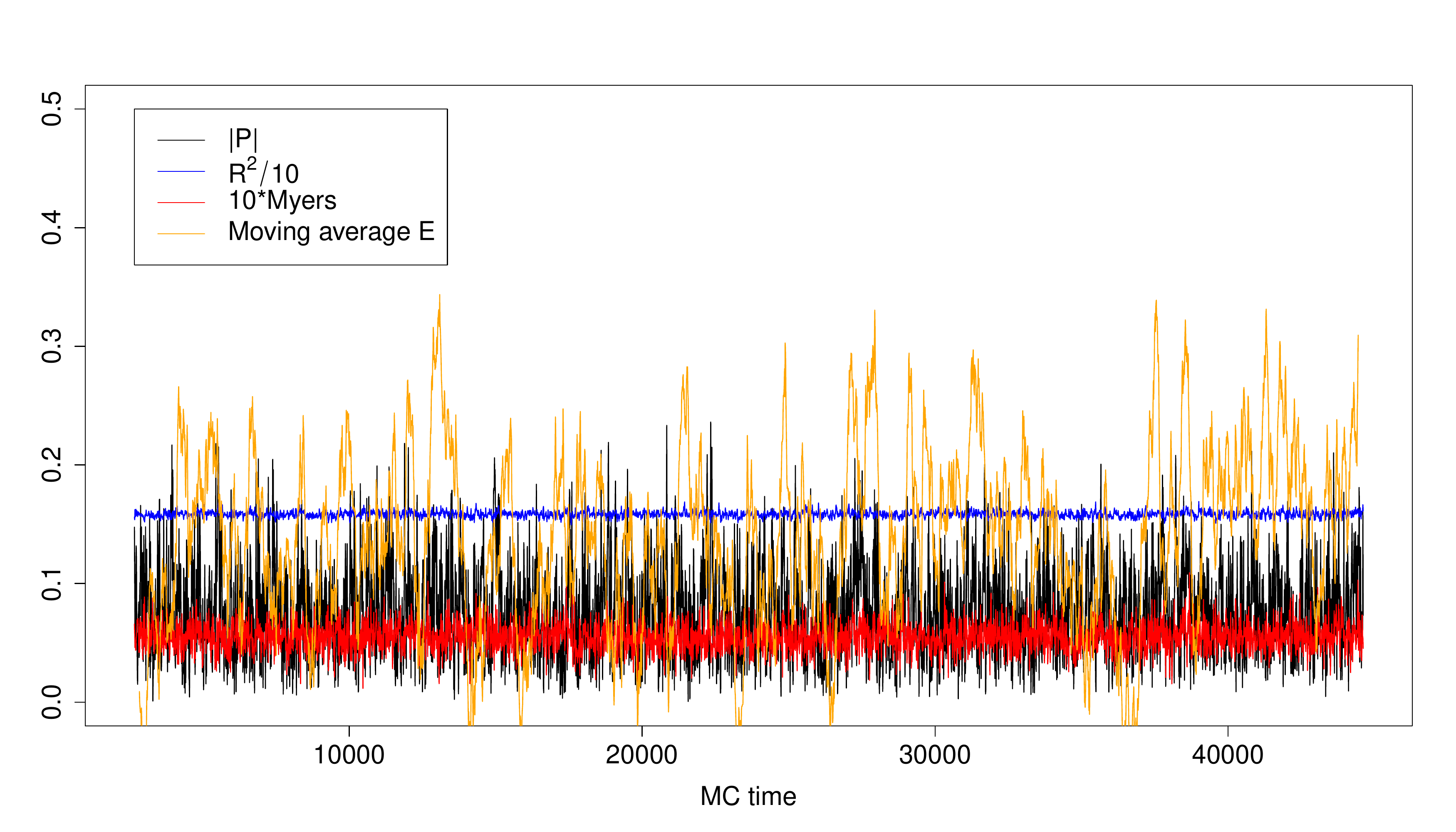}\\
	\includegraphics[scale=0.21]{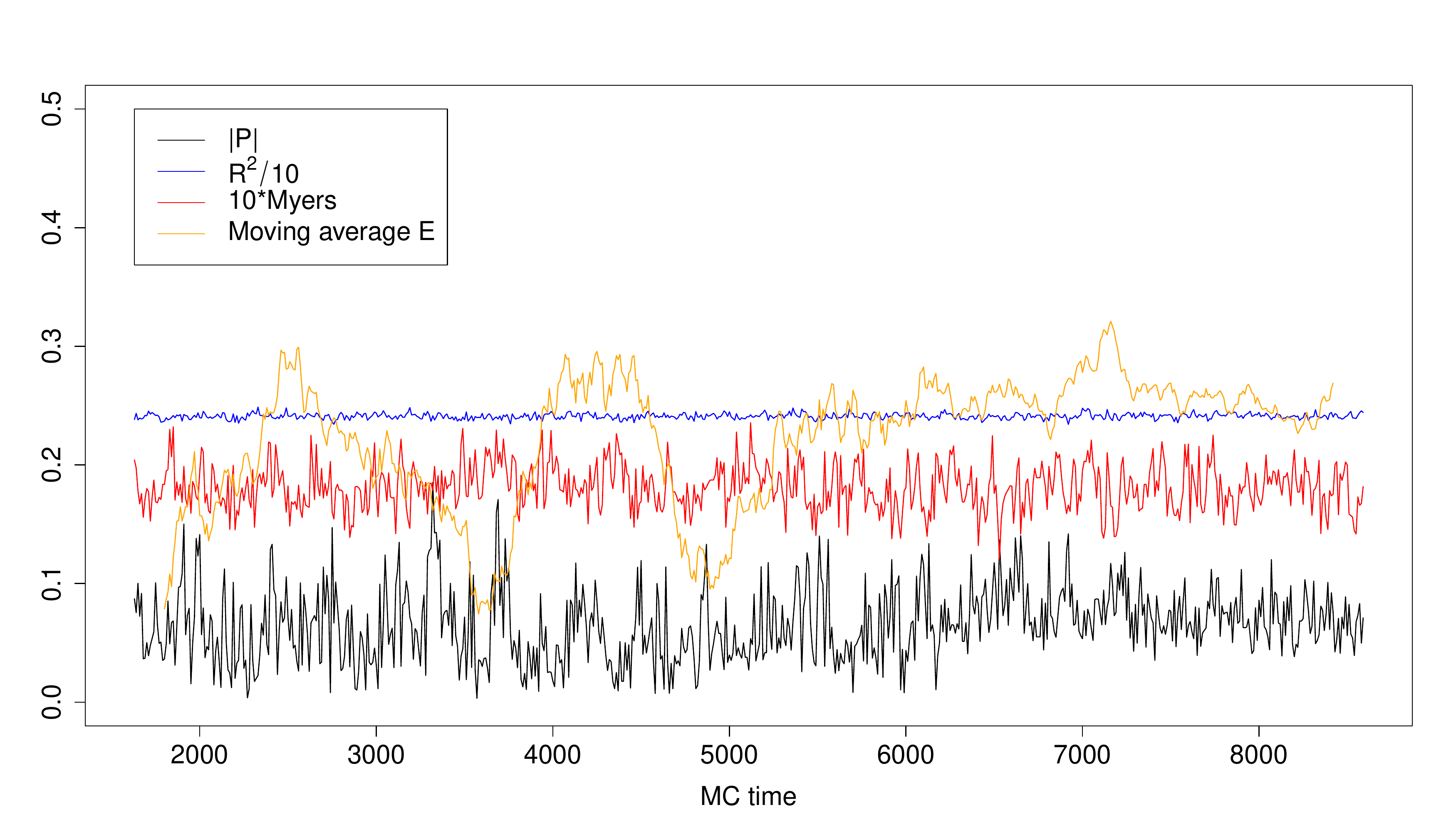}
	\includegraphics[scale=0.21]{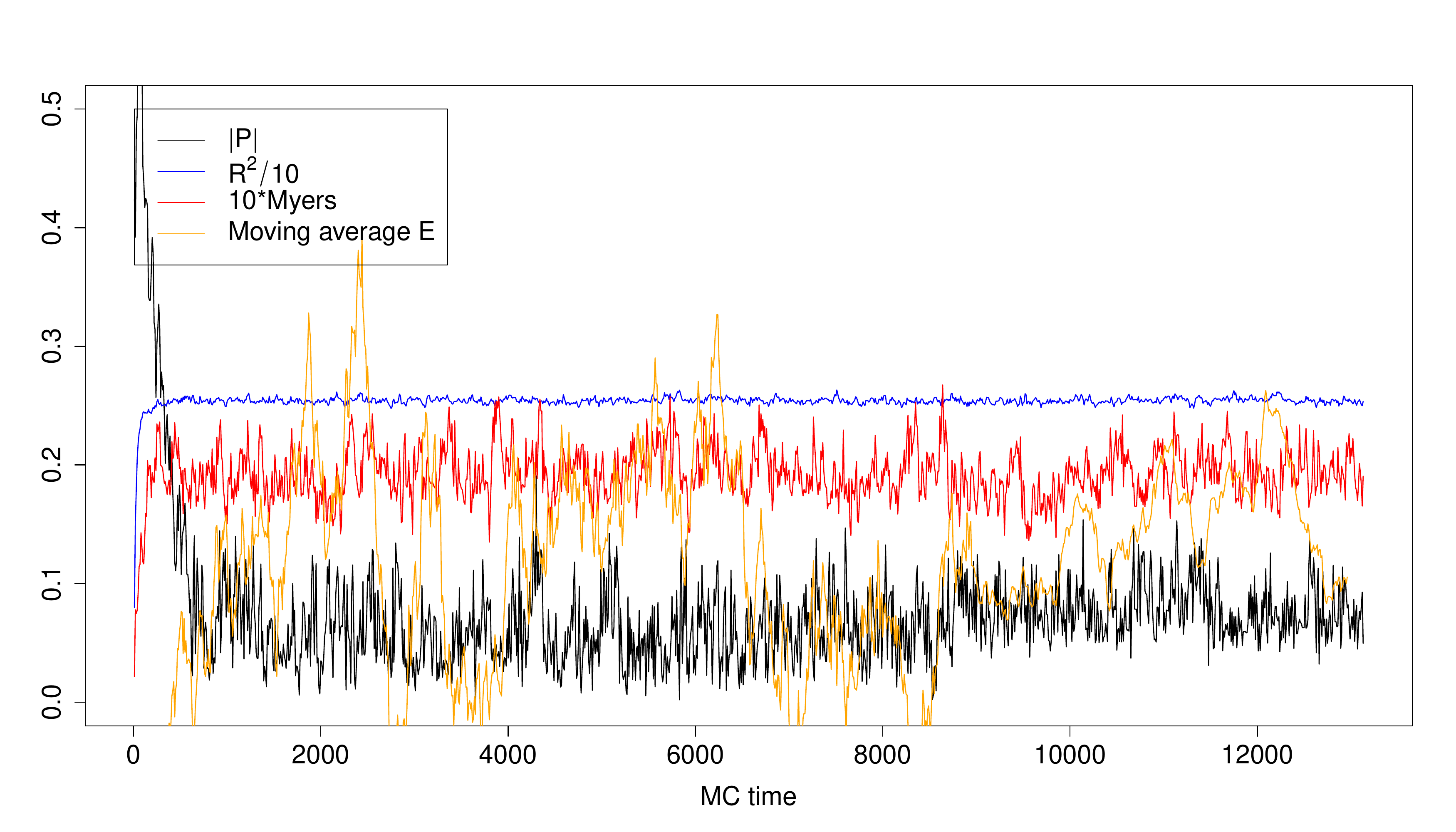}
	\caption{Typical Monte Carlo histories for gauged data. [Up]: from left to right $\mu=2, T=0.2, N=12, L=48$ and $\mu=4, T=0.4$ for same $N,L$. [Bottom]: from left to right $\mu=2, T=0.2, N=16, L=48$ and same $\mu,T$ for $N=16, L=96$.}
	\label{fig:MC_histories}
\end{figure}

Some representative Monte Carlo histories from the data we used are shown in Fig.~\ref{fig:MC_histories}. Similar Monte Carlo histories appear in all the range of $\mu$ and temperatures we have investigated indicating the high statistics analysis.  

\section{Miscellaneous results on BMN matrix model}

\subsection{Hamiltonian splitting in perturbative regime}\label{sec:Hamiltonian_splitting}
 When we Legendre transform \eqref{BMNaction} and take the large $\mu$ limit, the Hamiltonian splits into a free and an interacting part which decouple from each other
 \be 
 H=H_0+H_{\rm int},
 \ee 
 where 
 \begin{align}
 H_0=&\frac{N}{\lambda}\Tr\left[\left(\frac{\lambda}{N}\right)^2\frac{1}{2}\left(\Pi^M\right)^2+\frac{\mu^2}{2}\left(X^i\right)^2+\frac{\mu^2}{8}\left(X^a\right)^2-\frac{3\mu}{4}i\bar{\psi}^\alpha\gamma_{123}\psi_\alpha\right],\\
 H_{\rm int}=&\frac{N}{\lambda}\Tr\left[-\mu i\epsilon_{ijk}X^iX^jX^k+\frac{1}{4}[X^M,X^N]^2+\bar{\psi}^\alpha\gamma^M[X^M,\psi_\alpha]\right],
 \end{align}
 with $\Pi^M=\frac{\delta\mathcal{L}}{\delta \dot{X}^M}$ being the conjugate momenta for bosonic matrices. The $H_0$ terms construct the free $U(1)$ sector of the model while $H_{\rm int}$ denotes the interactive $SU(N)$ part. In the large $\mu$ limit, the interactive sector can be treated perturbatively while the free sector is claimed to be protected from contributions to all orders in $\mu$  \cite{Dasgupta2002, Dasgupta:2002ru, Kim:2002if, Kim:2002zg}.  In addition we can introduce harmonic oscillators defined by the operators
 \begin{align}\label{eq:oscillators_1}
 A_i:=&\sqrt{\frac{1}{\mu}}\left(\frac{\lambda}{N}\frac{\Pi_i}{\sqrt{2}}-\frac{i\mu}{\sqrt{2}}\sqrt{\frac{N}{\lambda}}X_i\right),\\
 \label{eq:oscillators_2}B_a:=&\sqrt{\frac{2}{\mu}}\left(\frac{\lambda}{N}\frac{\Pi_a}{\sqrt{2}}-\frac{i\mu}{2\sqrt{2}}\sqrt{\frac{N}{\lambda}}X_a\right)
 \end{align}
 which obey canonical commutation relations 
 \begin{align} 
 [A_i,A^\dagger_j]=\delta_{ij}\quad,\quad
 [B_a,B^\dagger_b]=\delta_{ab}.
 \end{align} 
 Then, the bosonic part of the free Hamiltonian results in 
 \be 
 H_0^{\rm bosonic}=\frac{N}{\lambda}\Tr\left[\mu A_i^\dagger A_i+\frac{\mu}{2}B_a^\dagger B_a\right] 
 \ee 
 
 \noindent
 Complexifying the real spinor matrices $\psi_\alpha$ using 
 \be 
 \psi^\pm=\mathcal{C}^\pm \psi\quad \text{where} \quad \mathcal{C}^\pm=\frac{1}{2}\left(\mathds{1}\pm i\gamma_{123}\right),
 \ee  
 yields the anticommutation relations 
 \be 
 \{\psi^{+\alpha},\psi^{-\beta}\}=\frac{1}{2}\left(\mathcal{C}^+\right)_{\alpha\beta}\quad,\quad \{\psi^{+\alpha},\psi^{+\beta}\}=0=\{\psi^{-\alpha},\psi^{-\beta}\}.
 \ee 
 We may now use the chirality property of the complexified fermions $(i\gamma_{123})\psi^\pm=\pm\psi^\pm$ accompanied with $\left(\mathcal{C}^\pm\right)^2=\mathcal{C}^\pm$ and $\mathcal{C}^+\mathcal{C}^-=0$. Recalling the splitting  $\psi=\psi^++\psi^-$ the fermionic part results in 
 \be 
 H_0^{\rm fermionic}=-\frac{3\mu}{4}\Tr\left[ {i\bar{\psi}^\alpha\gamma_{123}\psi_\alpha}\right]=\frac{3\mu}{2}\Tr\left[\psi^{+\alpha}\psi^{-\alpha}\right].
 \ee 
 Summing both we get the $U(1)$ free part of the Hamiltonian written in terms of bosonic and fermionic harmonic oscillators 
 \be
 H_0=\frac{N}{\lambda}\Tr\left[\mu A_i^\dagger A_i+\frac{\mu}{2}B_a^\dagger B_a+\frac{3\mu}{2}\psi^{+\alpha}\psi^{-\alpha}\right].
 \ee 
 In our conventions, the $SO(3)$ and $SO(6)$ sectors have mass $\mu$ and $\mu/2$ respectively while the fermions have mass $3\mu/4$ \footnote{We recall the splitting $\psi=\psi^++\psi^-$.}. The zero-energy ground state of this free Hamiltonian is denoted as $\ket{0}$ and is annihilated by 
 \be 
 A_i\ket{0}=0\quad,\quad B_a\ket{0}=0\quad,\quad \psi^{-\alpha}\ket{0}=0.
 \ee  
 The $U(1)$ free sector of the above Hamiltonian is spanned by excitations of operators of word-length one, e.g $\Tr[A_i^\dagger],\Tr[B_a^\dagger],\Tr[\bar{\psi}^\alpha]$ while the $SU(N)$ free sector is spanned by operators of word-length two and larger. The lightest mode is given by the $SO(6)$ part of the free Hamiltonian with lowest energy 
 \be \label{eq:lowest_adjoint}
 E_{SO(6)}=\frac{\mu}{2}.
 \ee  
 For the gauged model this is the first excited state created by $\Tr{B^\dagger_a}\ket{0}$, while the remaining excitations of the free sector are shown in Table~\ref{vacuadegentable}. For the ungauged model, one can simply act with $B^\dagger_a$ on $\ket{0}$ which results in the same energy. Six oscillators are yielding a six-fold degeneracy of this sector. In addition there are $n=3$ oscillators for the $SO(3)$ part with energy $E_{SO(3)}=\mu$ and $n=8$ fermionic oscillators with energy $E_{\rm fermions}=3\mu/4$. The ground state energy vanishes since due to supersymmetry we have 
 \be 
 3\mu-8\cdot\frac{3\mu}{4}+6\cdot\frac{\mu}{2}=0.
 \ee  
 The lowest adjoint mode of the gauged theory in the perturbative limit $\mu\to\infty$ is \eqref{eq:lowest_adjoint} and perturbation theory shows that it is protected at least to first order in $\mu$ \cite{Maldacena:2018vsr, Dasgupta:2002ru, Kim:2002if}.
 
 The spectrum of the free Hamiltonian has been studied in perturbation theory (with perturbative parameter $\sim1/\mu$, e.g \eqref{eq:dimensionless_coupling_mu}) in \cite{Kim:2003rza,Kim:2002if, Dasgupta:2002ru}  and their energy, representations and degeneracy are given in Table ~\ref{vacuadegentable}. It was conjectured that the free spectrum does not receive any perturbative corrections to all orders in $\mu^{-1}$. On the other hand, short representations of the $SU(N)$ sector can, in principle, combine and form multiplets, and indeed they may receive perturbative corrections. On top of that we may also note that there could be non-perturbative corrections \cite{ Dasgupta:2002ru}. A precise analysis of the form of non-perturbative and perturbative corrections for the construction of multiplets has not been done and, therefore, the energy correction can not be estimated precisely.  Whether or not this is something that can be justified analytically we do not know, because also the non-perturbative corrections can not be estimated at all. 
 
\begin{center}
	\begin{tabular}{|c|c|c|c|}
		\hline
		\text{state} & $SO(6)\times SO(3)$ \text{reps.}  & \text{energy} & \text{degeneracy}\\
		\hline 
		$\ket{0}$ &(\textbf{1,1})  & 0 & 1 \\
		$\Tr B^\dagger_a\ket{0} $ & (\textbf{6,1})  & $\frac{\mu}{2}$ & 6 \\
		$\Tr \psi^\dagger_{M\alpha}\ket{0} $ & (\textbf{$\bar{\textbf{4}}$,2})  & $\frac{3\mu}{4}$ & 8 \\
		$\Tr A^\dagger_i\ket{0}$ & (\textbf{1,3})  & $\mu$ & 3 \\
		$\Tr B^\dagger_a\Tr B^\dagger_b\ket{0} $ & \textbf{(1,1)+(20,1)}  & $\mu$ & 1+20 \\
		$\Tr \left(B^\dagger_aB^\dagger_b\right)-\frac{1}{N}\Tr B^\dagger_a\Tr B^\dagger_b\ket{0} $ & \textbf{(1,1)+(20,1)}  & $\mu$ & 1+20 \\
		\hline
	\end{tabular}
	\captionof{table}{Lowest energy states for the trivial background $X=0$, their representations and  degeneracy. The first five lines correspond to the $U(1)$ part of the model which is free. }
	\label{vacuadegentable}
\end{center}

\subsection{Representation algebra of the BMN model}\label{app:rep_algebra}
Finite $\mu$ corrections to $n_E$ are not known. We observed that they change with $\mu$ and here we can ask whether or not the works of \cite{Kim:2002zg} and \cite{Dasgupta:2002ru} could provide some insight. On the other hand, finite $\mu$ corrections to $C_{\rm adj}$ depend on the mixing of states, presumably coming from the interacting sector of the model whose precise form is also unknown. 

Let us highlight some of these results in the literature. The classification of the superalgebra in the plane wave limit of M-theory has been considered in \cite{Kim:2002zg}. Indeed, in this work and building on the results from Kac \cite{Kac}, it was shown that the complexification of the special unitary Lie superalgebras $\mathfrak{su}(2|4; 2,0)$ (for $\mu>0$) or $\mathfrak{su}(2|4;2,4)$ (for $\mu<0$) results in $\mathbf{A}(1,3)\cong \mathfrak{sl}(2,5)$. However, here we are always in a scenario of positive $\mu$, and hence the bosonic part of the algebra is given as $\mathfrak{su}(2,0)\oplus \mathfrak{su}(0,4)\cong \mathfrak{so}(3)\oplus \mathfrak{so}(6)$.

The peculiar supersymmetry of the BMN model manifests itself in the time dependence of the supersymmetry transformations. In the large $\mu$ limit the Hamiltonian splits into a free and interacting part (see Appendix~\ref{sec:Hamiltonian_splitting}). The commutator between the interacting Hamiltonian and the supersymmetric charge is proportional to the supersymmetric charge. This implies that the boson and fermion masses differ. In particular, the energy level difference given by the commutator between the supercharges $Q_\alpha$ and the Hamiltonian in the interacting $SU(N)$ sector
\be 
[H_{\rm int},Q_\alpha]=\frac{\mu}{4}Q_\beta\gamma_{\beta\alpha}^{123}+\Tr(\psi_\alpha \mathcal{G}),
\ee 
is $\frac{\mu}{4}$ in our conventions whenever the Gauss constraint is satisfied. The application of the supersymmetry charges $Q^{\pm}_{\alpha}$, where one performs a chirality split as 
\be 
Q^\pm=\mathcal{C}^\pm Q\quad,\quad \mathcal{C}^\pm=\frac{1}{2}\left(\mathds{1}\pm i\gamma_{123}\right),
\ee 
can be applied at most eight times for $Q^+$ or eight times for $Q^-$ changing the energy level by $Q^\pm\ket{state}=\mp\frac{\mu}{4}\ket{state}$. This results in a multiplet consisting of $2^8=256$ states with irreducible representations at each level. Thus, one can change irreducible representations by acting with $Q^\pm$ resulting in a shift of the energy at each level. This energy shift coincides with the difference between bosonic and fermionic masses $\mu-\frac{3\mu}{4}=\frac{\mu}{4}=\frac{3\mu}{4}-\frac{\mu}{2}$ in the $SO(3)$ and $SO(6)$ sectors respectively. The successful application of a supercharge $Q_\alpha$ on a state leads to changing one $u(1)\oplus su(2)\oplus su(4)$ irreducible representation to another\cite{Kim:2002zg}. 
The multiplets (energy levels) split into a specific positive integer given by $m$. The energy of each multiplet differs from the vacuum ($E_0$) as \cite{Kim:2002zg}
\be\label{eq:energy_shifts}
E=E_0+\frac{\mu m}{4}~,~\quad\text{with}\quad m=0,1,\cdots,8.
\ee 
Each of the zeroth and highest level multiplets correspond to an irreducible representation of $u(1)\oplus su(2)\oplus su(4)$, while others are in general reducible representations.
Therefore the maximum shift for the gauged theory comes from the highest energy level with $m=8$ corresponding to eight applications of $Q^-$.  For $\mu>0$, $m=0$ has the lowest energy $E=E_0$ representing the ground state.  This is the situation for the gauged model. 

On the other hand for the ungauged model, one can perform a shift in the Hamiltonian 
\be 
H^{\rm new}=H-\Tr\left(X^1\mathcal{G}\right),
\ee 
accompanied by the condition $(\gamma^1+\mathds{1})\epsilon=0$. This is a supersymmetric deformation of the ungauged theory and in \cite{Maldacena:2018vsr} it was shown that this shift compensates the gauge condition in the superalgebra resulting again in 
\be
[H^{\rm new},Q_\alpha]=\frac{\mu}{4}Q_\beta\gamma_{\beta\alpha}^{123}.
\ee 

We expect that in the ungauged model, extra degrees of freedom living on the boundary of the theory and interpreted as open strings \cite{Maldacena:2018vsr} (which in principle can reach deep into the bulk) may contribute in the partition function. These degrees of freedom transform in the adjoint representation. Therefore, one is led to ask what is the adjoint representation of the superalgebra of the BMN model. This question has been answered in \cite{Kim:2002zg} and the result is given in Table~\ref{AdjointofA}
\begin{center}
	\begin{tabular}{c|c}
		\text{Energy} & \text{Representations} \\
		\hline 
		$+\mu/4$ &(\textbf{2,4})  \\
		$0$ & $(\textbf{1,1)}\oplus(\textbf{3,1})\oplus(\textbf{1,15})$  \\
		$-\mu/4$ & $(\textbf{2},\bar{\textbf{4}})$   \\
		\hline
	\end{tabular}
	\captionof{table}{Adjoint representation of $\mathbf{A}(1,3)$}
	\label{AdjointofA}
\end{center}

\bibliographystyle{JHEP}
\bibliography{matrix-model}

\end{document}